\documentclass[acmsmall,screen]{acmart}

\usepackage{booktabs}
\usepackage{graphicx}
\usepackage{amsmath}
\usepackage{amssymb}
\usepackage{algorithmic}
\usepackage{algorithm}
\usepackage{subcaption}
\usepackage{url}
\usepackage{siunitx}
\usepackage{xspace}
\usepackage{listings}
\usepackage{microtype}
\usepackage{multirow}
\usepackage{multicol}
\usepackage{tabularx}
\usepackage{mathtools}
\usepackage{enumitem}
\usepackage{adjustbox}
\usepackage{wrapfig}

\setcopyright{rightsretained}
\acmPrice{}
\acmDOI{10.1145/3276493}
\acmYear{2018}
\copyrightyear{2018}
\acmJournal{PACMPL}
\acmVolume{2}
\acmNumber{OOPSLA}
\acmArticle{123}
\acmMonth{11}
\startPage{1}

\definecolor{todocolor}{rgb}{0.8,0,0}
\definecolor{keywordcolor}{rgb}{0.5,0,0.5}

\newcommand{\figref}[1]{Figure~\ref{fig:#1}}
\newcommand{\figsref}[2]{Figures~\ref{fig:#1} and~\ref{fig:#2}}
\newcommand{\threefigsref}[3]{Figures~\ref{fig:#1}, ~\ref{fig:#2}, and~\ref{fig:#3}}
\newcommand{\subfigref}[1]{(\subref{fig:#1})}

\newcommand{\secref}[1]{Section~\ref{sec:#1}}

\newcommand{\tabref}[1]{Table~\ref{tab:#1}}
\newcommand{\HIDE}[1]{}

\ifdefined\notodo
\newcommand{\TODO}[1]{}
\else
\newcommand{\TODO}[1]{{\color{todocolor}#1}}
\fi

\newcommand{\yes}{\checkmark}
\newcommand{\no}{}
\newcommand{\choice}{(\yes)}
\newcolumntype{R}[2]{%
  >{\adjustbox{angle=#1,lap=\width-(#2)}\bgroup}%
  l%
  <{\egroup}%
}

\definecolor{textgray}{gray}{0.4}
\lstdefinestyle{levelfunc}{
  language=C++,
  mathescape,
  frame=none,
  aboveskip=\medskipamount,
  belowskip=\medskipamount,
  columns=flexible,
  basicstyle=\fontsize{8}{8}\ttfamily,
  keywordstyle=\color{keywordcolor},
  commentstyle=\color{gray},
  showstringspaces=false,
}
\newcommand\code[1]{\lstinline[columns=fullflexible, mathescape=true, basicstyle=\ttfamily]|#1|}

\newcommand{\taco}{\code{taco}\xspace}

\SetEnumitemKey{twocol}{
  before=\raggedcolumns\setlength{\multicolsep}{\topsep}\begin{multicols}{2},
  after=\end{multicols}
}

\begin{document}

\title{Format Abstraction for Sparse Tensor Algebra Compilers}

\author{Stephen Chou}
\affiliation{
  \institution{MIT CSAIL}
  \streetaddress{32-G778, 32 Vassar Street}
  \city{Cambridge}
  \state{MA}
  \postcode{02139}
  \country{USA}
}
\email{s3chou@csail.mit.edu}

\author{Fredrik Kjolstad}
\affiliation{
  \institution{MIT CSAIL}
  \streetaddress{32-G778, 32 Vassar Street}
  \city{Cambridge}
  \state{MA}
  \postcode{02139}
  \country{USA}
}
\email{fred@csail.mit.edu}

\author{Saman Amarasinghe}
\affiliation{
  \institution{MIT CSAIL}
  \streetaddress{32-G744, 32 Vassar Street}
  \city{Cambridge}
  \state{MA}
  \postcode{02139}
  \country{USA}
}
\email{saman@csail.mit.edu}

\begin{abstract}

  This paper shows how to build a sparse tensor algebra compiler that is
  agnostic to tensor formats (data layouts).  We develop an interface that
  describes formats in terms of their capabilities and properties, and show how
  to build a modular code generator where new formats can be added as plugins.
  We then describe six implementations of the interface that compose to form
  the dense, CSR/CSF, COO, DIA, ELL, and HASH tensor formats and countless
  variants thereof.  With these implementations at hand, our code generator can
  generate code to compute any tensor algebra expression on any combination of
  the aforementioned formats.

  To demonstrate our technique, we have implemented it in the~\taco{} tensor
  algebra compiler.  Our modular code generator design makes it simple to add
  support for new tensor formats, and the performance of the generated code is
  competitive with hand-optimized implementations.  Furthermore, by
  extending~\taco{} to support a wider range of formats specialized for
  different application and data characteristics, we can improve end-user 
  application performance.  For example, if input data is provided in the COO
  format, our technique allows computing a single matrix-vector multiplication
  directly with the data in COO, which is up to 3.6$\times$ faster than by
  first converting the data to CSR.
  
\end{abstract}



\maketitle

\section{Introduction}
\label{sec:introduction}
Tensor algebra is a powerful tool to compute on multidimensional data, with
applications in machine learning~\cite{tensorflow}, data
analytics~\cite{anandkumartensor}, physical sciences~\cite{feynman1963}, and
engineering~\cite{Kolecki2002b}.  Tensors generalize matrices to any number of
dimensions and are often large and sparse, meaning most components are zeros.
To efficiently compute with sparse tensors requires exploiting their sparsity
in order to avoid unnecessary computations.

Recently, \citet{kjolstad2017} proposed \taco{}, a compiler for sparse tensor
algebra.  \taco{} takes as input a tensor algebra expression in high-level
index notation and generates efficient imperative code that computes the
expression.  Their compiler technique supports tensor operands stored in any
format where each dimension can be described as dense or sparse.  This
encompasses tensors stored in dense arrays as well as variants of the
compressed sparse row (CSR) format.

These are, however, only some of the tensor formats that are used in practice.
Important formats that are not supported by \taco{} include the coordinate
(COO) format, the diagonal (DIA) format, the ELLPACK (ELL) format, and hash
maps (HASH), along with countless blocked and high-dimensional variants and
compositions of these.  Each format is important for different reasons.  COO is
the natural way to enumerate sparse tensors and is often the format in which
users provide data~\cite{frostt-format}.  It is not the most performant format,
but when a tensor will only be used once, it is often more efficient to compute
directly with COO than to convert the data to another format.  ELL exposes
vectorization opportunities for SpMV and is useful for matrices that contain a
bounded number of nonzeros per row, such as matrices from well-formed meshes.
DIA has the added benefit of being very compact and is suited to matrices that
compute stencils on Eulerian grids and images.  Hash maps support random access
without having to explicitly store zeros and can be useful for multiplications
involving very sparse operands.  An application may need any, or even several,
of these formats, making it important to support computing with all and any
combination of them.

\begin{figure}
  \begin{minipage}{0.32\linewidth}
    \centering
    \includegraphics{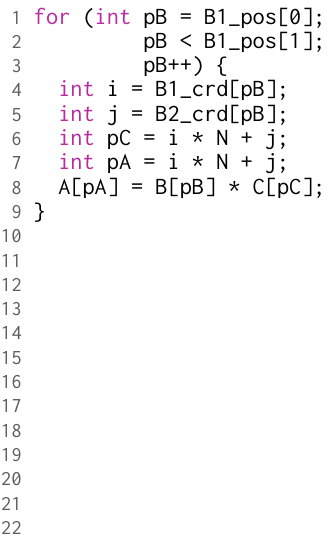}
    \subcaption{
      $B$ is COO, $C$ is dense array
    }
    \label{fig:elwise-multiplication-coo-dense}
  \end{minipage}
  \hfill
  \begin{minipage}{0.33\linewidth}
    \centering
    \includegraphics{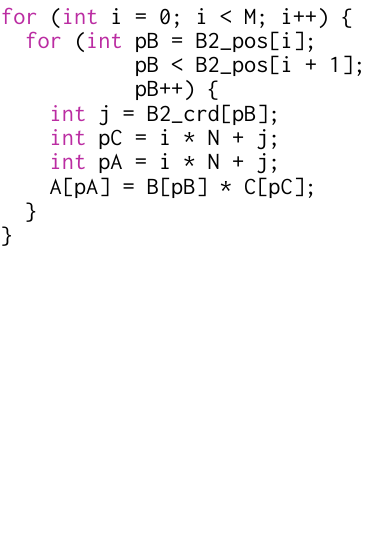}
    \subcaption{
      $B$ is CSR, $C$ is dense array
    }
    \label{fig:elwise-multiplication-csr-dense}
  \end{minipage}
  \hfill
  \begin{minipage}{0.33\linewidth}
    \centering
    \includegraphics{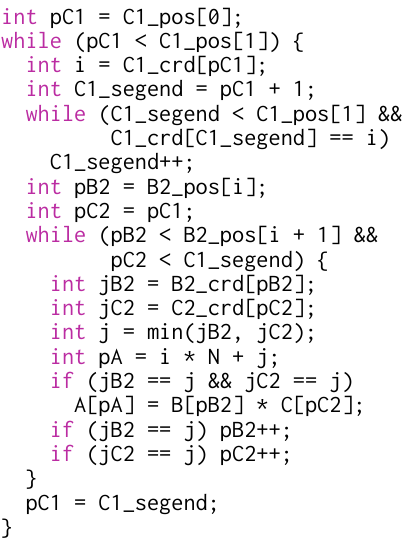}
    \subcaption{
      $B$ is CSR, $C$ is COO
    }
    \label{fig:elwise-multiplication-csr-coo}
  \end{minipage}
  \caption{
    Code to compute the component-wise product of two matrices stored in
    varying formats.
  }
  \label{fig:elwise-multiplication-code}
\end{figure}

The approach of \citeauthor{kjolstad2017}~\shortcite{kjolstad2017} was
hard-coded for sparse dimensions that are compressed using the same arrays as
CSR.  By contrast, COO matrices store full coordinates, while DIA implicitly
encodes coordinates of nonzeros using very different data structures.  To
enable efficiently computing with any format, a tensor algebra compiler must
emit distinct code to cheaply iterate over each format.  COO, for instance,
needs a single loop that iterates over row and column dimensions together
(\figref{elwise-multiplication-coo-dense}), whereas CSR needs two nested loops
that each iterate over a dimension (\figref{elwise-multiplication-csr-dense}).
To efficiently compute with operands in multiple formats, a compiler must emit
code that concurrently iterates over the operands.
\figref{elwise-multiplication-csr-coo} shows, however, that this is not a
straightforward combination of code that individually iterate over the operand
formats.  Rather, a compiler must also emit distinct code for each combination
of formats to obtain good performance in all cases.  Because the number of
combinations is exponential in the number of formats though, one cannot readily
extend the approach of~\citet{kjolstad2017} to support additional formats by
directly hard-coding for them in the code generator.  We instead need a different 
approach that is modular with respect to formats.

We generalize the recent work on tensor algebra compilation to support a much
wider range of disparate tensor formats.  We describe a level-based abstraction
that captures how to efficiently access data structures for encoding tensor
dimensions, but that hides the specifics behind a fixed interface.  We develop
six per-dimension formats, all of which expose this common interface, that
compose to express all the tensor formats mentioned above.  Two of these
per-dimension formats---\emph{dense} and \emph{compressed}---were hard-coded
into \taco{}.  The other four---\emph{singleton}, \emph{range}, \emph{offset},
and \emph{hashed}---are new to this work and compose to express variants of
COO, DIA, ELL, HASH, and countless other tensor formats not supported
by~\citet{kjolstad2017}.
We then present a code generation algorithm that, guided by our abstraction,
generates efficient tensor algebra kernels that are optimized for operands
stored in any mix of formats.  The result is a powerful system that lets users
mix and match formats to suit their application and data, and which can be
readily extended to support new formats without modifying the code
generator.
In summary, our contributions are:
\begin{description}

  \item[Levelization] We survey many known tensor formats 
    (\secref{tensor-formats}) and show that they can be represented as
    hierarchical compositions of just six per-dimension level formats
    (\figref{formats-by-dimensions}).

  \item[Level abstraction] We describe an abstraction for level formats that
    hides the details of how a level format encodes a tensor dimension behind a
    common interface, which describes how to access the tensor dimension and
    exposes its properties (\secref{storage-abstraction}).

  \item[Modular code generation] We present a code generation technique that
    emits code to efficiently compute on tensors stored in any combination of
    formats, which reasons only about capabilities and properties of level
    formats and is not hard-coded for any specific format
    (\secref{code-generation}).

\end{description}

To evaluate the technique, we implemented it as an extension to the~\taco{}
tensor algebra compiler~\cite{kjolstad2017}.  We find that our technique emits
code that has performance competitive with existing sparse linear and tensor
algebra libraries.  Our technique also supports a much wider range of sparse
tensor formats than other existing libraries as well as the approach
of~\citet{kjolstad2017}.  This lets us improve the end-to-end performance of
applications that use~\taco{}.  For instance, our extension enables computing a
single matrix-vector multiplication directly with data provided in the COO
format, which is up to 3.6$\times$ faster than by first converting the data to
CSR (\secref{evaluation}).

\section{Tensor Storage Formats}
\label{sec:tensor-formats}
There exist many formats for storing sparse tensors that are used in practice.
Each format is ideal under specific circumstances, but none is universally
superior.  The ideal format depends on the structure and sparsity of the data,
the computation, and the hardware.  It is thus desirable to support computing
with as many formats as possible.  This is, however, made difficult by the need
for specialized code for every combination of operand tensor formats, even for
the same computation.

\subsection{Survey of Tensor Formats}
\label{sec:existing-tensor-formats}

\begin{figure*}
  \begin{minipage}[b]{0.28\linewidth}
    \centering
    \includegraphics[scale=0.25]{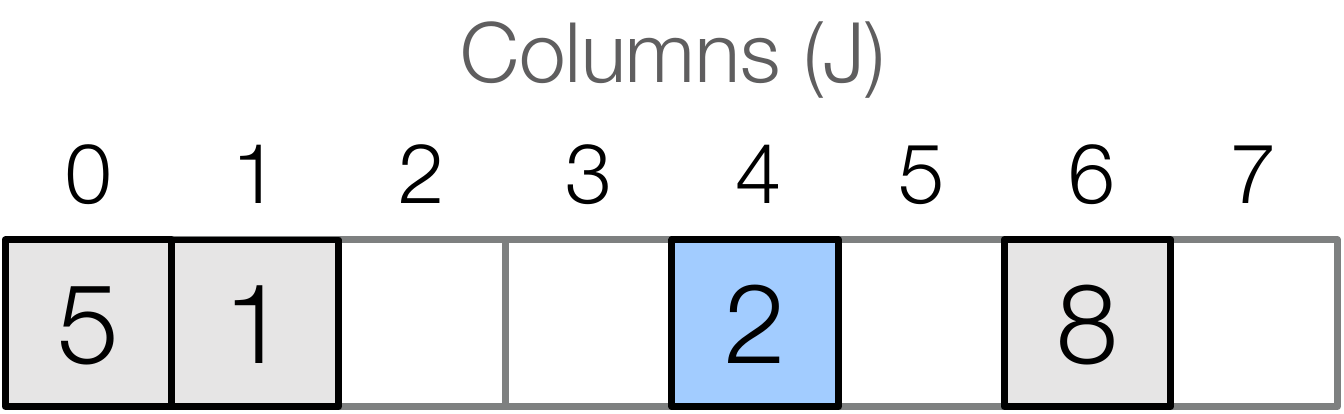}
    \subcaption{An 8-vector}
    \label{fig:vector-example}
    \vspace*{-2mm}
  \end{minipage}
  \begin{minipage}[b]{0.27\linewidth}
    \centering
    \includegraphics[scale=0.25]{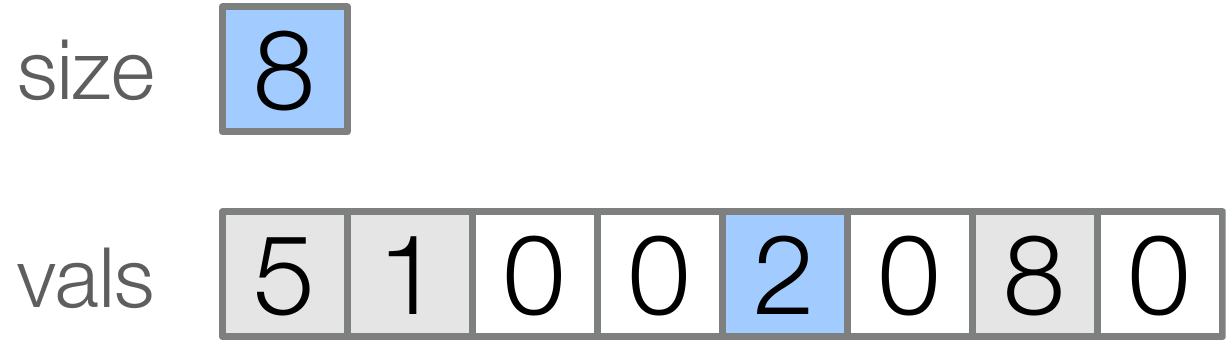}
    \subcaption{Dense array}
    \label{fig:vector-example-dense}
    \vspace*{-2mm}
  \end{minipage}
  \begin{minipage}[b]{0.21\linewidth}
    \centering
    \includegraphics[scale=0.25]{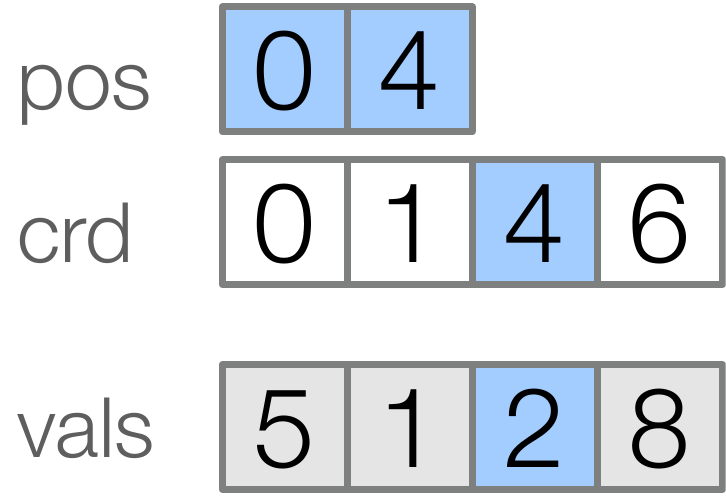}
    \subcaption{Sparse vector}
    \label{fig:vector-example-sparse}
    \vspace*{-2mm}
  \end{minipage}
  \begin{minipage}[b]{0.22\linewidth}
    \centering
    \includegraphics[scale=0.25]{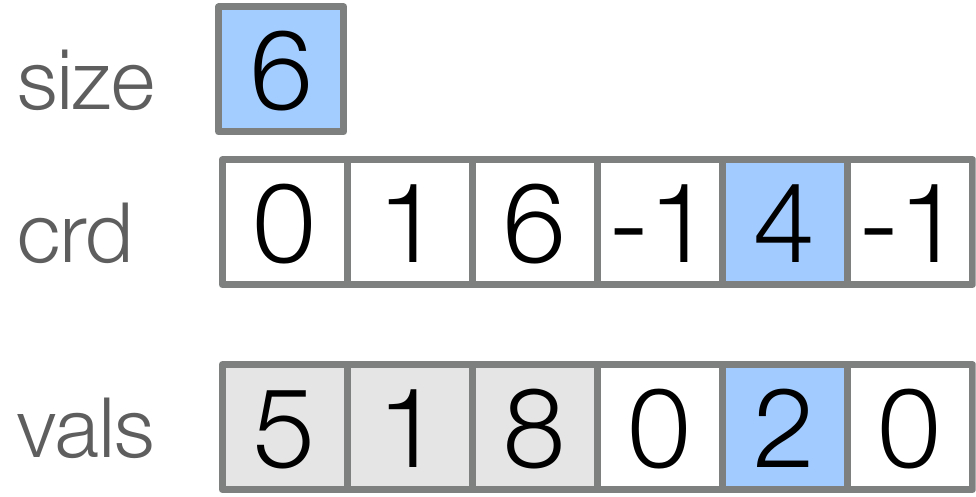}
    \subcaption{Hash map}
    \label{fig:vector-example-hashed}
    \vspace*{-2mm}
  \end{minipage}
  \vspace*{1mm}
  \rule{\linewidth}{0.4pt}
  \begin{minipage}[b]{0.25\linewidth}
    \centering
    \includegraphics[scale=0.25]{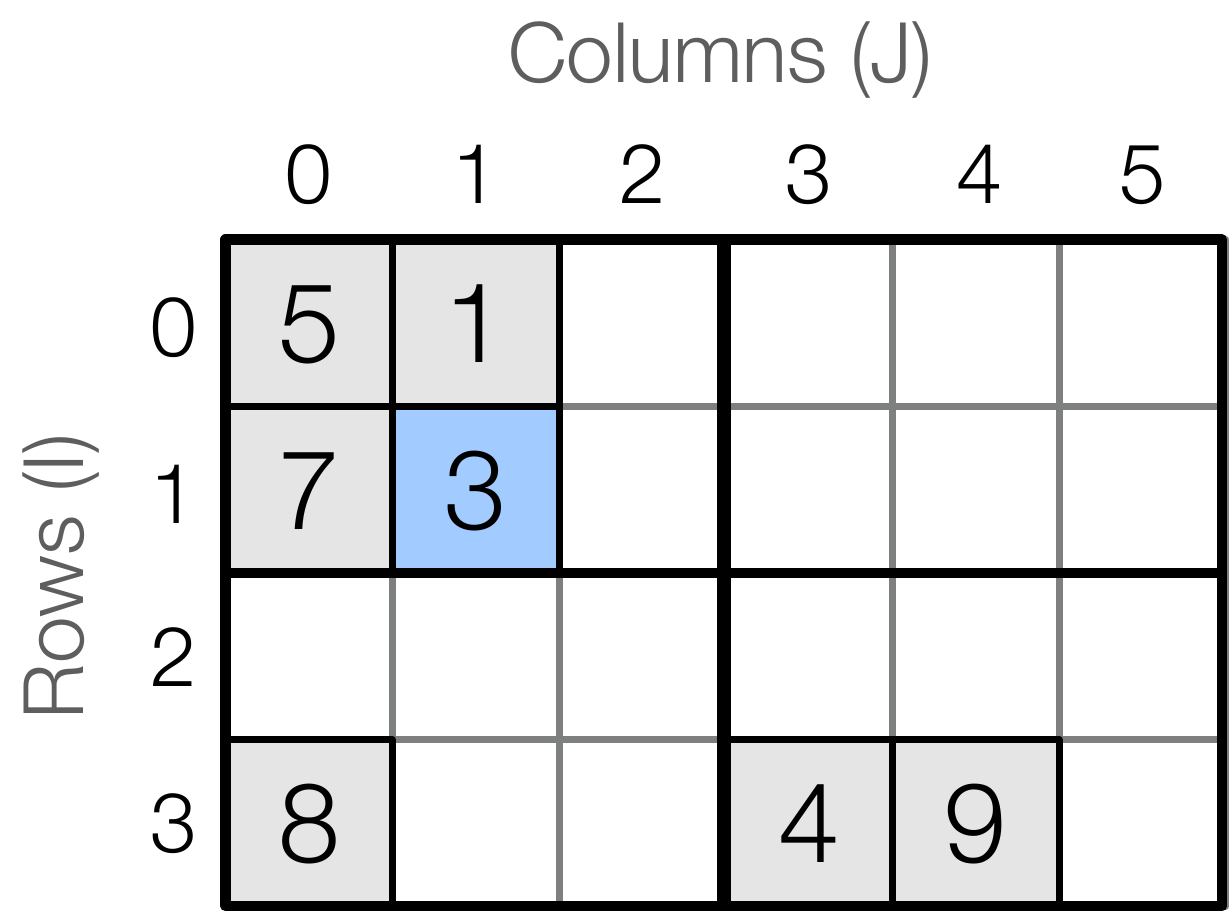}
    \subcaption{
      A 4$\times$6 matrix
    }
    \label{fig:matrix-example}
    \medskip
  \end{minipage}
  \begin{minipage}[b]{0.29\linewidth}
    \centering
    \includegraphics[scale=0.25]{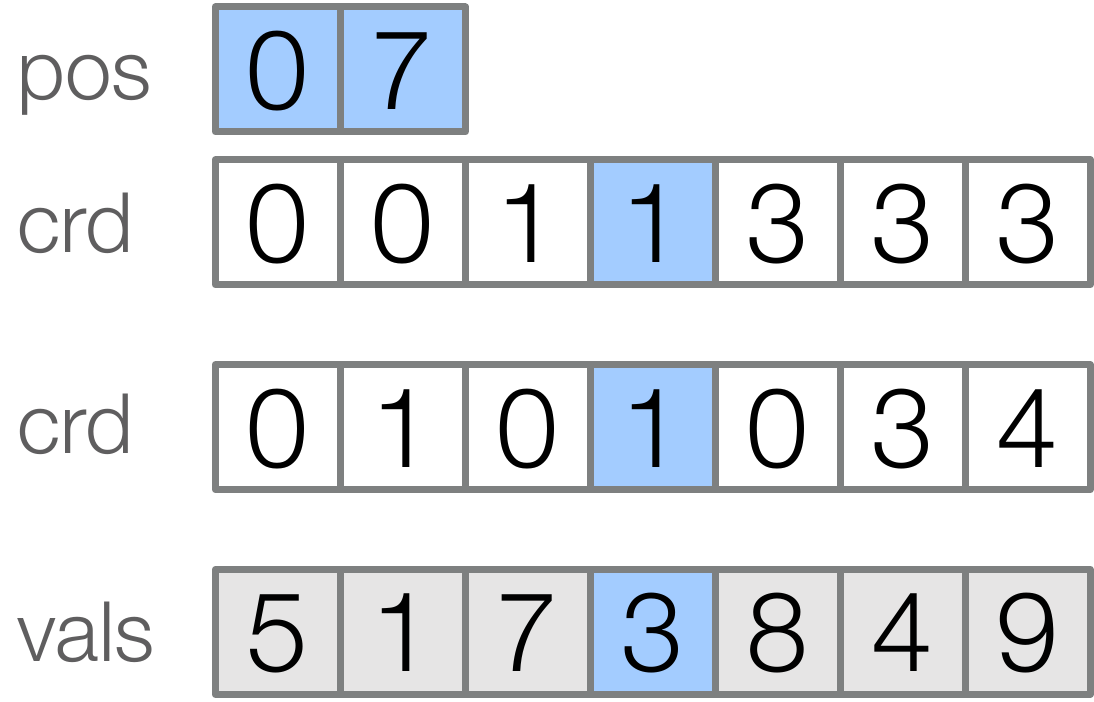}
    \subcaption{
      COO
    }
    \label{fig:matrix-example-coo}
    \medskip
  \end{minipage}
  \begin{minipage}[b]{0.22\linewidth}
    \centering
    \includegraphics[scale=0.25]{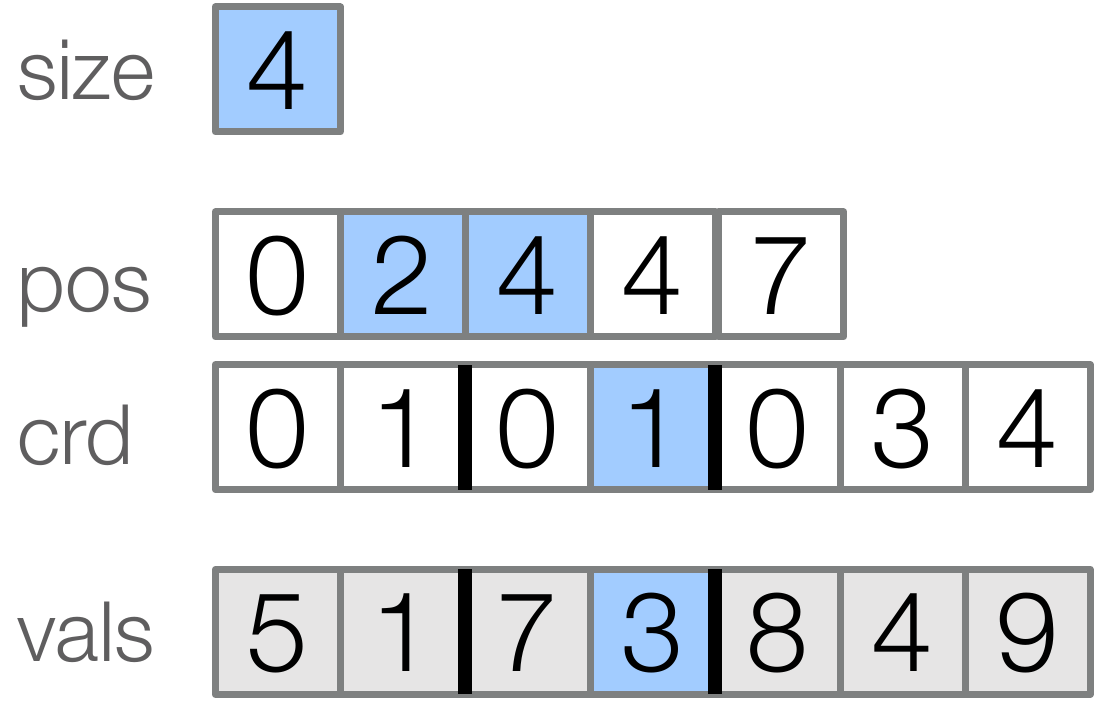}
    \subcaption{
      CSR
    }
    \label{fig:matrix-example-csr}
    \medskip
  \end{minipage}
  \begin{minipage}[b]{0.22\linewidth}
    \centering
    \includegraphics[scale=0.25]{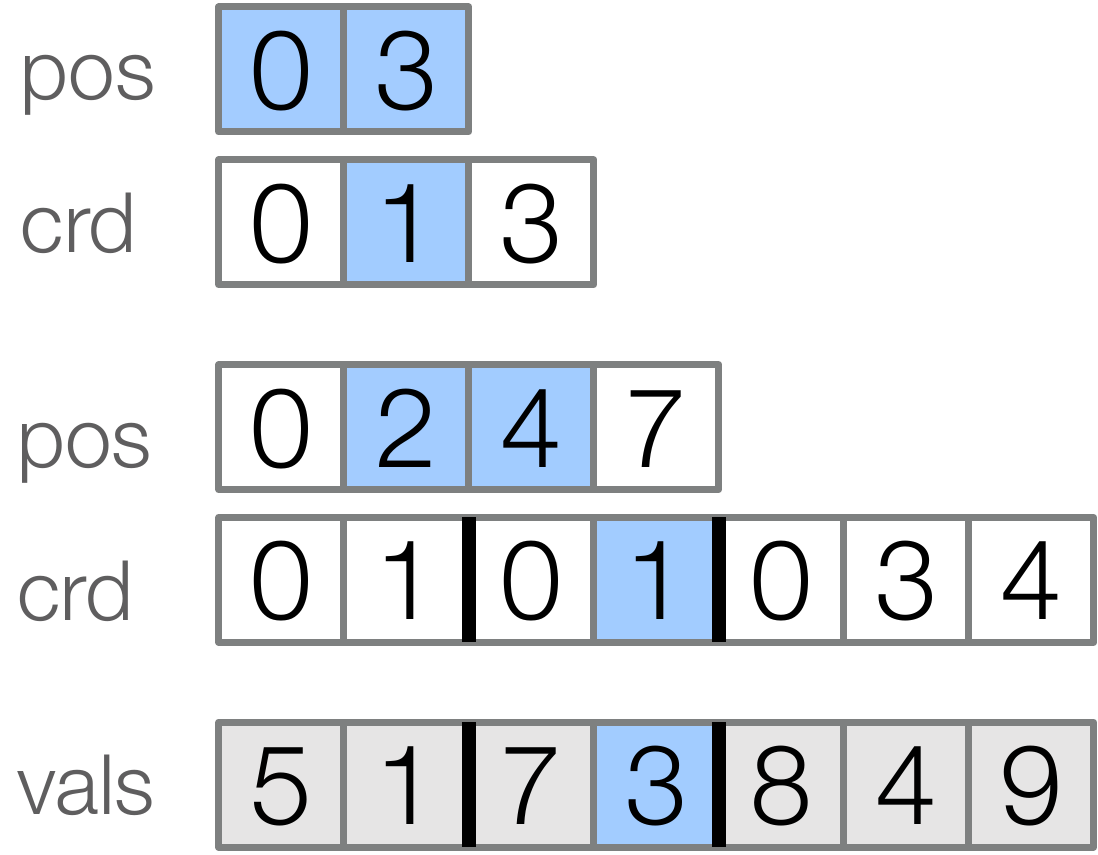}
    \subcaption{
      DCSR
    }
    \label{fig:matrix-example-dcsr}
    \medskip
  \end{minipage}
  \begin{minipage}[b]{0.32\linewidth}
    \centering
    \includegraphics[scale=0.25]{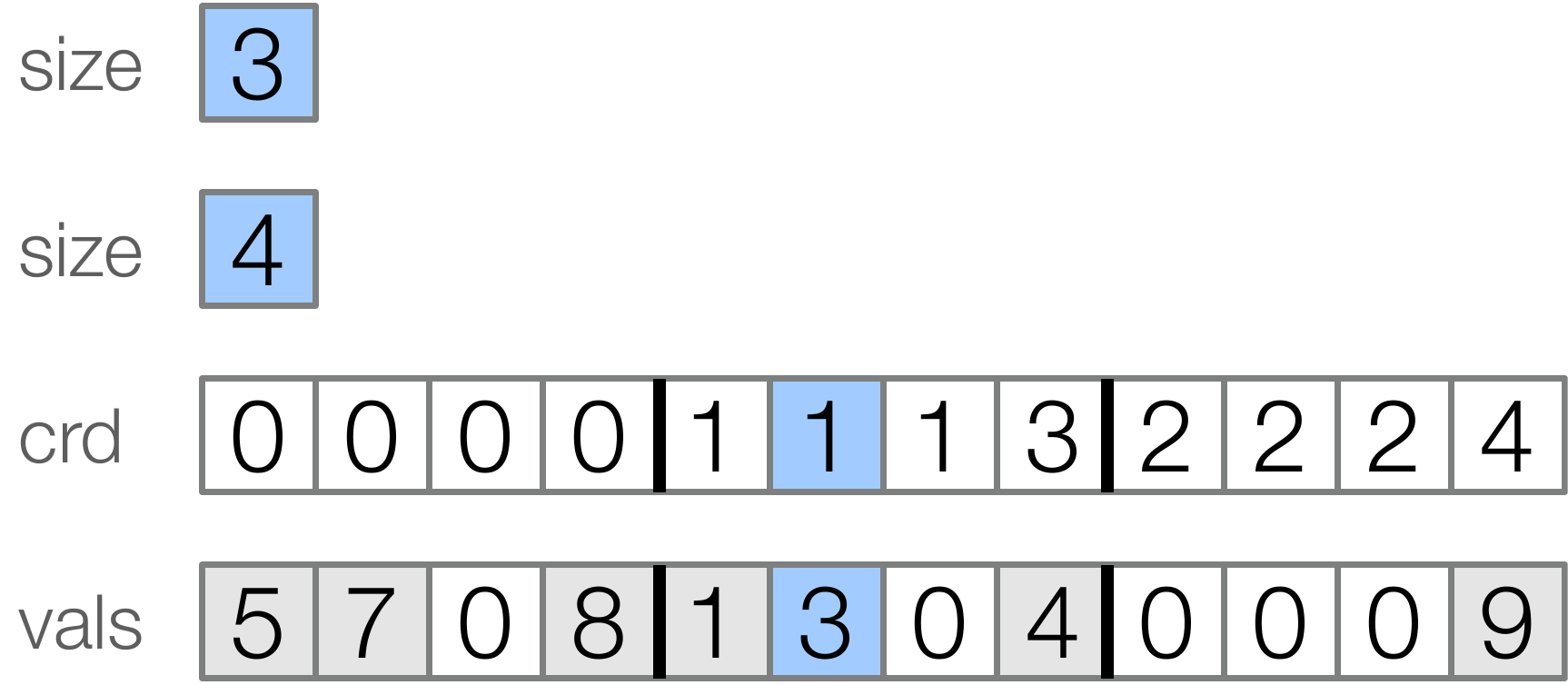}
    \subcaption{
      ELL
    }
    \label{fig:matrix-example-ell}
    \vspace*{-2mm}
  \end{minipage}
  \begin{minipage}[b]{0.24\linewidth}
    \centering
    \includegraphics[scale=0.25]{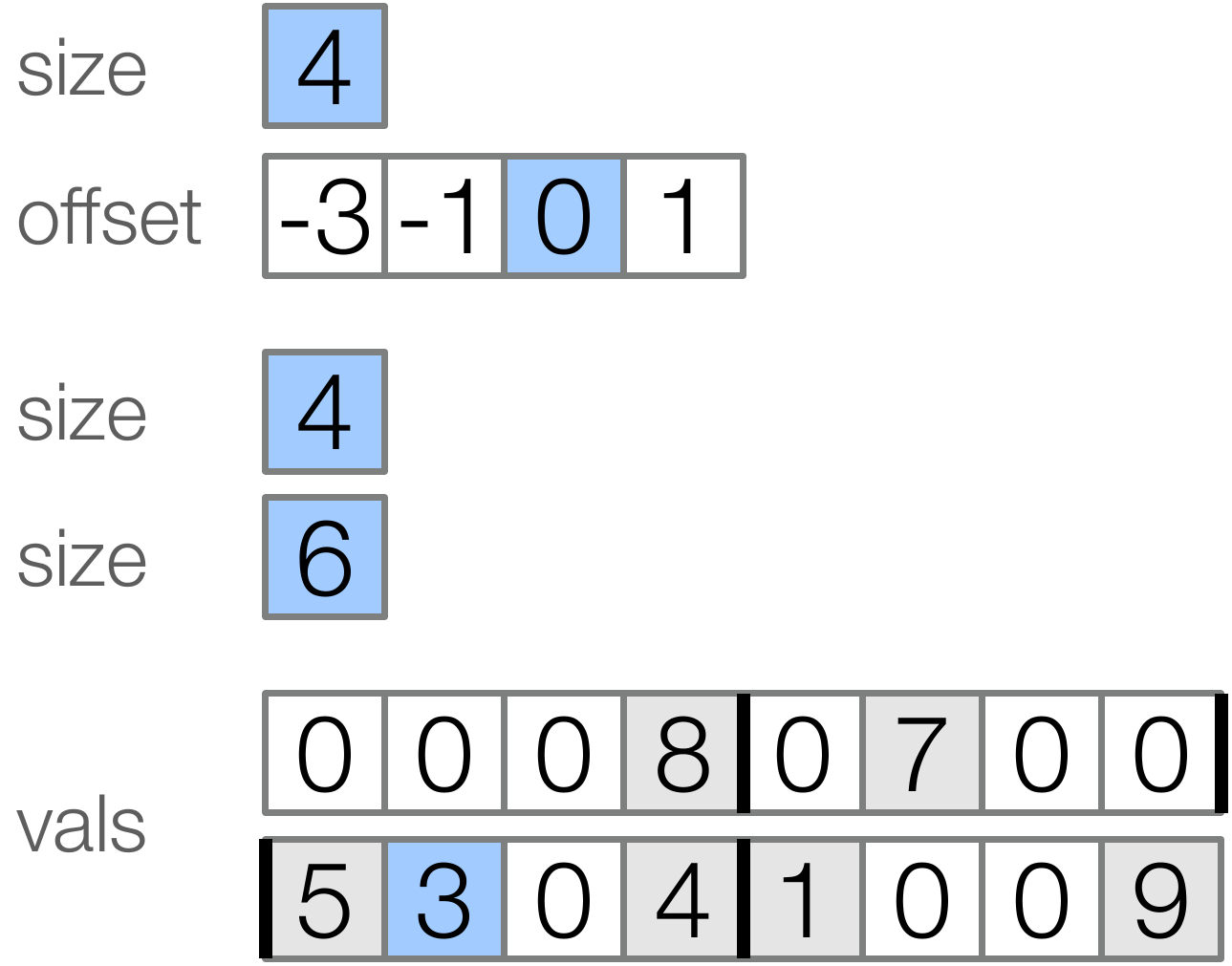}
    \subcaption{
      DIA
    }
    \label{fig:matrix-example-dia}
    \vspace*{-2mm}
  \end{minipage}
  \begin{minipage}[b]{0.22\linewidth}
    \centering
    \medskip
    \includegraphics[scale=0.25]{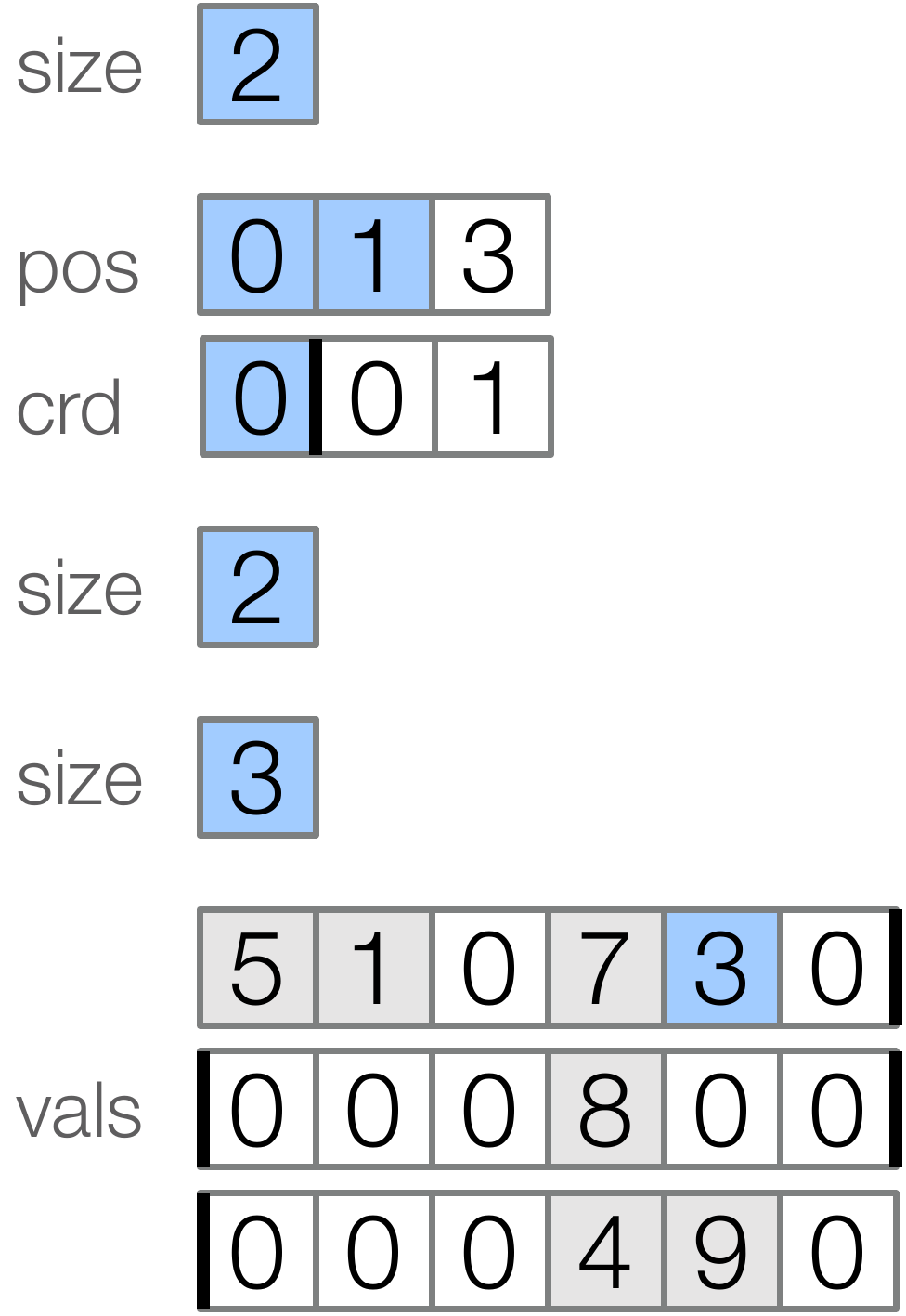}
    \subcaption{
      BCSR
    }
    \label{fig:matrix-example-bcsr}
    \vspace*{-2mm}
  \end{minipage}
  \begin{minipage}[b]{0.20\linewidth}
    \centering
    \includegraphics[scale=0.25]{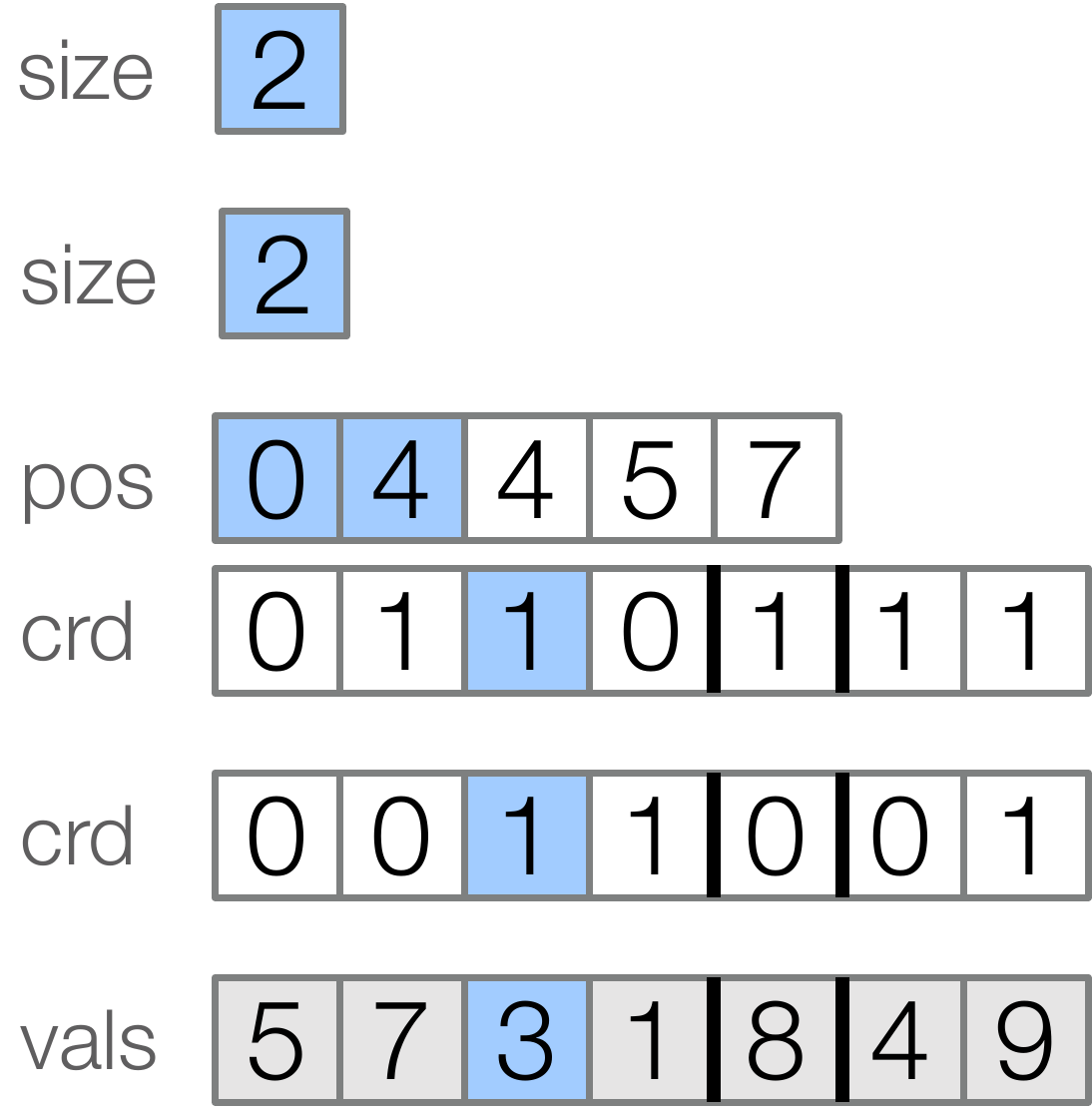}
    \subcaption{
      CSB
    }
    \label{fig:matrix-example-csb}
    \vspace*{-2mm}
  \end{minipage}
  \vspace*{1.0mm}
  \rule{\linewidth}{0.4pt}
  \begin{minipage}[b]{0.21\linewidth}
    \centering
    \includegraphics[scale=0.25]{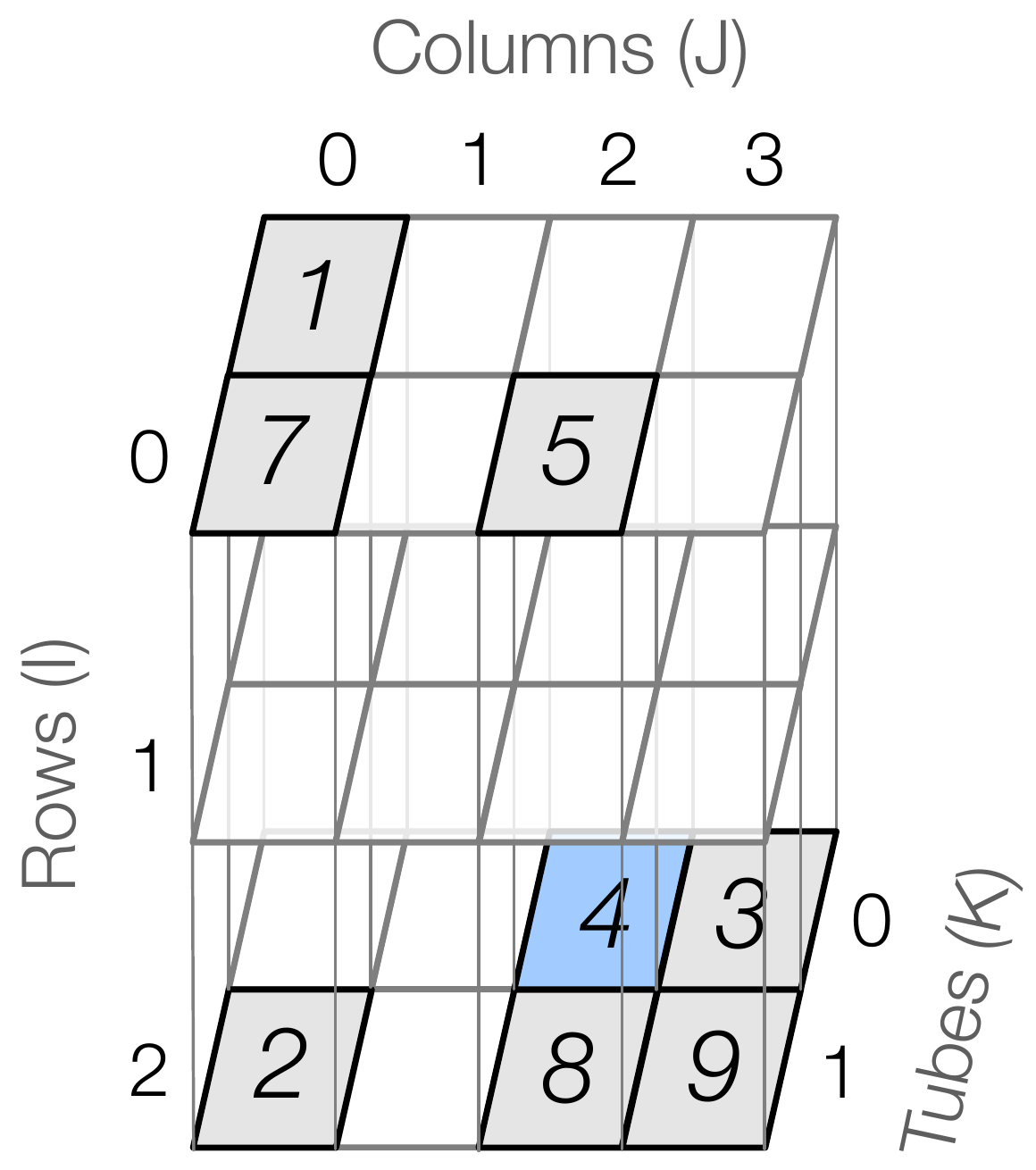}
    \subcaption{
      A 3$\times$4$\times$2 tensor
    }
    \label{fig:tensor-example}
  \end{minipage}
  \begin{minipage}[b]{0.24\linewidth}
    \centering
    \includegraphics[scale=0.25]{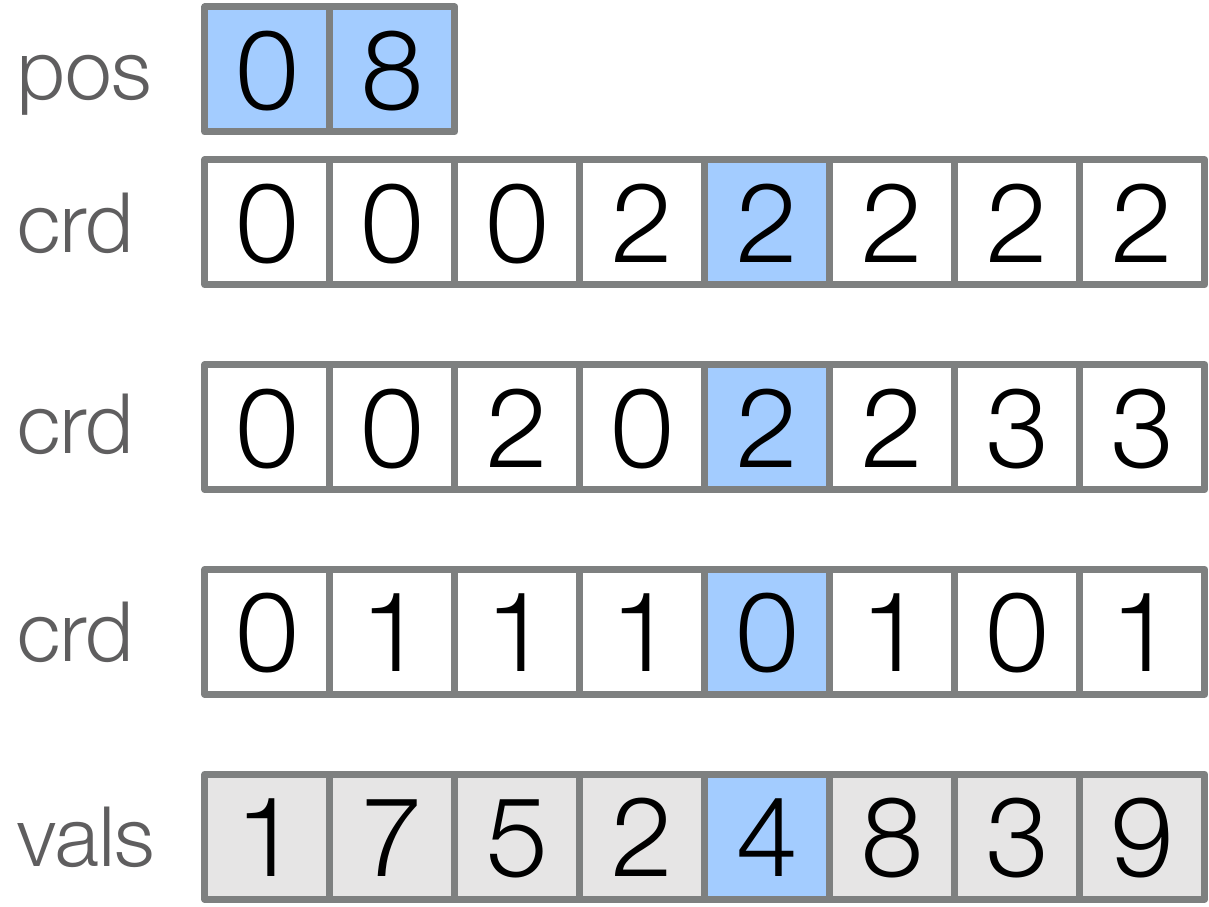}
    \subcaption{
      COO
    }
    \label{fig:tensor-example-coo}
  \end{minipage}
  \begin{minipage}[b]{0.24\linewidth}
    \centering
    \includegraphics[scale=0.25]{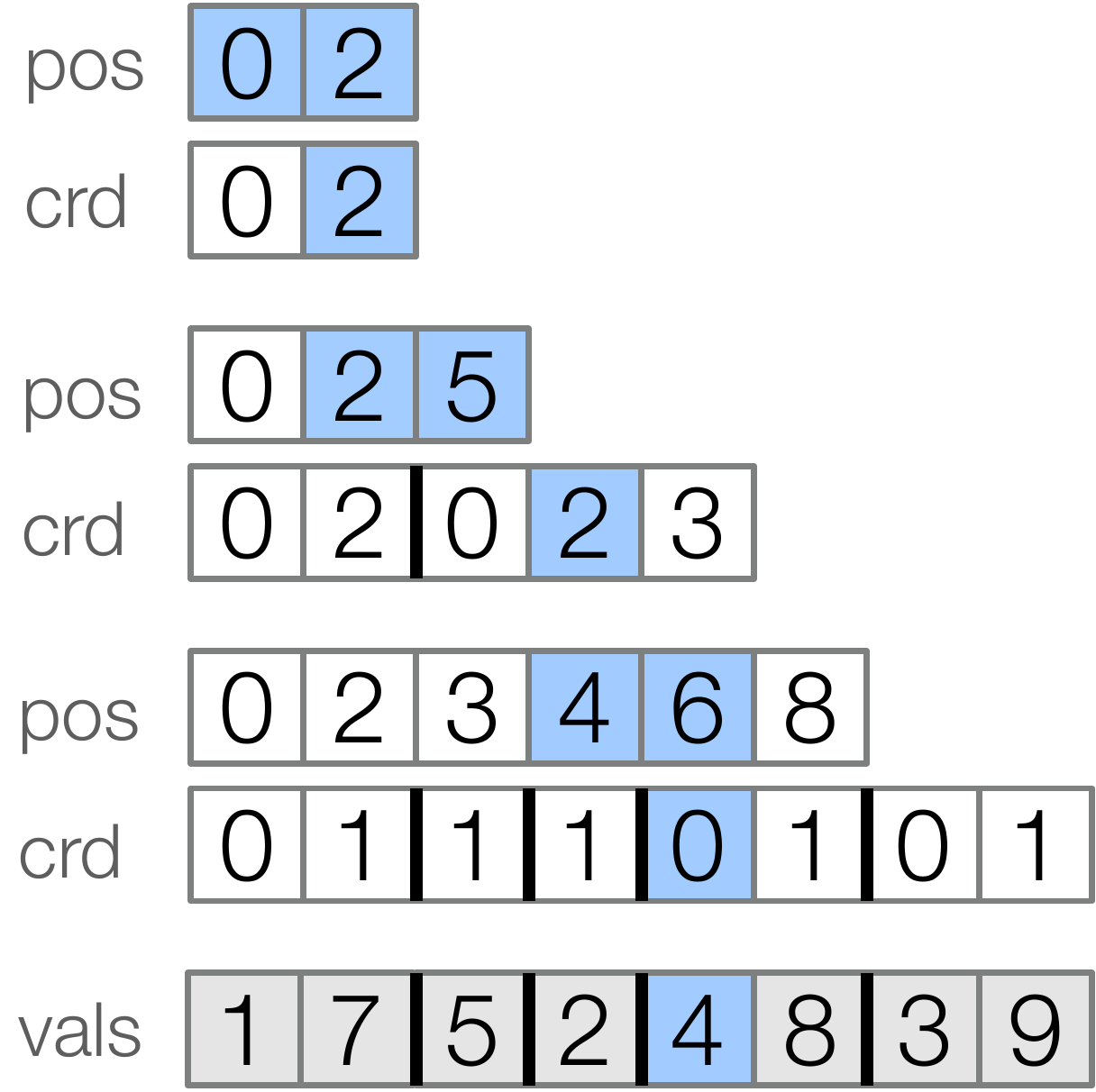}
    \subcaption{
      CSF
    }
    \label{fig:tensor-example-csf}
  \end{minipage}
  \begin{minipage}[b]{0.29\linewidth}
    \centering
    \includegraphics[scale=0.25]{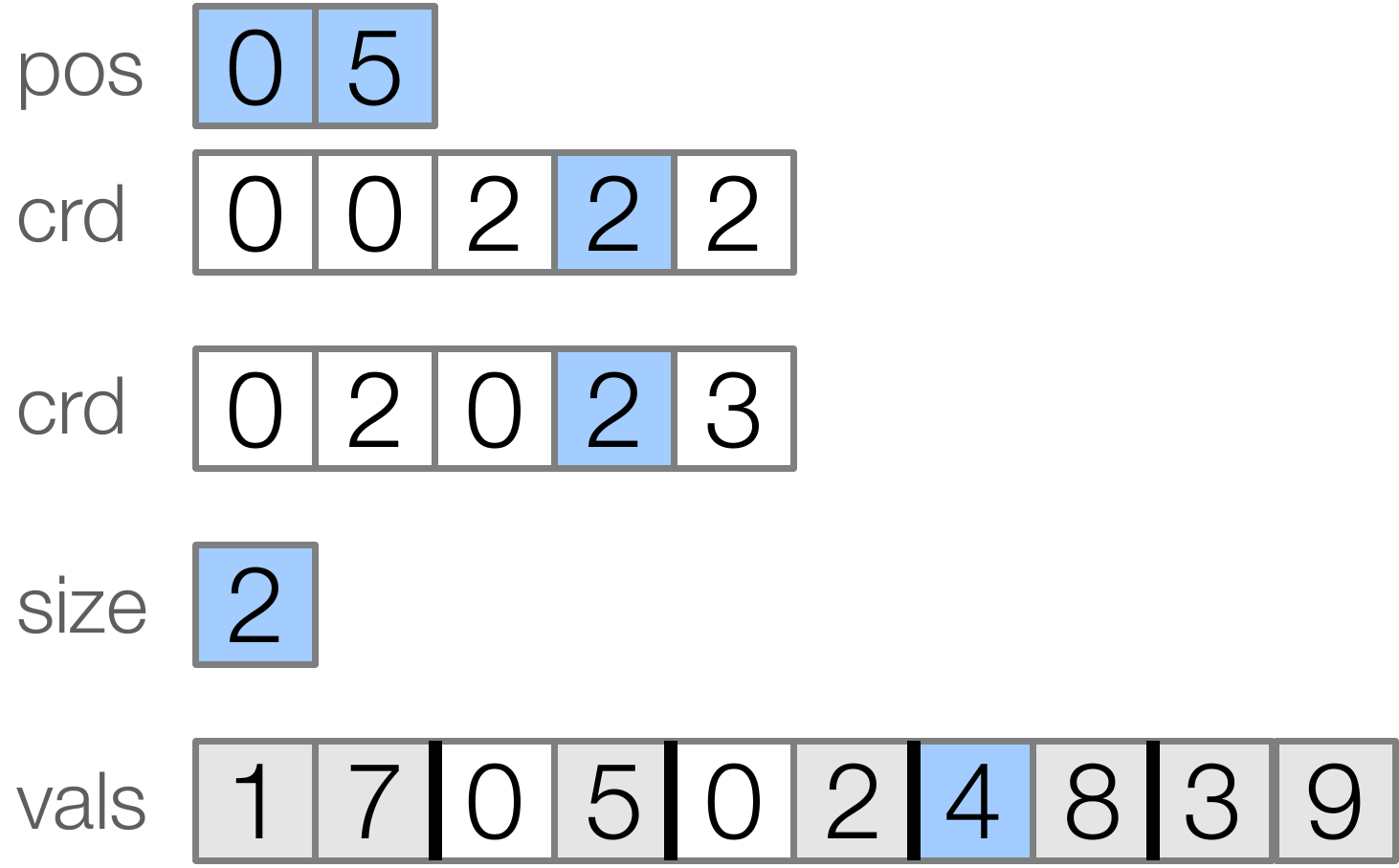}
    \subcaption{
      Mode-generic 
    }
    \label{fig:tensor-example-mode-generic}
  \end{minipage}
  \caption{
    Identical tensors stored in various formats.  Array elements shaded blue
    encode the same nonzero.
  }
  \label{fig:tensor-formats}
\end{figure*}

Several examples of tensor storage formats from the literature are shown
in~\figref{tensor-formats}.  A straightforward way to store an $n$th-order
tensor (i.e., a tensor with $n$ dimensions) is to use an $n$-dimensional dense
array, which explicitly stores all tensor components including zeros.
\figref{vector-example-dense} shows dense storage for an 1st-order tensor (a
vector).  A desirable feature of dense arrays is that the value at any
coordinate can be accessed in constant time.  Storing a sparse tensor in a
dense array, however, is inefficient as a lot of memory is wasted to store
zeros.  Furthermore, performance is lost computing with these zeros even though
they do not meaningfully contribute to the result.  For tensors with many large
dimensions, it may even be impossible to use a dense array due to lack of
memory.  

The simplest way to efficiently store a sparse tensor is to keep a list of its
nonzero coordinates and values
(\threefigsref{vector-example-sparse}{matrix-example-coo}{tensor-example-coo}).
This is typically known as the coordinate (COO) format~\cite{bader2007}.  In
contrast to dense arrays, COO tensors consume only $\Theta(\operatorname{nnz})$
memory.  In addition, many common file formats for storing tensors, such as the
Matrix Market exchange format~\cite{mm-format} and the FROSTT sparse tensor
format~\cite{frostt-format}, closely mirror the COO format.  This minimizes
preprocessing cost as inserting nonzero values and their coordinates only
requires appending them to the \code{crd} and \code{vals} arrays.

Unlike dense arrays though, the COO format does not provide efficient random
access.  This, as we will see in~\secref{code-generation}, limits the
performance of multiplicative operations.  Hash maps (HASH) eliminate this
drawback by storing tensor coordinates in a randomly accessible hash table
(\figref{vector-example-hashed}).  However, hash maps do not support
efficiently iterating over stored nonzeros in order, which in turn restricts
the performance of additive operations.

The COO format also redundantly stores row coordinates that are compressed out
in the compressed sparse row (CSR) format for sparse matrices/2nd-order
tensors~(\figref{matrix-example-csr}).  Compressing out redundant row
coordinates increases performance for computations that are typically memory
bandwidth-bound, such as sparse matrix-vector multiplication (SpMV).  In
\figref{matrix-example-coo}, for instance, the row coordinate is duplicated for
the last three nonzeros in the same row.  CSR removes the redundant row
coordinates, using an auxiliary array (\code{pos}
in~\figref{matrix-example-csr}) to keep track of which nonzeros belong to each
row.  The doubly compressed sparse row (DCSR) format~\cite{dcsc} achieves
additional compression for hypersparse matrices by only storing the rows that
contain nonzeros~(\figref{matrix-example-dcsr}).  For higher-order tensors,
\citeauthor{smith2015tensor}~\shortcite{smith2015tensor} describe a
generalization of CSR, called compressed sparse fiber (CSF), that compresses
every dimension~(\figref{tensor-example-csf}).  Tensors stored in any of these
compressed formats, however, are costly to assemble or modify.

Many important applications work with tensors whose nonzero components form a
regular pattern.  Matrices that encode vertex-edge connectivity of well-formed
unstructured meshes, for instance, have a bounded number of nonzero components
per row.  This is exploited by the ELLPACK (ELL) format, which stores the same
number of components for each row (\figref{matrix-example-ell})~\cite{ell}.
Thus, it only has to store the column coordinates and nonzero values in the
implicitly indexed row positions, which are stored contiguously in memory
making it possible to efficiently vectorize SpMV~\cite{DAzevedo2005}.  If 
nonzeros are further restricted to a few dense diagonals, then their
coordinates can be computed from the offsets of the diagonals.  This pattern is
common in grid and image applications, and allows the diagonal (DIA) format to
forgo storing the column coordinates altogether
(\figref{matrix-example-dia})~\cite{saad2003}.  However, for matrices that do
not conform to assumed structures, structured tensor formats may needlessly 
store many zeros and thus actually degrade performance.

The block compressed sparse row (BCSR) format~\cite{bcsr} generalizes CSR by
storing a dense block of nonzeros in the \texttt{vals} array for every nonzero
coordinate~(\figref{matrix-example-bcsr}).  This reduces storage and exposes
opportunities for vectorization, and is ideal for inherently blocked matrices
from FEM applications.  The mode-generic sparse tensor format, proposed
by \citeauthor{Baskaran2012}~\shortcite{Baskaran2012}, generalizes the idea of
BCSR to higher-order tensors~(\figref{tensor-example-mode-generic}).  It stores
a tensor as a sparse collection of any-order dense blocks, with the coordinates 
of the blocks stored in COO (i.e., the \code{crd} arrays).
By contrast, the compressed sparse block (CSB) format, proposed by
\citeauthor{bulucc2009}~\shortcite{bulucc2009}, represents a matrix as a dense
collection of sparse blocks stored in COO~(\figref{matrix-example-csb}).

\subsection{Computing with Disparate Tensor Formats}
\label{sec:computing-with-disparate-formats}

The existence of so many disparate tensor formats makes it challenging to
support efficiently computing with all of them.  As we have seen, different
formats may use vastly dissimilar data structures to encode tensor coordinates
and thus need very different code to iterate over them.  Iterating over a dense
matrix's column dimension, for instance, simply entails looping over all
possible coordinates along the dimension.  Efficiently iterating over a CSR
matrix's column dimension, by contrast, requires looping over its \code{crd}
array and dereferencing it at each position to access the column coordinates
(\figref{elwise-multiplication-csr-dense}, lines 2--5).  Furthermore, to
efficiently compute with tensors stored in different combinations of formats
can require completely different strategies for simultaneously iterating over
multiple formats.  For example, code to compute the component-wise product of
CSR matrix $B$ with dense matrix $C$ simply has to iterate over $B$ and pick
out corresponding nonzeros from $C$ (\figref{elwise-multiplication-csr-dense},
line 6).  Computing the same operation with COO matrix $C$, however, requires
vastly different code as neither CSR nor COO supports efficient random access
into the column dimension.  Instead, code to compute the component-wise product
has to simultaneously co-iterate over and merge the \code{crd} arrays that
encode $B$ and $C$'s column coordinates
(\figref{elwise-multiplication-csr-coo}, lines 8--20).

To obtain good performance with all tensor formats, a code generator therefore
needs to be able to emit specialized code for any combination of distinct
formats.  The approach of~\citet{kjolstad2017} manages this by effectively
hard-coding a distinct strategy for computing with each format combination into
the code generator.  (More precisely, it hard-codes for combinations of the two
formats that may be used to encode tensor dimensions.)  However, this approach
does not scale with the number of supported formats.  In particular, let $F$
denote a set of formats that use distinct data structures to encode nonzeros.
Each subset of $F$ represents a combination of distinct formats that can be
used to store operands in a tensor algebra expression.  Supporting all formats
in $F$ would thus require individually hard-coding a code generation strategy
for every subset of $F$, of which there are $\Theta(2^{|F|})$.  This
exponential blow-up effectively prevented \taco{} from supporting more
disparate tensor formats like COO and DIA, necessitating a more scalable
solution.

\section{Tensor Storage Abstraction}
\label{sec:storage-abstraction}
As we have seen, a tensor algebra compiler must generate specialized code for
every possible combination of supported formats in order to obtain good
performance.  Since the number of combinations is exponential in the number of
formats though, it is infeasible to exhaustively hard-code support for every
possible format.  In this section, we show how common variants of all the
tensor formats examined in~\secref{tensor-formats} can be expressed as
compositions of just six per-dimension formats.  We will also present an
abstraction that captures shared capabilities and properties of per-dimension
formats.  The abstraction generalizes common patterns of accessing tensor
storage and guides a format-agnostic code generation algorithm that we describe
in~\secref{code-generation}.

\subsection{Coordinate Hierarchies}
\label{sec:coordinate-hierarchies}

\begin{figure*}
  \begin{minipage}[t]{.255\linewidth}
    \centering
    \includegraphics[scale=0.55]{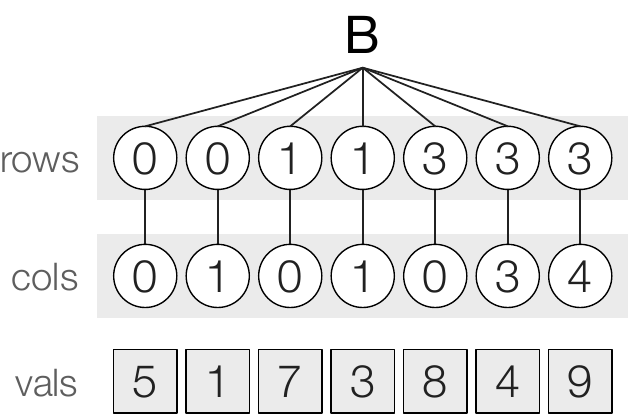}
    \subcaption{
      COO
    }
    \label{fig:matrix-hierarchy-coo}
  \end{minipage}
  \begin{minipage}[t]{.255\linewidth}
    \centering
    \includegraphics[scale=0.55]{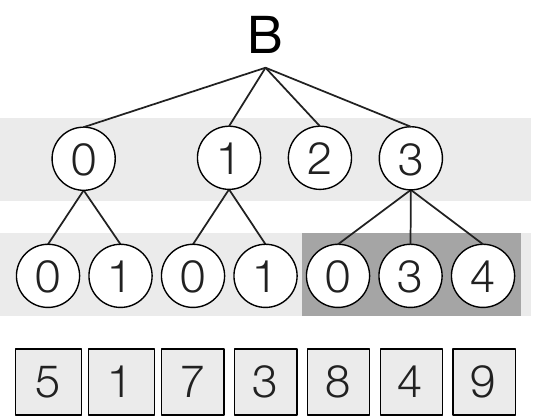}
    \subcaption{
      CSR
    }
    \label{fig:matrix-hierarchy-csr}
  \end{minipage}
  \begin{minipage}[t]{.47\linewidth}
    \centering
    \includegraphics[scale=0.55]{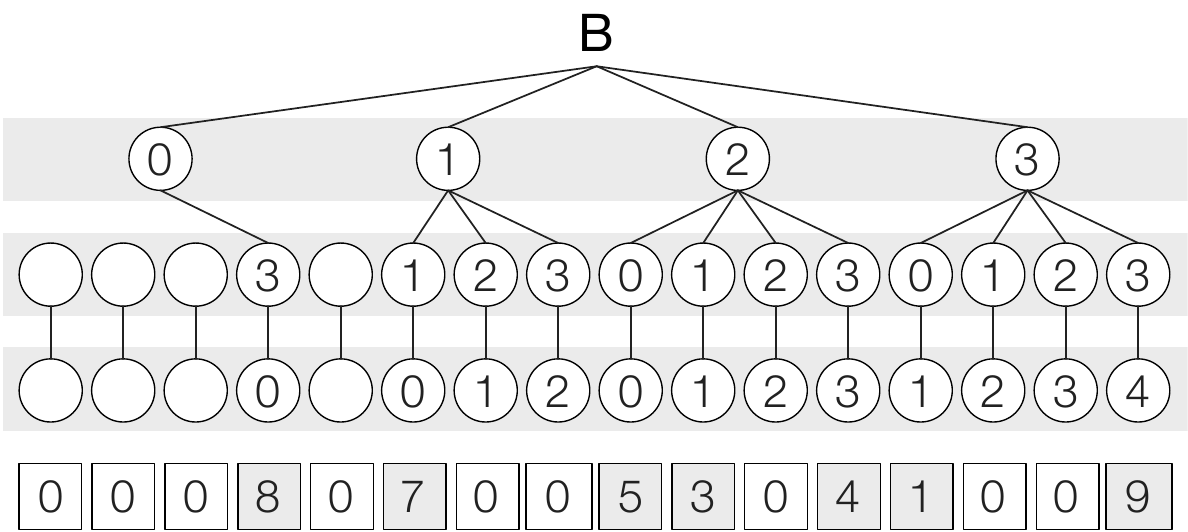}
    \subcaption{
      DIA
    }
    \label{fig:matrix-hierarchy-dia}
  \end{minipage}
  \caption{
    Coordinate hierarchies for the same matrix, shown
    in~\figref{matrix-example}, stored in different formats.  The levels
    labeled rows and cols encode the matrix's row and column coordinates
    respectively.  A coordinate hierarchy's structure reflects how the
    underlying storage format encodes a tensor's nonzeros.
  }
  \label{fig:coordinate-hierarchies}
\end{figure*}

The per-dimension formats can be understood by viewing tensor storage as a
hierarchy of coordinates, where each hierarchy level encodes the coordinates
along one tensor dimension.  Each path from the root to a leaf in this
hierarchy encodes the coordinates in each dimension of one tensor component.
\figref{coordinate-hierarchies} shows examples of coordinate hierarchies for a
matrix stored in different formats.  The matrix component values are shown at
the bottom of each hierarchy.  In~\figref{matrix-hierarchy-csr}, for instance,
the rightmost path represents the tensor component $B(3,4)$ with the value $9$.
As we will see in~\secref{code-generation}, representing tensor storage
hierarchically lets a code generator decompose any computation into the simpler
problem of merging coordinate hierarchy levels.

The structure of a coordinate hierarchy reflects the encoding of nonzeros in
memory and lets the code generator reason about how to iterate over tensors
without knowing the specific tensor format.  For example, the coordinate
hierarchy for a COO matrix (\figref{matrix-hierarchy-coo}) consists of
coordinate chains that encode each nonzero.  This chain structure reflects that
the COO format stores the complete coordinates of each nonzero.  By contrast,
the coordinate hierarchy for a CSR matrix (\figref{matrix-hierarchy-csr}) is
tree-structured and components in the same row share a row coordinate parent.
This tree structure reflects that the CSR format removes redundant row
coordinates using the auxiliary \code{pos} array.

A coordinate hierarchy has one level (shaded light gray
in~\figref{coordinate-hierarchies}) for every tensor dimension, and
per-dimension \emph{level formats} describe how to store coordinate hierarchy
levels in memory.  Each position (a node) in a level may encode some coordinate
(the number in the node) along the corresponding tensor dimension.
Alternatively, a position may contain an unlabeled node, which encodes no
coordinate and reflects padding in the underlying physical storage.  The
unlabeled nodes in~\figref{matrix-hierarchy-dia}, for instance, represent
segments of each diagonal that are out of the bounds of valid matrix
coordinates.  Typically, different levels in a coordinate hierarchy will be
stored in separate data structures in memory, though the abstraction also
permits levels to share the same underlying data structure.  As will be
evident, this can be useful for some tensor formats such as DIA, which uses the
same array (i.e., \code{offset}) to encode both the row and column levels
in~\figref{matrix-hierarchy-dia}.

Each node in a level may also be connected to a parent in the previous level.
Coordinates that share the same parent are referred to as \emph{siblings}.  The
coordinates highlighted in dark gray in~\figref{matrix-hierarchy-csr}, for
instance, are siblings that share the parent row coordinate $3$.  A
coordinate's \emph{ancestors} refer to the set of coordinates that are encoded
by the path from its parent to the root.

\begin{figure*}
  \begin{minipage}[b]{\linewidth}
    \centering
    \includegraphics[scale=0.35]{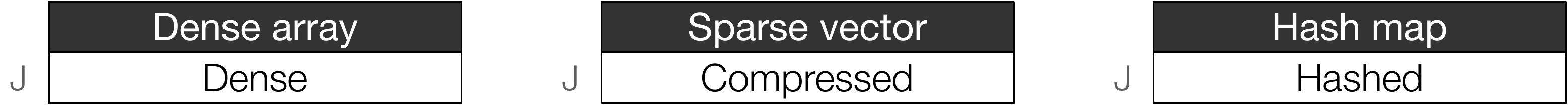}
    \subcaption{Vector storage formats}
    \label{fig:vector-formats-by-dimensions}
  \end{minipage}
  \begin{minipage}[b]{\linewidth}
    \centering
    \bigskip
    \smallskip
    \includegraphics[scale=0.35]{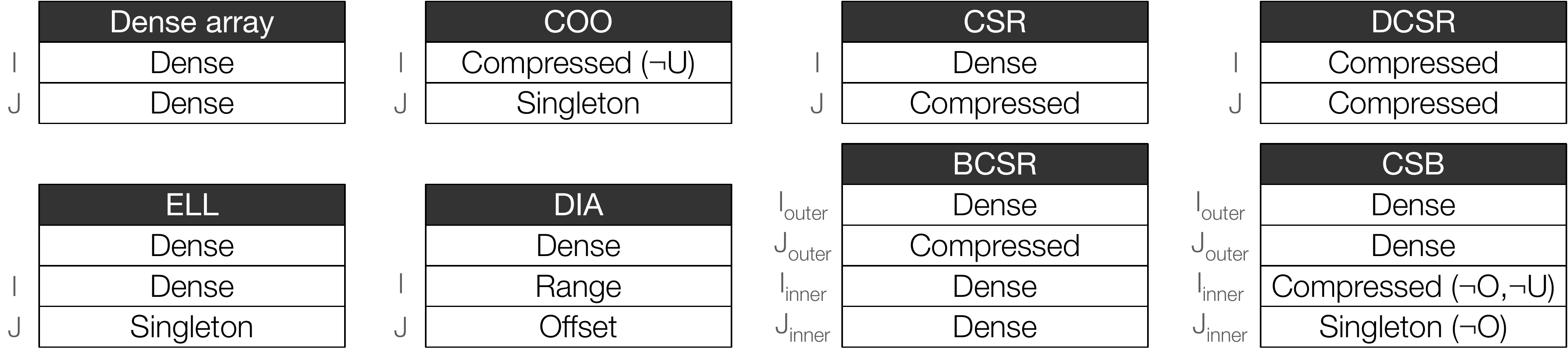}
    \subcaption{Matrix storage formats}
    \label{fig:matrix-formats-by-dimensions}
  \end{minipage}
  \begin{minipage}[b]{\linewidth}
    \centering
    \bigskip
    \smallskip
    \includegraphics[scale=0.35]{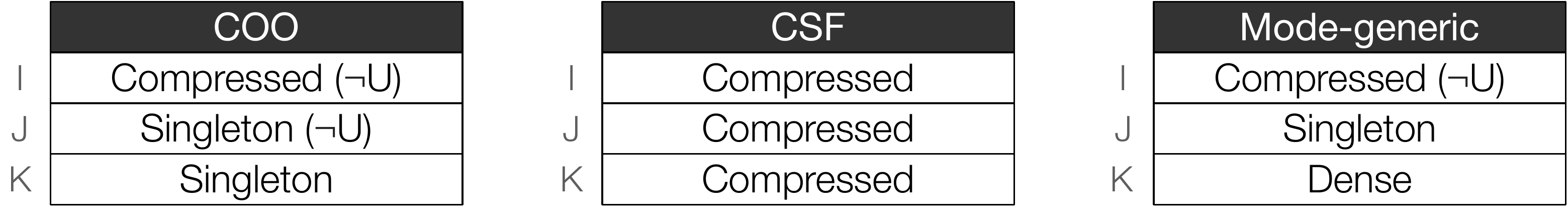}
    \subcaption{3rd-order tensor storage formats}
    \label{fig:tensor-formats-by-dimensions}
  \end{minipage}
  \caption{
    Common tensor formats composed from per-dimension level formats.  We cast
    structured matrix formats as higher-order tensor formats.  The label beside
    a level identifies the tensor dimension it represents.  Unless otherwise
    stated, all levels other than hashed are ordered and unique
    (see~\secref{level-properties}); hashed levels are unordered and unique.
    ($\neg$O) denotes an unordered level and ($\neg$U) denotes a non-unique
    level.
  }
  \label{fig:formats-by-dimensions}
\end{figure*}

We propose six level formats that suffice to represent all the tensor formats
described in~\secref{tensor-formats}, though many more are possible within our
framework.  Each level format (or \emph{level type}) can encode all the nodes
in a level along with the edges connecting them to their parents. Some of these
level formats implicitly encode coordinates (e.g., as an interval), while
others explicitly store them (e.g., in a segmented vector).  At a high level,
given a parent coordinate in the $(i-1)$-th level of a coordinate hierarchy,
the six level formats encode its children coordinates in the $i$-th level as
follows:
\smallskip
\begin{description}
  \begin{minipage}[t]{.775\linewidth}
  \item[Dense] levels store the size of the corresponding dimension ($N$) and
    encode the coordinates in the interval $[0,N)$.
    \figref{matrix-hierarchy-csr} shows the row dimension of a CSR matrix
    encoded as a dense level with the array on the right.
  \end{minipage}%
  \hfill
  \begin{minipage}[t]{.215\linewidth}
  \vspace*{-2.7mm}
  \begin{figure}[H]
      \centering
      \includegraphics[scale=0.25]{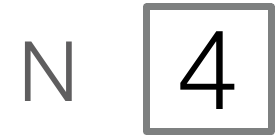}
  \end{figure}
  \end{minipage}
\end{description}
\begin{description}
  \begin{minipage}[t]{.775\linewidth}
  \item[Compressed] levels store coordinates in a segment of the \code{crd}
    array, with the segment bounds stored in the \code{pos} array.
    \figref{matrix-hierarchy-csr} shows the column dimension of a CSR matrix
    encoded as a compressed level with the arrays on the right.  Given a parent
    coordinate 1, for instance, the level encodes two child coordinates 0 and
    1, stored in \code{crd} between positions $\texttt{pos[1]} = 2$ (inclusive) 
    and $\texttt{pos[2]} = 4$ (exclusive).
  \end{minipage}%
  \hfill
  \begin{minipage}[t]{.215\linewidth}
  \vspace*{1.6mm}
  \begin{figure}[H]
      \centering
      \includegraphics[scale=0.25]{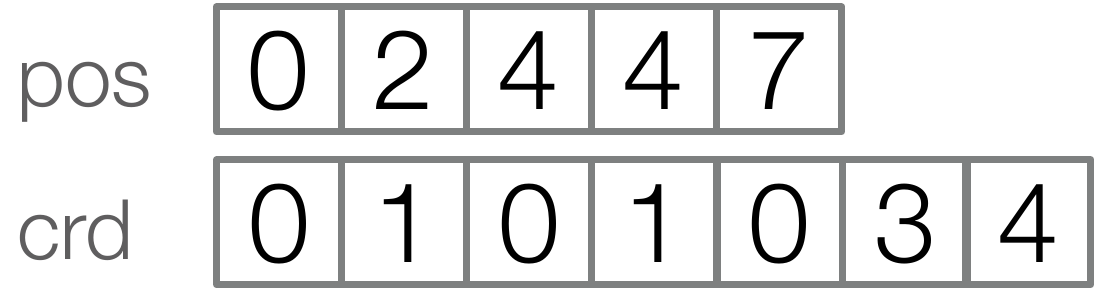}
  \end{figure}
  \end{minipage}
\end{description}
\begin{description}
  \begin{minipage}[t]{.775\linewidth}
  \item[Singleton] levels store a single coordinate with no sibling in the
    \code{crd} array.  \figref{matrix-hierarchy-coo} shows the column dimension
    of a COO matrix encoded as a singleton level with the array on the right.
  \end{minipage}%
  \hfill
  \begin{minipage}[t]{.215\linewidth}
  \vspace*{-2.7mm}
  \begin{figure}[H]
      \centering
      \includegraphics[scale=0.25]{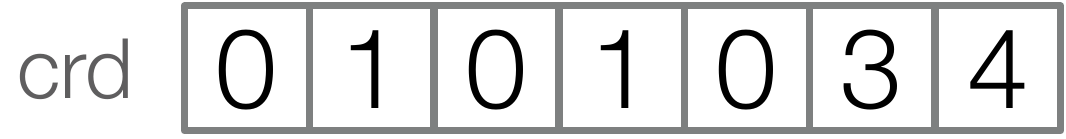}
  \end{figure}
  \end{minipage}
\end{description}
\begin{description}
  \begin{minipage}[t]{.775\linewidth}
  \item[Range] levels encode the coordinates in an interval with bounds 
    computed from an offset and from dimension sizes $N$ and $M$.
    \figref{matrix-hierarchy-dia} shows the row dimension of a DIA matrix
    encoded as a range level with the arrays on the right.  Given a parent
    coordinate 1, the level encodes coordinates between $\max(0,
    -\texttt{offset[1]}) = 1$ and $\min(4, 6 - \texttt{offset[1]}) = 4$.
  \end{minipage}%
  \hfill
  \begin{minipage}[t]{.215\linewidth}
  \vspace*{-2.2mm}
  \begin{figure}[H]
      \centering
      \includegraphics[scale=0.25]{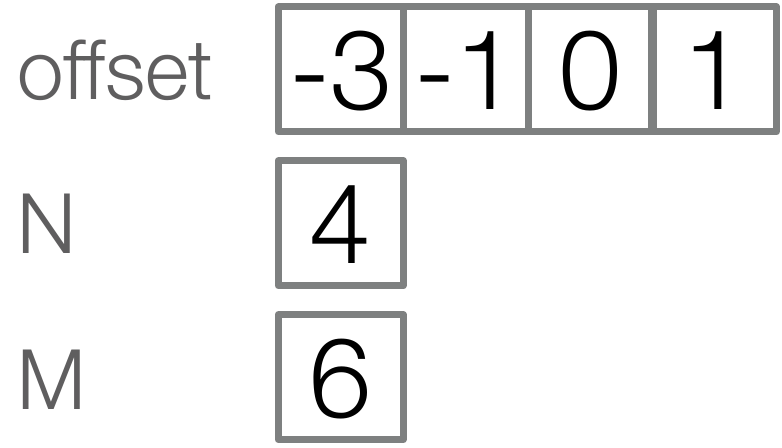}
  \end{figure}
  \end{minipage}
\end{description}
\begin{description}
  \begin{minipage}[t]{.775\linewidth}
  \item[Offset] levels encode a single coordinate with no sibling, shifted from 
    the parent coordinate by a value in the \code{offset} array.
    \figref{matrix-hierarchy-dia} shows the column dimension of a DIA matrix
    encoded as an offset level with the array on the right.  Given a parent
    coordinate 3 and an offset index 1, for instance, the level encodes the
    coordinate $3 + \texttt{offset[1]} = 2$.
  \end{minipage}%
  \hfill
  \begin{minipage}[t]{.215\linewidth}
  \vspace*{1.6mm}
  \begin{figure}[H]
      \centering
      \includegraphics[scale=0.25]{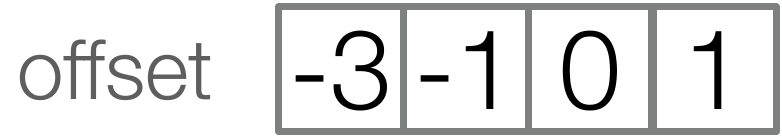}
  \end{figure}
  \end{minipage}
\end{description}
\begin{description}
  \begin{minipage}[t]{.775\linewidth}
  \item[Hashed] levels store coordinates in a segment, of size $W$, of a hash 
    map (\code{crd}).  The arrays on the right, with empty buckets marked by
    $-1$, encode the column dimension of a hash map row vector as a hashed
    level.
  \end{minipage}%
  \hfill
  \begin{minipage}[t]{.215\linewidth}
  \vspace*{-4.4mm}
  \begin{figure}[H]
      \centering
      \includegraphics[scale=0.25]{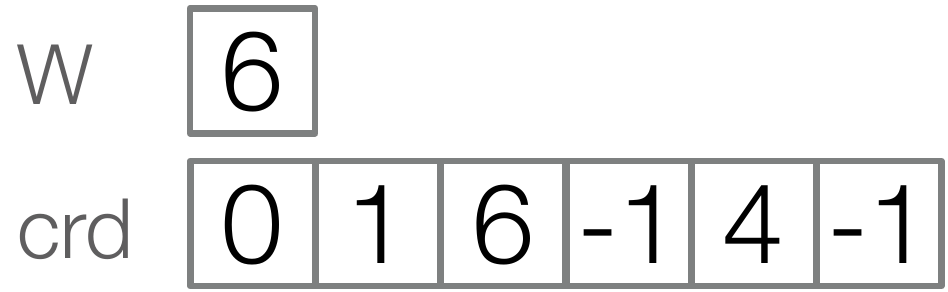}
  \end{figure}
  \end{minipage}
\end{description}
\tabref{level-functions} presents the precise semantics of the level formats,
encoded in \emph{level functions} that each level format implements and that
fully describe how data structures associated with the level format can be
interpreted as a coordinate hierarchy level.  (\secref{level-capabilities}
describes level functions in more depth.)  \figref{formats-by-dimensions} shows
how these level formats can be composed to express all the tensor formats
surveyed in~\secref{tensor-formats}.  The same level formats, however, may
be combined in other ways to express additional tensor formats.  For
instance, a variant of DIA for matrices that have only sparsely-filled
diagonals can be expressed as the combination (dense, compressed,
offset), which replaces the range level that implicitly assumes diagonals are
densely filled.  We cast structured matrix formats, like BCSR, as formats for
higher-order tensors; the added dimensions expose more complex tensor
structures.

The code generator in~\secref{code-generation} emits code that accesses and
modifies coordinate hierarchy levels through an abstract level interface, which
ensures the code generator is not tied to specific formats.  This makes it
extensible and maintainable as adding support for new formats does not require
any change to the code generation algorithm.  The abstract interface to a
coordinate hierarchy level consists of \emph{level capabilities} and
\emph{properties}.  Level capabilities instruct the code generator on how to
iterate over and index into levels, as well as on how to add coordinates to a
level.  Properties of a level let the compiler emit optimized loops that
exploit tensor attributes to increase computational performance.
\tabref{level-capabilities-and-properties} identifies the capabilities and
properties of each level type.

\begin{table}
  \caption {
    Supported capabilities and properties of each level type.  V, P, I, and A
    indicate that a level type supports coordinate value iteration, coordinate
    position iteration, insert, and append respectively.  A \choice{} indicates
    that a level type can be configured to either possess or not possess a
    particular property.
  }
  {\small
  \begin{tabularx}{\linewidth}{Xcccccccc}
    \toprule
    \multicolumn{1}{c}{\multirow{2}{*}{Level Type}} & \multicolumn{3}{c}{Capabilities} & \multicolumn{5}{c}{Properties} \\
    \cmidrule(lr){2-4}
    \cmidrule(lr){5-9}
    & Iteration & Locate & Assembly & Full & Ordered & Unique & Branchless & Compact \\
    \midrule
    Dense       & V & \yes & I   & \yes    & \choice & \choice & \no    & \yes    \\
    Range       & V & \no  & \no & \no     & \choice & \choice & \no    & \no     \\
    Compressed  & P & \no  & A   & \choice & \choice & \choice & \no    & \yes    \\
    Singleton   & P & \no  & A   & \choice & \choice & \choice & \yes   & \yes    \\
    Offset      & P & \no  & \no & \no     & \choice & \choice & \yes   & \no     \\
    Hashed      & P & \yes & I   & \choice & \no     & \choice & \no    & \no     \\
    \bottomrule
  \end{tabularx}
  }
  \label{tab:level-capabilities-and-properties}
\end{table}

\subsection{Level Access Capabilities}
\label{sec:level-capabilities}

Every coordinate hierarchy level provides a set of \emph{capabilities} that can
be used to access or modify its coordinates.  Each capability is exposed as a
set of \emph{level functions} with a fixed interface that a level must
implement to support the capability.
\tabref{level-capabilities-and-properties} identifies the capabilities that 
each level format supports, and~\tabref{level-functions} shows the
level functions that implement the access capabilities.

Level capabilities provide an abstraction for manipulating physical indices
(tensor storage) in a format-agnostic manner.  As an example, the column
dimension of a CSR matrix is a compressed level, which provides the coordinate
position iteration capability.  This capability is exposed as two level
functions, \code{pos_bounds} and \code{pos_access}, as shown
in~\tabref{level-functions}.  To access the column coordinates in dark gray
in~\figref{matrix-hierarchy-csr}, we first determine their range by calling
\code{pos_bounds} with the position of row coordinate 3 as input.  We then call
\code{pos_access} for each position in this range to get the column coordinate
values.  Under the hood, \code{pos_bounds} indexes the \code{pos} array to
locate the \code{crd} array segment that stores the column coordinates, while
\code{pos_access} retrieves each of the coordinates from \code{crd}.  These
level functions fully describe how to access CSR indices efficiently, while
hiding details such as the existence of the \code{pos} and \code{crd} arrays
from the caller.

The abstract interface to a coordinate hierarchy level exposes three different
access capabilities: \emph{coordinate value iteration}, \emph{coordinate
position iteration}, and \emph{locate}.  Every level must provide coordinate
value iteration or coordinate position iteration and may optionally also
provide the locate capability.  \tabref{level-capabilities-and-properties}
identifies the access capabilities supported by each level type.  Our code
generation algorithm will emit code that calls level functions to access
physical tensor storage, which ensures the algorithm is not tied to any
specific types of data structures.

\begin{table}
  \caption{
    Level functions defined for each of the six level types listed
    in~\secref{coordinate-hierarchies}.  \secref{level-capabilities} describes
    how these level functions implement the access capabilities supported by
    each level type, as identified
    in~\tabref{level-capabilities-and-properties}.
  }
  {\small
  \begin{tabularx}{\columnwidth}{lXX}
    \toprule
    \multicolumn{1}{c}{Level Type} & \multicolumn{2}{c}{Level Function Definitions} \\
    \midrule
    \multirow{3}{*}{Dense} & \vspace*{-0.43cm} {
\begin{lstlisting}[style=levelfunc]
coord_bounds(i$_1$, $\dots$, i$_{k-1}$):
  return <0, N$_k$>
\end{lstlisting}} & \vspace*{-0.43cm} {
\begin{lstlisting}[style=levelfunc]
coord_access(p$_{k-1}$, i$_1$, $\dots$, i$_k$):
  return <p$_{k-1}$ * N$_k$ + i$_k$, true>
\end{lstlisting}} \\ [-1.25\normalbaselineskip]
    \cline{2-3}
    & \vspace*{-0.33cm} {
\begin{lstlisting}[style=levelfunc]
locate(p$_{k-1}$, i$_1$, $\dots$, i$_k$):
  return <p$_{k-1}$ * N$_k$ + i$_k$, true>
\end{lstlisting}} & \\ [-1.25\normalbaselineskip] 
    \hline
    \multirow{3}{*}{Range} & \vspace*{-0.33cm} {
\begin{lstlisting}[style=levelfunc]
coord_bounds(i$_1$, $\dots$, i$_{k-1}$):
  return <max(0, -offset[i$_{k-1}$]), 
           min(N$_k$, M$_k$ - offset[i$_{k-1}$])>
\end{lstlisting}} & \vspace*{-0.33cm} {
\begin{lstlisting}[style=levelfunc]
coord_access(p$_{k-1}$, i$_1$, $\dots$, i$_k$):
  return <p$_{k-1}$ * N$_k$ + i$_k$, true>
\end{lstlisting}} \\ [-1.25\normalbaselineskip]
    \hline
    \multirow{2}{*}{Compressed} & \vspace*{-0.33cm} {
\begin{lstlisting}[style=levelfunc]
pos_bounds(p$_{k-1}$):
  return <pos[p$_{k-1}$], pos[p$_{k-1}$ + 1]>
\end{lstlisting}} & \vspace*{-0.33cm} {
\begin{lstlisting}[style=levelfunc]
pos_access(p$_k$, i$_1$, $\dots$, i$_{k-1}$):
  return <crd[p$_k$], true>
\end{lstlisting}} \\ [-1.25\normalbaselineskip]
    \hline
    \multirow{2}{*}{Singleton} & \vspace*{-0.33cm} {
\begin{lstlisting}[style=levelfunc]
pos_bounds(p$_{k-1}$):
  return <p$_{k-1}$, p$_{k-1}$ + 1>
\end{lstlisting}} & \vspace*{-0.33cm} {
\begin{lstlisting}[style=levelfunc]
pos_access(p$_k$, i$_1$, $\dots$, i$_{k-1}$):
  return <crd[p$_k$], true>
\end{lstlisting}} \\ [-1.25\normalbaselineskip]
    \hline
    \multirow{2}{*}{Offset} & \vspace*{-0.33cm} {
\begin{lstlisting}[style=levelfunc]
pos_bounds(p$_{k-1}$):
  return <p$_{k-1}$, p$_{k-1}$ + 1>
\end{lstlisting}} & \vspace*{-0.33cm} {
\begin{lstlisting}[style=levelfunc]
pos_access(p$_k$, i$_1$, $\dots$, i$_{k-1}$):
  return <i$_{k-1}$ + offset[i$_{k-2}$], true>
\end{lstlisting}} \\ [-1.25\normalbaselineskip]
    \hline
    \multirow{9}{*}{Hashed} & \vspace*{-0.33cm} {
\begin{lstlisting}[style=levelfunc]
pos_bounds(p$_{k-1}$):
  return <p$_{k-1}$ * W$_k$, (p$_{k-1}$ + 1) * W$_k$>
\end{lstlisting}} & \vspace*{-0.33cm} {
\begin{lstlisting}[style=levelfunc]
pos_access(p$_k$, i$_1$, $\dots$, i$_{k-1}$):
  return <crd[p$_k$], crd[p$_k$] != -1>
\end{lstlisting}} \\ [-1.25\normalbaselineskip]
    \cline{2-3}
    & \vspace*{-0.33cm} {
\begin{lstlisting}[style=levelfunc]
locate(p$_{k-1}$, i$_1$, $\dots$, i$_k$):
  int p$_k$ = i$_k$ % W$_k$ + p$_{k-1}$ * W$_k$
  if (crd[p$_k$] != i$_k$ && crd[p$_k$] != -1) {
    int end = p$_k$
    do {
      p$_k$ = (p$_k$ + 1) % W$_k$ + p$_{k-1}$ * W$_k$
    } while (crd[p$_k$] != i$_k$ && crd[p$_k$] != -1 && p$_k$ != end)
  }
  return <p$_k$, crd[p$_k$] == i$_k$>
\end{lstlisting}} & \\ [-1.25\normalbaselineskip] 
    \bottomrule
  \end{tabularx}
  }
  \label{tab:level-functions}
\end{table}

\begin{figure*}
  \begin{minipage}[b]{0.49\linewidth}
    \centering
    \includegraphics[scale=0.25]{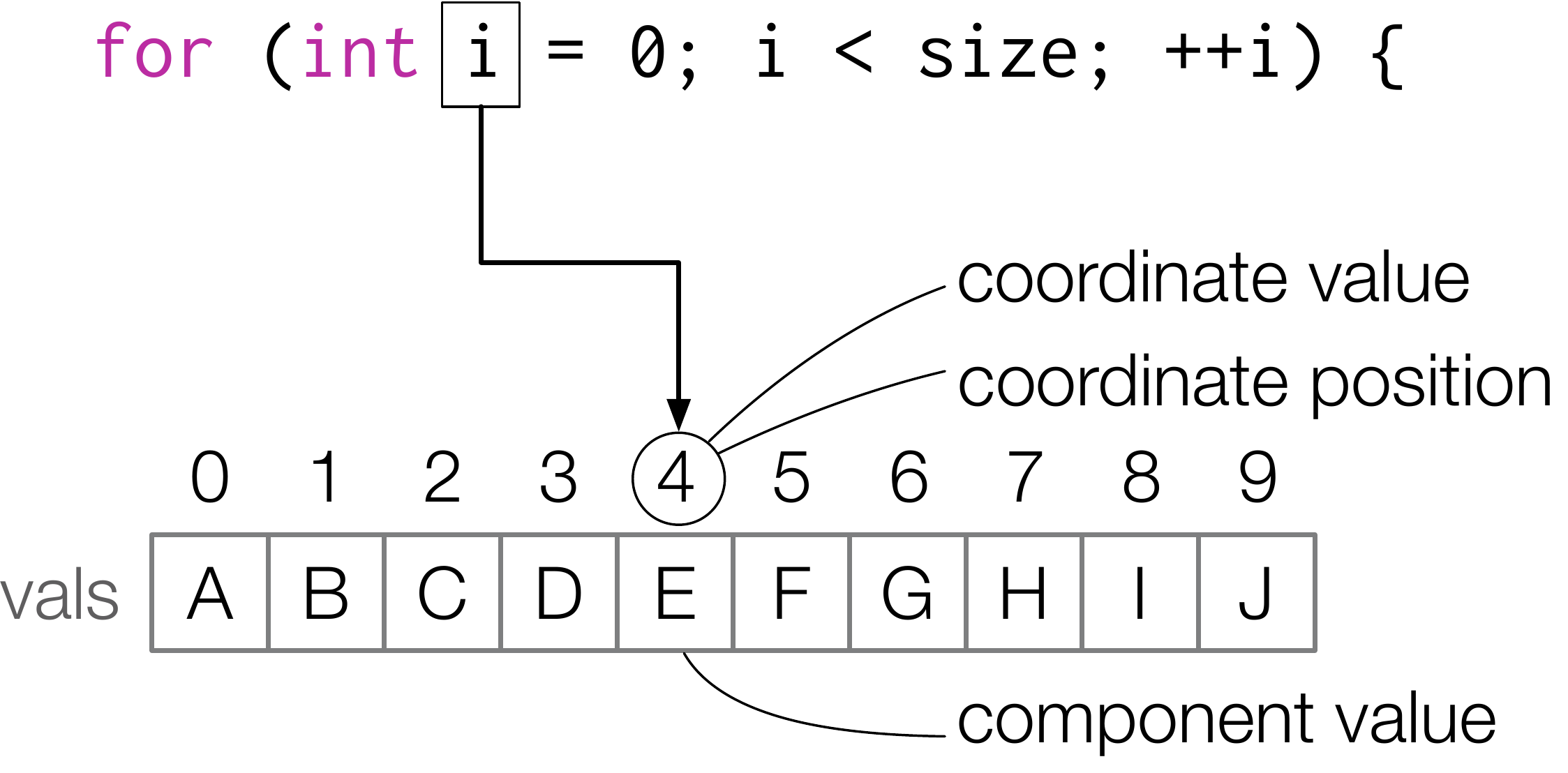}
    \subcaption {\label{fig:dense-iteration}
      Iterating over a dense array
      \vspace{-1.5mm}
    }
  \end{minipage}
  \begin{minipage}[b]{0.49\linewidth}
    \centering
    \includegraphics[scale=0.25]{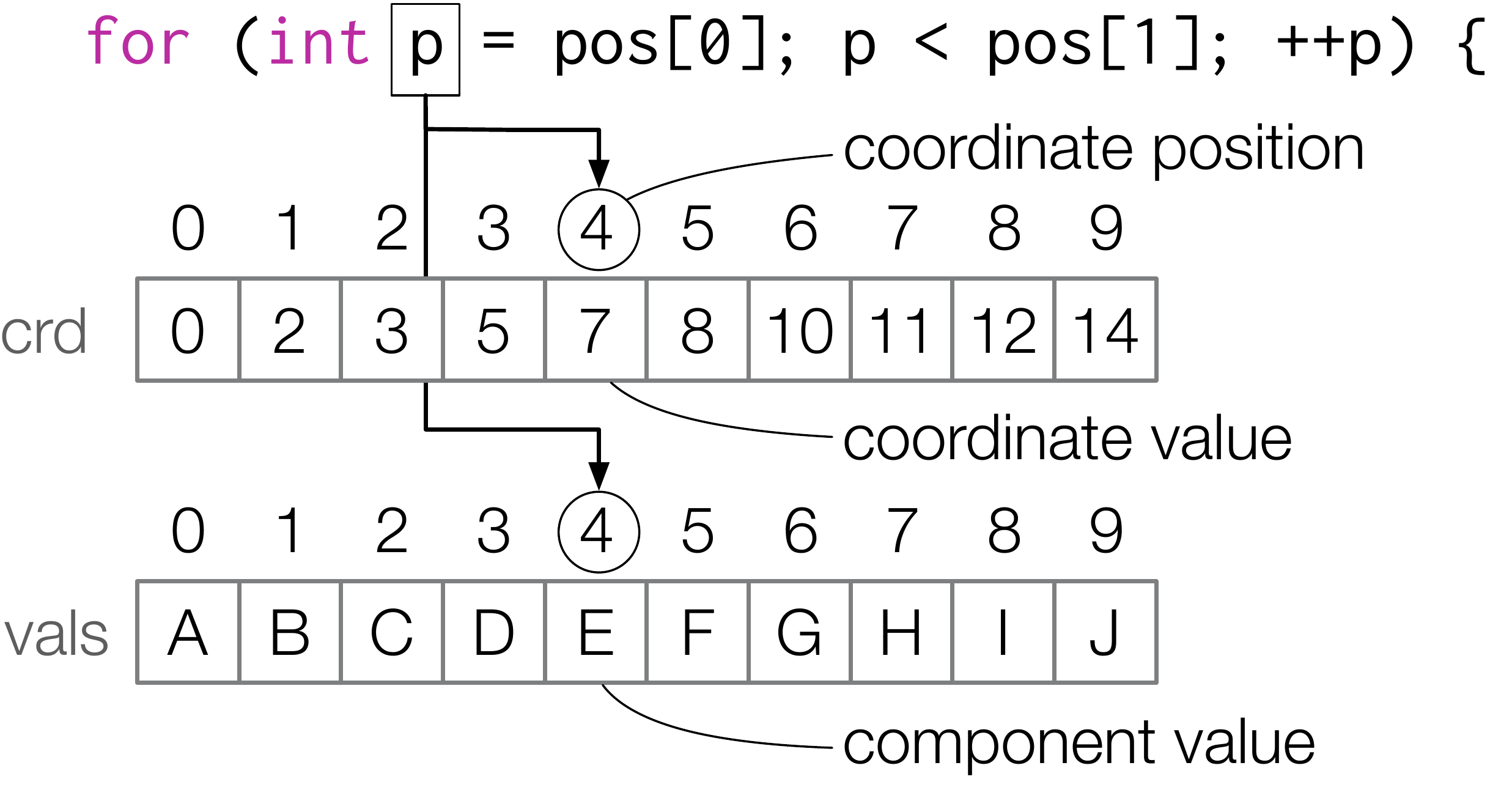}
    \subcaption {\label{fig:sparse-iteration}
      Iterating over a sparse vector
      \vspace{-1.5mm}
    } 
  \end{minipage}
  \begin{minipage}[t]{0.49\linewidth}
    \centering
    \begin{lstlisting}[language=C++,mathescape,xleftmargin=\parindent]
ibegin$_k$, iend$_k$ = coord_bounds(i$_1$, $...$, i$_{k-1}$);
for (i$_k$ = ibegin$_k$; i$_k$ < iend$_k$; ++i$_k$) {
  p$_k$, found = coord_access(p$_{k-1}$, i$_1$, $...$, i$_k$);
  if (found) {
    // coords and values dominated by i$\color{gray}_k$ at pos 
    // p$\color{gray}_k$ encode subtensor $\color{gray}B(i_1, ..., i_k, :, ..., :)$
  }
}
    \end{lstlisting}
    \vspace{-2mm}
    \subcaption {\label{fig:coord-val-iteration-semantics}
      Coordinate value iteration
    }
  \end{minipage}
  \hfill
  \begin{minipage}[t]{0.49\linewidth}
    \centering
    \begin{lstlisting}[language=C++,mathescape,xleftmargin=\parindent]
pbegin$_k$, pend$_k$ = pos_bounds($p_{k-1}$);
for (p$_k$ = pbegin$_k$; p$_k$ < pend$_k$; ++p$_k$) {
  i$_k$, found = pos_access(p$_k$, i$_1$, $...$, i$_{k-1}$);
  if (found) {
    // coords and values dominated by i$\color{gray}_k$ at pos
    // p$\color{gray}_k$ encode subtensor $\color{gray}B(i_1, ..., i_k, :, ..., :)$
  }
}
    \end{lstlisting}
    \vspace{-2mm}
    \subcaption {\label{fig:coord-pos-iteration-semantics}
      Coordinate position iteration
    }
  \end{minipage}
  \caption{
    To iterate over a dense vector, we loop over its \emph{coordinates} and use
    them to index into the \code{vals} array. To iterate over a sparse vector, 
    we loop over \emph{coordinate positions} and use them to access the
    \code{crd} and \code{vals} arrays.  Coordinate value iteration and
    coordinate position iteration generalize these patterns.
  }
  \label{fig:iteration-examples}
\end{figure*}

\subsubsection*{Coordinate Value Iteration}

The coordinate value iteration capability directly iterates over coordinates.
It generalizes the method in~\figref{dense-iteration} for iterating over a
dense vector and is exposed as two level functions.  The first returns an
iterator over coordinates of a coordinate hierarchy level (\code{coord_bounds})
and the second accesses the position of each coordinate (\code{coord_access}):
\begin{lstlisting}[language=C++,mathescape,xleftmargin=\parindent]
coord_bounds(i$_1$, $\dots$, i$_{k-1}$) -> <ibegin$_k$, iend$_k$>
coord_access(p$_{k-1}$, i$_1$, $\dots$, i$_k$) -> <p$_k$, found>
\end{lstlisting}
More precisely, given a list of ancestor coordinates ($i_1, \dots, i_{k-1}$),
\code{coord_bounds} returns the bounds of an iterator over coordinates that may
have those ancestors.  For each coordinate $i_k$ within those bounds,
\code{coord_access} either returns the position of a child of $p_{k-1}$ that
encodes $i_k$ and returns \code{found} as true, or alternatively returns
\code{found} as false if the coordinate does not actually exist.  These
functions can be used to iterate tensor coordinates in the general case as
demonstrated in~\figref{coord-val-iteration-semantics}.  
In practice though,
the code in ~\figref{coord-val-iteration-semantics} can be optimized by
removing the conditional if we know that an implementation of
\code{coord_access} always returns \code{found} as true.

\subsubsection*{Coordinate Position Iteration}

The coordinate position iteration capability, on the other hand, iterates over
coordinate positions.  It generalizes the method in~\figref{sparse-iteration}
for iterating over a sparse vector and is also exposed as two level functions.
The first returns an iterator over positions in a level (\code{pos_bounds}) and
the second accesses the coordinate encoded at each position
(\code{pos_access}):
\begin{lstlisting}[language=C++,mathescape,xleftmargin=\parindent]
pos_bounds(p$_{k-1}$) -> <pbegin$_k$, pend$_k$>
pos_access(p$_k$, i$_1$, $\dots$, i$_{k-1}$) -> <i$_k$, found>
\end{lstlisting}
More precisely, given a coordinate at position $p_{k-1}$, \code{pos_bounds}
returns the bounds of an iterator over positions that may have $p_{k-1}$ as
their parent.  For each position $p_k$ in those bounds, \code{pos_access}
returns the coordinate encoded at that position, or alternatively returns
\code{found} as false if $p_k$ is not actually a child of $p_{k-1}$ or does not
encode a coordinate (i.e., if $p_k$ is unlabeled).  These functions can be used
to iterate tensor coordinates with code like that shown
in~\figref{coord-pos-iteration-semantics}; this code shares a similar structure
to that for coordinate value iteration but has the roles of $i_k$ and $p_k$
reversed.

\subsubsection*{Locate}

The locate capability provides random access into a coordinate hierarchy level
through a function that computes the position of a coordinate:
\begin{lstlisting}[language=C++,mathescape,xleftmargin=\parindent]
locate(p$_{k-1}$, i$_1$, $\dots$, i$_k$) -> <p$_k$, found>
\end{lstlisting}
\code{locate} has similar semantics as \code{coord_access}.  Given a coordinate
$i_{k-1}$ at position $p_{k-1}$, \code{locate} attempts to locate among its
children the coordinate $i_k$.  If \code{locate} finds $i_k$, then it returns
$i_k$'s position $p_k$ and returns \code{found} as true; otherwise it returns
\code{found} as false.  Traversing a path in a coordinate hierarchy to access a
single tensor component can be done by successively calling \code{locate} at
every level.  As we will see in~\secref{code-generation}, having operands with
efficient implementations of the locate capability leads to code that avoids
iterating over every nonzero in those operands.

\subsection{Level Properties}
\label{sec:level-properties}

A coordinate hierarchy level may also declare up to five \emph{properties}:
\emph{full}, \emph{ordered}, \emph{unique}, \emph{branchless}, and
\emph{compact}.  These properties describe characteristics of a level, such as
whether coordinates are arranged in order, and are invariants that are
explicitly enforced or implicitly assumed by the underlying physical index.
The column dimension of a sorted CSR matrix, for instance, is both ordered and
unique (\figref{formats-by-dimensions}), which means it stores every coordinate
just once and in increasing order.  Our code generation technique relies on
these level properties to emit optimized code.

\tabref{level-capabilities-and-properties} identifies the properties of each
level type.  Some coordinate hierarchy levels may be configured with a
property, depending on the application.  Configurable properties reflect
invariants that are not tied to how a physical index encodes coordinates.  For
example, the \code{crd} array in compressed levels typically store coordinates
in order when used in the CSR format, but the same data structure can also
store coordinates out of order.  \figref{formats-by-dimensions} shows how level
types with configurable properties can be configured to represent tensor
formats.

\subsubsection*{Full}

A level is full if every collection of coordinates that share the same
ancestors encompasses all valid coordinates along the corresponding tensor
dimension.  For instance, a level that represents a CSR matrix's row dimension
(\figref{matrix-hierarchy-csr}) encodes every row coordinate and is thus full.
By contrast, a level that represents the same CSR matrix's column dimension is
not full as it only stores the coordinates of nonzero components.

\subsubsection*{Unique}

A level is unique if no collection of coordinates that share the same ancestors
contains duplicates.  For example, a level that represents a CSR matrix's row
dimension necessarily encodes every coordinate just once and is thus unique.
By contrast, a level that represents a COO matrix's row dimension
(\figref{matrix-hierarchy-coo}) can store a coordinate more than once and is
thus not unique.

\subsubsection*{Ordered}

A level is ordered if coordinates that share the same ancestors are ordered in
increasing value, coordinates with different ancestors are ordered
lexicographically by their ancestors, and duplicates are ordered by their
parents' positions.  For example, a level that represents a sorted CSR matrix's
column dimension stores coordinates in increasing order and is thus ordered.  A
level that represents a hash map vector, however, is not as coordinates are
stored in hash order instead.

\subsubsection*{Branchless}

A level is branchless if no coordinate has a sibling and each coordinate in
the previous level has a child.  For example, the coordinate hierarchy for a
COO matrix consists strictly of chains of coordinates, making the lower level
branchless.  By contrast, a level that represents a CSR matrix's column
dimension can have multiple coordinates with the same parent and is thus not
branchless.

\subsubsection*{Compact}

A level is compact if no two coordinates are separated by an unlabeled node
that does not encode a coordinate.  For instance, a level that represents a CSR
matrix's column dimension encodes coordinates in one contiguous range of
positions and is thus compact.  A level that represents a hash map vector,
however, is not as it can have unlabeled positions that reflect empty buckets.

\subsection{Level Output Assembly Capabilities}
\label{sec:output-assembly}

The capabilities described in \secref{level-capabilities} iterate over and
access coordinate hierarchy levels.  A level may also provide \emph{insert} and
\emph{append} capabilities for adding new coordinates to the level.  These
capabilities let us assemble, in a format-agnostic manner, the data structures
that store the result of a computation.
\tabref{level-capabilities-and-properties} identifies the assembly capabilities
that various level types support, and~\tabref{assembly-functions} shows the
level functions that implement those capabilities.

\begin{table}
  \caption{
    Definitions of level functions that implement assembly capabilities for
    various level types.
  }
  \centering
  {\small
  \begin{tabularx}{\columnwidth}{lp{5.2cm}X}
    \toprule
    \multicolumn{1}{c}{Level Type} & \multicolumn{2}{c}{Level Function Definitions} \\
    \midrule
    \multirow{3}{*}{Dense} &
    \vspace*{-0.43cm} {
\begin{lstlisting}[style=levelfunc]
insert_coord(p$_k$, i$_k$):
  // do nothing
\end{lstlisting}}
    &
    \vspace*{-0.43cm} {
\begin{lstlisting}[style=levelfunc]
insert_init(sz$_{k-1}$, sz$_k$):
  // do nothing
\end{lstlisting}}
    \\ [-1.25\normalbaselineskip] &
    \vspace*{-0.33cm} {
\begin{lstlisting}[style=levelfunc]
size(sz$_{k-1}$):
  return sz$_{k-1}$ * N$_k$ 
\end{lstlisting}}
    &
    \vspace*{-0.33cm} {
\begin{lstlisting}[style=levelfunc]
insert_finalize(sz$_{k-1}$, sz$_{k}$):
  // do nothing
\end{lstlisting}}
    \\ [-1.25\normalbaselineskip]
    \hline
    \multirow{9}{*}{Compressed} &
    \vspace*{-0.33cm} {
\begin{lstlisting}[style=levelfunc]
append_coord(p$_k$, i$_k$):
  crd[p$_k$] = i$_k$
\end{lstlisting}}
    &
    \vspace*{-0.33cm} {
\begin{lstlisting}[style=levelfunc]
append_init(sz$_{k-1}$, sz$_k$):
  for (int p$_{k-1}$ = 0; p$_{k-1}$ <= sz$_{k-1}$; ++p$_{k-1}$) {
    pos[p$_{k-1}$] = 0
  }
\end{lstlisting}}
    \\ [-1.25\normalbaselineskip] &
    \vspace*{-0.33cm} {
\begin{lstlisting}[style=levelfunc]
append_edges(p$_{k-1}$, pbegin$_k$, pend$_k$):
  pos[p$_{k-1}$ + 1] = pend$_k$ - pbegin$_k$
\end{lstlisting}}
    &
    \vspace*{-0.33cm} {
\begin{lstlisting}[style=levelfunc]
append_finalize(sz$_{k-1}$, sz$_{k}$):
  int cumsum = pos[0]
  for (int p$_{k-1}$ = 1; p$_{k-1}$ <= sz$_{k-1}$; ++p$_{k-1}$) {
    cumsum += pos[p$_{k-1}$]
    pos[p$_{k-1}$] = cumsum
  }
\end{lstlisting}}
    \\ [-1.25\normalbaselineskip] 
    \hline
    \multirow{4}{*}{Singleton} & 
    \vspace*{-0.33cm} {
\begin{lstlisting}[style=levelfunc]
append_coord(p$_k$, i$_k$):
  crd[p$_k$] = i$_k$
\end{lstlisting}}
    &
    \vspace*{-0.33cm} {
\begin{lstlisting}[style=levelfunc]
append_init(sz$_{k-1}$, sz$_k$):
  // do nothing
\end{lstlisting}}
    \\ [-1.25\normalbaselineskip] &
    \vspace*{-0.33cm} {
\begin{lstlisting}[style=levelfunc]
append_edges(p$_{k-1}$, pbegin$_k$, pend$_k$):
  // do nothing
\end{lstlisting}}
    &
    \vspace*{-0.33cm} {
\begin{lstlisting}[style=levelfunc]
append_finalize(sz$_{k-1}$, sz$_{k}$):
  // do nothing
\end{lstlisting}}
    \\ [-1.25\normalbaselineskip] 
    \hline
    \multirow{5}{*}{Hashed} &
    \vspace*{-0.33cm} {
\begin{lstlisting}[style=levelfunc]
insert_coord(p$_k$, i$_k$):
  crd[p$_k$] = i$_k$
\end{lstlisting}}
    &
    \vspace*{-0.33cm} {
\begin{lstlisting}[style=levelfunc]
insert_init(sz$_{k-1}$, sz$_k$):
  for (int p$_k$ = 0; p$_k$ < sz$_k$; ++p$_k$) {
    crd[p$_k$] = -1
  }
\end{lstlisting}}
    \\ [-1.25\normalbaselineskip] &
    \vspace*{-0.33cm} {
\begin{lstlisting}[style=levelfunc]
size(sz$_{k-1}$):
  return sz$_{k-1}$ * W$_k$ 
\end{lstlisting}}
    &
    \vspace*{-0.33cm} {
\begin{lstlisting}[style=levelfunc]
insert_finalize(sz$_{k-1}$, sz$_{k}$):
  // do nothing
\end{lstlisting}}
    \\ [-1.25\normalbaselineskip] 
    \bottomrule
  \end{tabularx}
  }
  \label{tab:assembly-functions}
\end{table}

\subsubsection*{Insert Capability}

Inserts coordinates at any position and is exposed as four level functions:

\begin{minipage}{.49\linewidth}
\begin{lstlisting}[language=C++,mathescape,xleftmargin=\parindent]
insert_coord(p$_k$, i$_k$) -> void
size(sz$_{k-1}$) -> sz$_k$
\end{lstlisting}
\end{minipage}
\hfill
\begin{minipage}{.49\linewidth}
\begin{lstlisting}[language=C++,mathescape,xleftmargin=\parindent]
insert_init(sz$_{k-1}$, sz$_k$) -> void
insert_finalize(sz$_{k-1}$, sz$_{k}$) -> void
\end{lstlisting}
\end{minipage}
The level function \code{insert_coord} inserts a coordinate $i_k$ into an
output level at position $p_k$ given by \code{locate}, and requires that the
level provide the locate capability.  The level function \code{insert_init}
initializes the data structures that encode an output level, while
\code{insert_finalize} performs any post-processing required after all
coordinates have been inserted.  Both take, as inputs, the sizes of the level
being initialized or finalized ($sz_k$) and its parent level ($sz_{k-1}$).  For
a level that provides the insert capability, its size is computed as a function
of its parent's size by the level function \code{size}.  For a level that
supports append, its size is the number of coordinates that have been appended.

\subsubsection*{Append Capability}

Appends coordinates to a level and is also exposed as four level functions:

\begin{minipage}{.49\linewidth}
\begin{lstlisting}[language=C++,mathescape,xleftmargin=\parindent]
append_coord(p$_k$, i$_k$) -> void
append_edges(p$_{k-1}$, pbegin$_k$, pend$_k$) -> void
\end{lstlisting}
\end{minipage}
\hfill
\begin{minipage}{.49\linewidth}
\begin{lstlisting}[language=C++,mathescape,xleftmargin=\parindent]
append_init(sz$_{k-1}$, sz$_k$) -> void
append_finalize(sz$_{k-1}$, sz$_{k}$) -> void
\end{lstlisting}
\end{minipage}
The level function \code{append_coord} appends a coordinate $i_k$ to the end of
an output level ($p_k$).  The function \code{append_edges} inserts edges that
connect all coordinates between positions \code{pbegin$_k$} and \code{pend$_k$}
to the coordinate at position $p_{k-1}$ in the previous level.  This enables 
attaching appended coordinates to the rest of the coordinate hierarchy.  In
contrast to the insert capability, the append capability requires result
coordinates to be appended in order.  \code{append_init} and
\code{append_finalize} serve identical purposes as \code{insert_init} and
\code{insert_finalize} and take the same arguments.

Following the semantics of the append capability, we can, for instance,
assemble the \code{crd} array of a CSR output matrix by repeatedly calling
\code{append_coord} as defined for compressed level types
(see~\tabref{assembly-functions}), with the coordinates of every result nonzero
as arguments.  Similarly, the \code{pos} array can be assembled with calls to
\code{append_init} at the start of the computation, \code{append_edges} after
the nonzeros of each row have been computed, and \code{append_finalize} at the
very end.

\section{Code Generation}
\label{sec:code-generation}
This section describes a code generation algorithm that emits efficient code to
compute with tensors stored in any combination of formats.  The algorithm
supports any tensor format expressible as compositions of level formats, which
includes all those described in~\secref{tensor-formats} and many others.  Our
algorithm extends the code generation technique proposed
by~\citeauthor{kjolstad2017}~\shortcite{kjolstad2017} to handle many more
disparate formats by only reasoning about capabilities and properties of level
formats.  This approach makes the complexity of our algorithm independent from
the number of formats supported.  Thus, support for new formats can be added
without modifying the code generator.

\subsection{Background}
\label{sec:background}

The code generation algorithm
of~\citeauthor{kjolstad2017}~\shortcite{kjolstad2017} takes as input a tensor
algebra expression in tensor index notation, which describes how each component
in the output is to be computed in terms of components in the operands.  Matrix
multiplication, for example, is expressed in this notation as $A_{ij}=\sum_k
B_{ik} C_{kj}$, which makes explicit that each component $A_{ij}$ in the
result is the inner product of the $i$-th row of $B$ and the
$j$-th column of $C$.  Similarly, matrix addition can be expressed as $A_{ij} =
B_{ij} + C_{ij}$.  Computing an expression in index notation requires merging
its operands---that is to say, iterating over the joint iteration space of the
operands---dimension by dimension.  For instance, code to add sparse matrices
must iterate over only rows that have nonzeros in either matrix and, for each
row, iterate over components that are nonzero in either matrix.  Additive
computations must iterate over the union of the operand nonzeros (i.e., a union
merge), while multiplicative ones must iterate over the intersection of the
operand nonzeros (i.e., an intersection merge).

\begin{figure}
  \begin{subfigure}[t]{0.32\linewidth}
    \centering
    \includegraphics[scale=0.25]{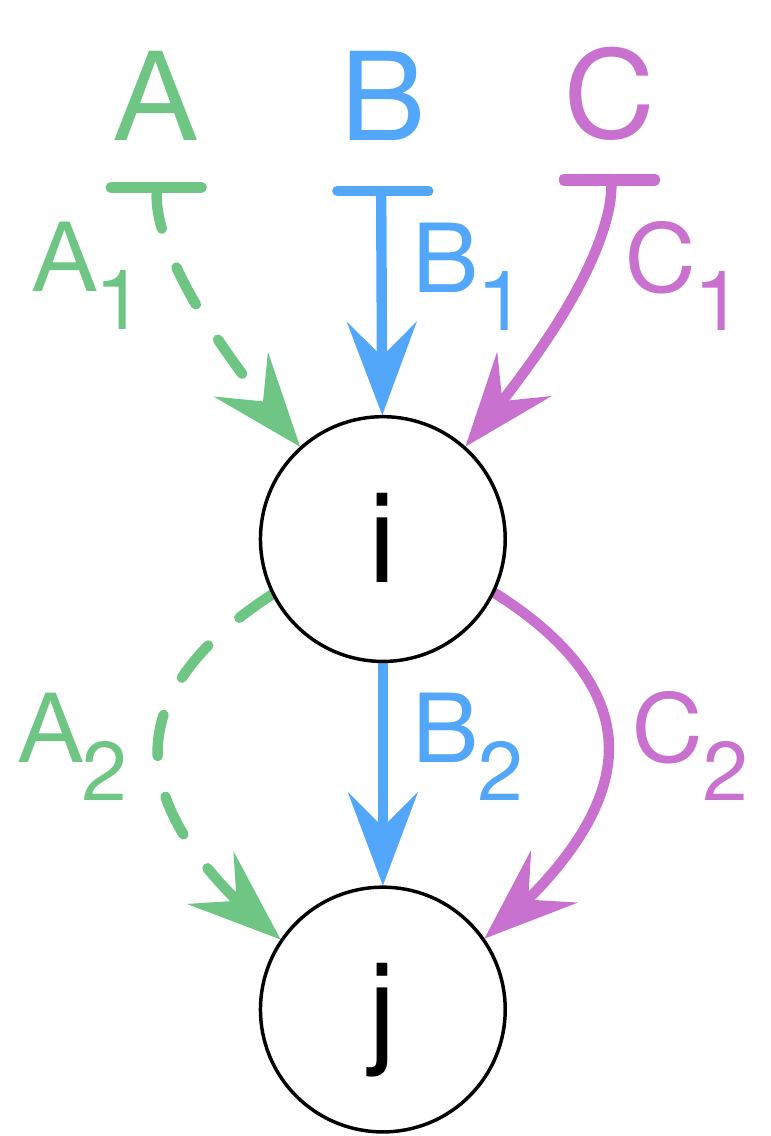}
    \caption {\label{fig:iteration-graph-matrix-add}
      Iteration graph
    }
  \end{subfigure}
  \hfill
  \begin{subfigure}[t]{0.32\linewidth}
    \centering
    \includegraphics[scale=0.112]{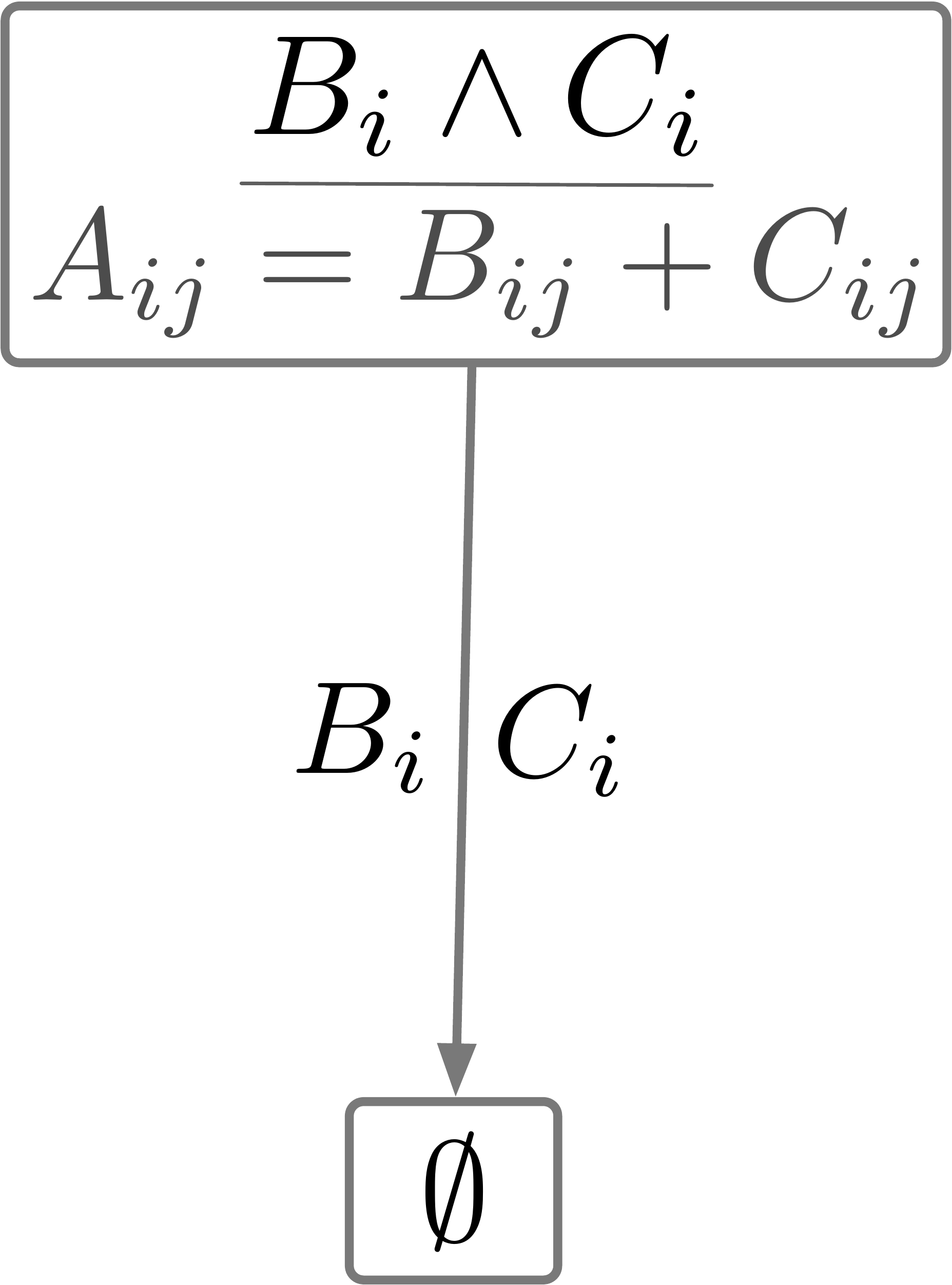}
    \caption {\label{fig:merge-lattice-matrix-add-i}
      Merge lattice for $i$
    }
  \end{subfigure}
  \hfill
  \begin{subfigure}[t]{0.34\linewidth}
    \centering
    \includegraphics[scale=0.112]{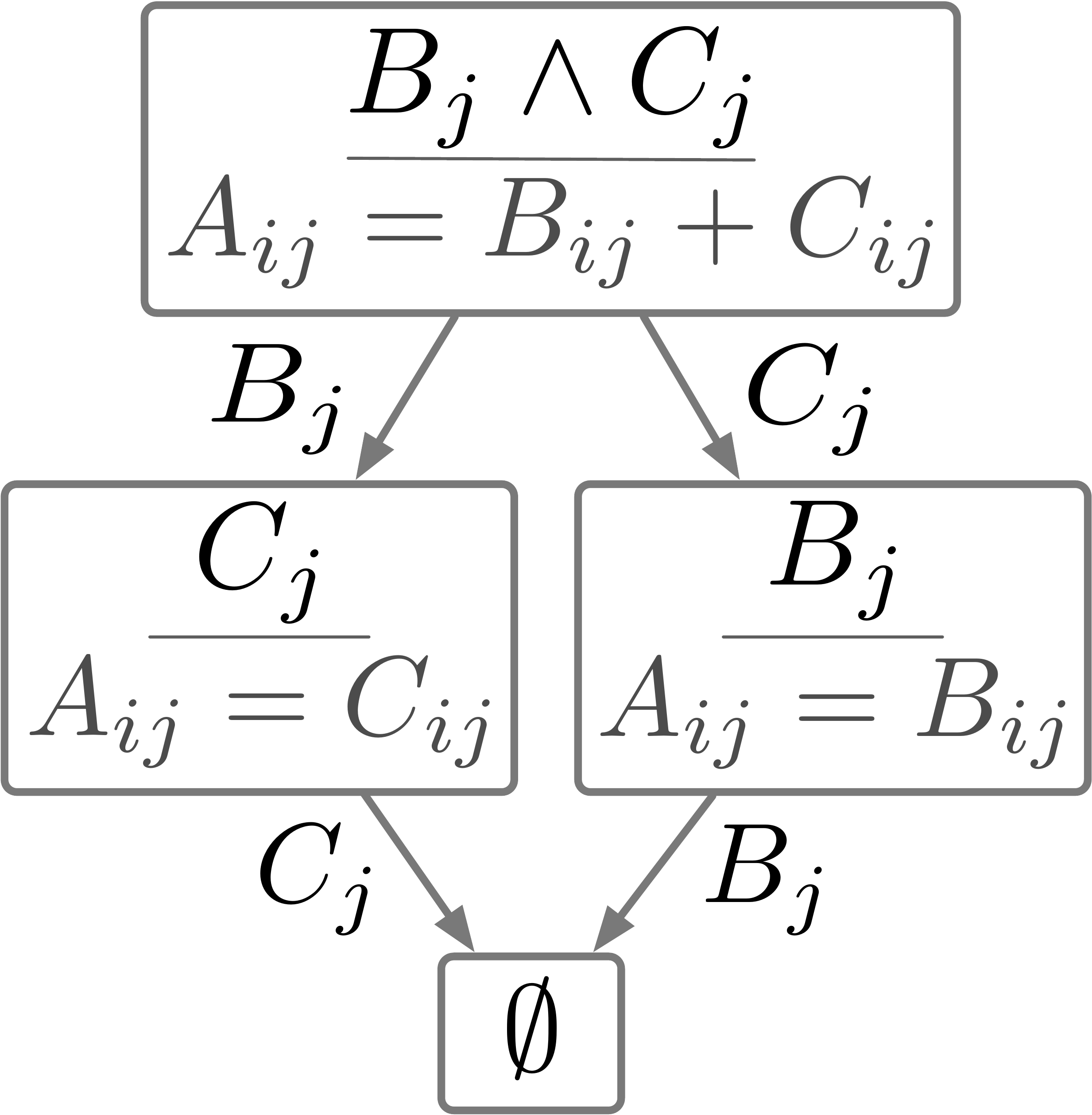}
    \caption {\label{fig:merge-lattice-matrix-add-j}
      Merge lattice for $j$
    }
  \end{subfigure}
  \caption {\label{fig:plus-iteration-graph-and-lattices}
    Iteration graph and optimized merge lattices for sparse matrix addition
    $A_{ij} = B_{ij} + C_{ij}$, where $B$ is a CSR matrix and $C$ is a COO
    matrix that is guaranteed to contain no empty row.
  }
\end{figure}

The proper order in which to iterate over dimensions of the joint iteration
space is determined from an iteration graph.  The iteration graph for a tensor
algebra expression $e$ consists of a set of index variables that appear in $e$
and a set of directed tensor paths that represent accesses into input and
output tensors.  Each tensor path connects index variables that are used in a
corresponding tensor access and is ordered based on the order of index
variables in the access expression and the order of dimensions in the accessed
tensor.  Determining the order in which to iterate over the joint iteration
space's dimensions reduces to ordering index variables into a hierarchy, such
that every tensor path edge goes from an index variable higher up to one lower
down.  As an example, \figref{iteration-graph-matrix-add} shows the iteration
graph for matrix addition with a CSR matrix and a COO matrix as inputs and a
row-major dense matrix as output.  From this iteration graph, we can determine
that computing the operation requires iterating over the row dimension before
the column dimension.

For each dimension, indexed by some index variable $v$, in the joint iteration
space, its corresponding merge lattice describes what loops are needed to fully
merge all input tensor dimensions that $v$ indexes into.  Each point in the
ordered lattice encodes a set of input tensor dimensions indexed by $v$ that
may contain nonzeros and that need to be simultaneously merged in one loop.
Each lattice point also encodes a sub-expression to be computed in the
corresponding loop.  Every path from the top lattice point to the bottom
lattice point represents a sequence of loops that might have to be executed at
runtime in order to fully merge the inputs.
\figsref{merge-lattice-matrix-add-i}{merge-lattice-matrix-add-j}, for instance,
show merge lattices for matrix addition with a CSR matrix $B$ and a COO matrix
$C$ that has no empty row.  To fully merge $B$ and $C$'s column dimensions, we
would start by running the loop that corresponds to the top lattice point
in~\figref{merge-lattice-matrix-add-j}, computing $A_{ij} = B_{ij} + C_{ij}$ in
each iteration.  This incrementally merges the two operands until one (e.g.,
$B$) has been fully merged into the output.  Then, to also fully merge the
other operand (i.e., $C$) into the output, we would only have to run the loop
that corresponds to the middle-left lattice point
in~\figref{merge-lattice-matrix-add-j}, which computes $A_{ij} = C_{ij}$ in
each iteration in order to copy the remaining nonzeros from $C$ to the output.

\subsection{Property-Based Merge Lattice Optimizations}
\label{sec:merge-lattice-optimizations}

Optimizations on merge lattices that simplify them to yield optimized code were
described by \citeauthor{kjolstad2017}~\shortcite{kjolstad2017}.  Those were,
however, all hard-coded to dense and compressed level formats.  We reformulate
the optimizations with respect to properties and capabilities of coordinate
hierarchy levels, so that they can be applied to other level types.  In
particular, given a merge lattice for index variable $v$, our algorithm removes
any lattice point that does not merge every full tensor dimension (i.e., a
dimension represented by a full coordinate hierarchy level) that $v$ indexes
into.  This is valid since full tensor dimensions are supersets of any sparse
dimension, so once we finish iterating over and merge a full dimension we must
have also visited every coordinate in the joint iteration space.  Applying
this optimization gives us the optimized merge lattice shown
in~\figref{merge-lattice-matrix-add-i}, which contains just a single lattice
point even though the computed operation requires a union merge.

Our algorithm also optimizes merging of any number of full dimensions by
emitting code to co-iterate over only those that do not support the locate
capability, with the rest accessed via calls to \code{locate}.  Depending on
whether the operands are unordered, this can reduce the complexity of the
co-iteration and thus the merge, assuming \code{locate} runs in constant time.

\subsection{Level Iterator Conversion}
\label{sec:iterator-conversion}

As we will see in~\secref{merging-hierarchy-levels}, efficient algorithms exist
for merging coordinate hierarchy levels that are ordered and unique, as well as
for intersection merges where unordered levels provide the locate capability.
\emph{Level iterator conversion} turns iterators over unordered and non-unique
levels without the locate capability into iterators with any desired
properties.  The aforementioned algorithms can then be used in conjunction to
merge any coordinate hierarchy levels.

In the rest of this subsection, we describe two types of iterator conversion,
\emph{deduplication} and \emph{reordering}, that can be composed to extend
support to any combination of coordinate hierarchy levels.  The flowchart on
the right in~\figref{merging-strategies} identifies, for each operand in an
intersection merge, the iterator conversions that are needed to merge the
operands.  Our code generation algorithm emits code to perform these necessary
iterator conversions on the fly at runtime.

\subsubsection*{Deduplication}

Duplicate coordinates complicate merging because it results in code that
repeatedly visit the same points in the iteration space.  Deduplication removes
duplicates from iterators over ordered and non-unique levels using a
deduplication loop.  Lines 5--7 in~\figref{elwise-multiplication-csr-coo} shows
an example of a deduplication loop that scans ahead and aggregates duplicate
coordinates, resulting in an iterator that enumerates unique coordinates.

\begin{figure}
  \centering
  \begin{subfigure}{0.48\linewidth}
    \centering
    \includegraphics[scale=0.6]{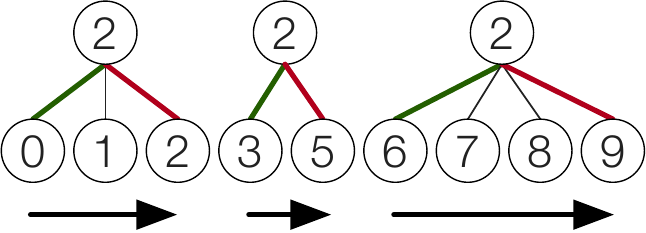}
    \caption {\label{fig:unchained-iterators}
      Separate iterators over the duplicates' children.
    }
  \end{subfigure}
  \hfill
  \begin{subfigure}{0.48\linewidth}
    \centering
    \includegraphics[scale=0.6]{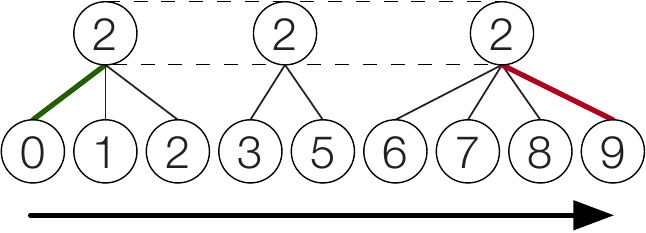}
    \caption {\label{fig:chained-iterators}
      Chained iterator over the duplicates' children.
    }
  \end{subfigure}
  \caption[Iterator chaining]{
    Iterator chaining chains the iterators over the children of duplicate
    coordinates~\subfigref{unchained-iterators} into a single iterator over all
    the children~\subfigref{chained-iterators}.  The arrows represent iterators
    with start and end bounds as green and red edges.
  }
  \label{fig:iterator-chaining}
\end{figure}

When non-unique levels are at the bottom of coordinate hierarchies, our
technique emits deduplication loops that sum the values of duplicate
coordinates.  Otherwise, the emitted deduplication loop combines the iterators
over the duplicate coordinates' children into a single iterator.  In general,
this requires a scratch array to store the child coordinates.
\figref{iterator-chaining} shows how to avoid the scratch array by chaining
together the iterators over the children.  This, however, requires that the
child level support position iteration and that the child and parent levels be
both ordered and compact.  With iterator chaining, the starting bound of the
first set of children and the ending bound of the last set of children become
the bounds of the chained iterator.  The resulting iterator provides the same
interface as a regular coordinate position iterator and can thus participate in
merging without a scratch array.  \figref{elwise-multiplication-csr-coo} shows
iterator chaining used to iterate over columns of a COO matrix.

\subsubsection*{Reordering}

A precondition for code to co-iterate over levels is that coordinates are
enumerated in order.  Reordering uses a scratch array to store an ordered copy
of an unordered level and replaces iterators over the unordered level with
iterators over the ordered copy.  The code generation algorithm can then emit
code that merges unordered levels by co-iterating over the ordered copies.

\subsection{Level Merge Code}
\label{sec:merging-hierarchy-levels}

\begin{figure}
  \centering
  \includegraphics[width=\linewidth]{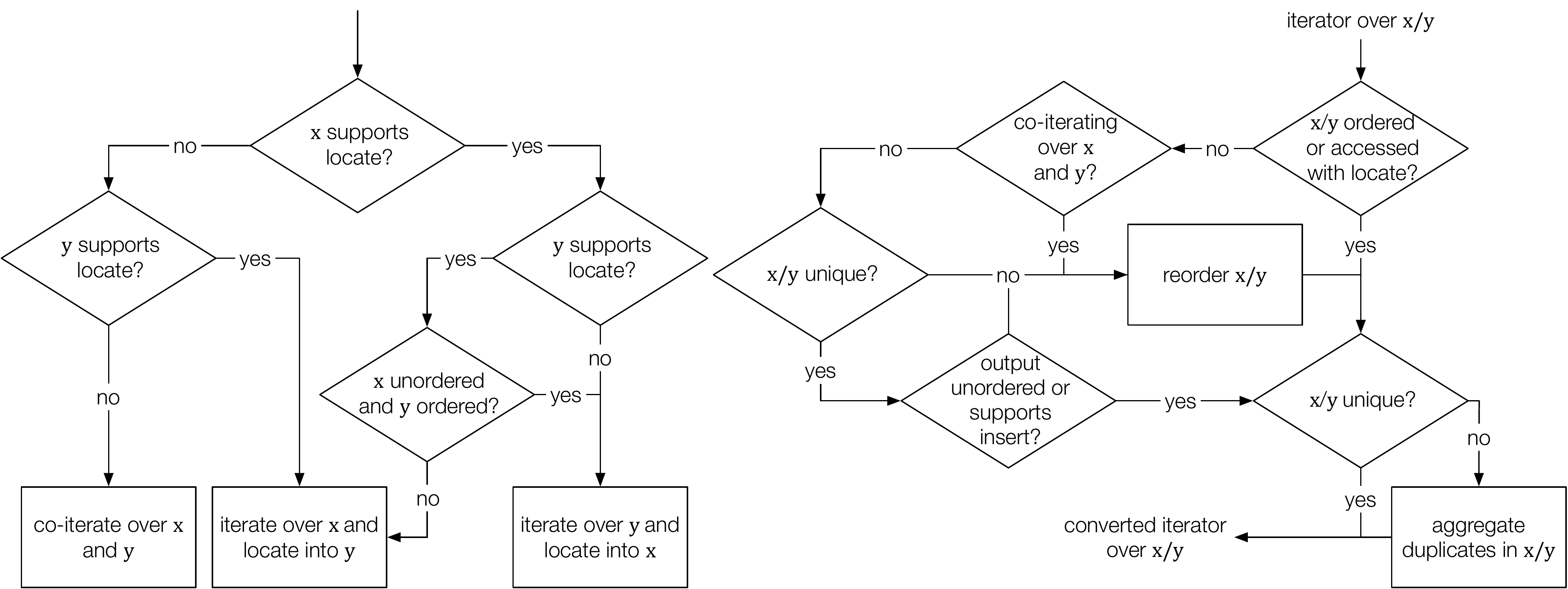}
  \caption{
    The most efficient strategies for computing the intersection merge of two
    vectors $x$ and $y$, depending on whether they support the locate
    capability and whether they are ordered and unique.  The sparsity
    structure of $y$ is assumed to not be a strict subset of the sparsity
    structure of $x$.  The flowchart on the right describes, for each
    operand, what iterator conversions are needed at runtime to compute the
    merge.
  }
  \label{fig:merging-strategies}
\end{figure}

As a result of how we defined coordinate hierarchies
in~\secref{storage-abstraction}, merging dimensions of tensor operands is
equivalent to merging the coordinate hierarchy levels that represent those
dimensions.  The most efficient method for merging levels depends on the
properties and supported capabilities of the merged levels.  Consider, for
instance, the component-wise multiplication of two vectors $x$ and $y$, which
requires iterating over the intersection of coordinate hierarchy levels that
encode their nonzeros.  \figref{merging-strategies} shows the asymptotically
most efficient strategies for computing the intersection merge depending on
whether the inputs are ordered or unique and whether they support the locate
capability.  These are the same strategies that our code generation algorithm
selects.

If neither input vector supports the locate capability, we can co-iterate over
the coordinate hierarchy levels that represent those vectors and compute a new
output component whenever we encounter nonzero components in both inputs that
share the same coordinate.  Lines 8--20
in~\figref{elwise-multiplication-csr-coo} shows another example of this method
applied to merge the column dimensions of a CSR matrix and a COO matrix.  This
method generalizes the two-way merge algorithm used in merge sort~\cite[Chapter
5.2.4]{knuth1973} to compute union or intersection merges of any number of
inputs.  Like the two-way merge algorithm though, it depends on being able to
enumerate input coordinates uniquely and in order.  Nonetheless, this method
can be used in conjunction with the level iterator conversions described
in~\secref{iterator-conversion} to merge any coordinate hierarchy levels
regardless of their properties, as~\figref{merging-strategies} demonstrates
how.

If one of the input vectors, $y$, supports the locate capability (e.g., it is a
dense array), we can instead just iterate over the nonzero components of $x$
and, for each component, locate the component with the same coordinate in $y$.
Lines 2--9 in~\figref{elwise-multiplication-csr-dense} shows another example of
this method applied to merge the column dimensions of a CSR matrix and a dense
matrix.  This alternative method reduces the merge complexity from
$O(\operatorname{nnz}(x) + \operatorname{nnz}(y))$ to
$O(\operatorname{nnz}(x))$ assuming \code{locate} runs in constant time.
Moreover, this method does not require enumerating the coordinates of $y$ in
order.  We do not even need to enumerate the coordinates of $x$ in order, as
long as there are no duplicates and we do not need to compute output components
in order (e.g., if the output supports the insert capability).  This method is
thus ideal for computing intersection merges of unordered levels.

We can generalize and combine the two methods described above to compute
arbitrarily complex merges involving unions and intersections of any number of
tensor operands.  At a high level, any merge can be computed by co-iterating
over some subset of its operands and, for every enumerated coordinate, locating
that same coordinate in all the remaining operands with calls to
\code{locate}.  Which operands need to be co-iterated can be identified
recursively from the expression $expr$ that we want to compute.  In
particular, for each subexpression $e = e_1 \, \, op \, \, e_2$ in $expr$, let
$Coiter(e)$ denote the set of operand coordinate hierarchy levels that need to
be co-iterated in order to compute $e$. If $op$ is an operation that requires a
union merge (e.g., addition), then computing $e$ requires co-iterating over all
the levels that would have to be co-iterated in order to separately compute
$e_1$ and $e_2$; in other words, $Coiter(e) = Coiter(e_1) \cup Coiter(e_2)$.
On the other hand, if $op$ is an operation that requires an intersection merge
(e.g., multiplication), then the set of coordinates of nonzeros in the result
$e$ must be a subset of the coordinates of nonzeros in either operand $e_1$ or
$e_2$.  Thus, in order to enumerate the coordinates of all nonzeros in the
result, it suffices to co-iterate over all the levels merged by just one of the
operands.  Without loss of generality, this lets us compute $e$ without having
to co-iterate over levels merged by $e_2$ that can instead be accessed with
locate; in other words, $Coiter(e) = Coiter(e_1) \cup (Coiter(e_2) \,
\backslash \, LocateCapable(e_2))$, where $LocateCapable(e_2)$ denotes the set
of levels merged by $e_2$ that support the locate capability.

\subsection{Code Generation Algorithm}
\label{sec:code-generation-algorithm}

\begin{figure}
  \begin{minipage}{0.65\linewidth}
    \centering
    \subcaptionbox{Algorithm to generate tensor algebra code.\label{fig:codegen-algorithm}}{
      \hspace*{-4mm}
      \includegraphics{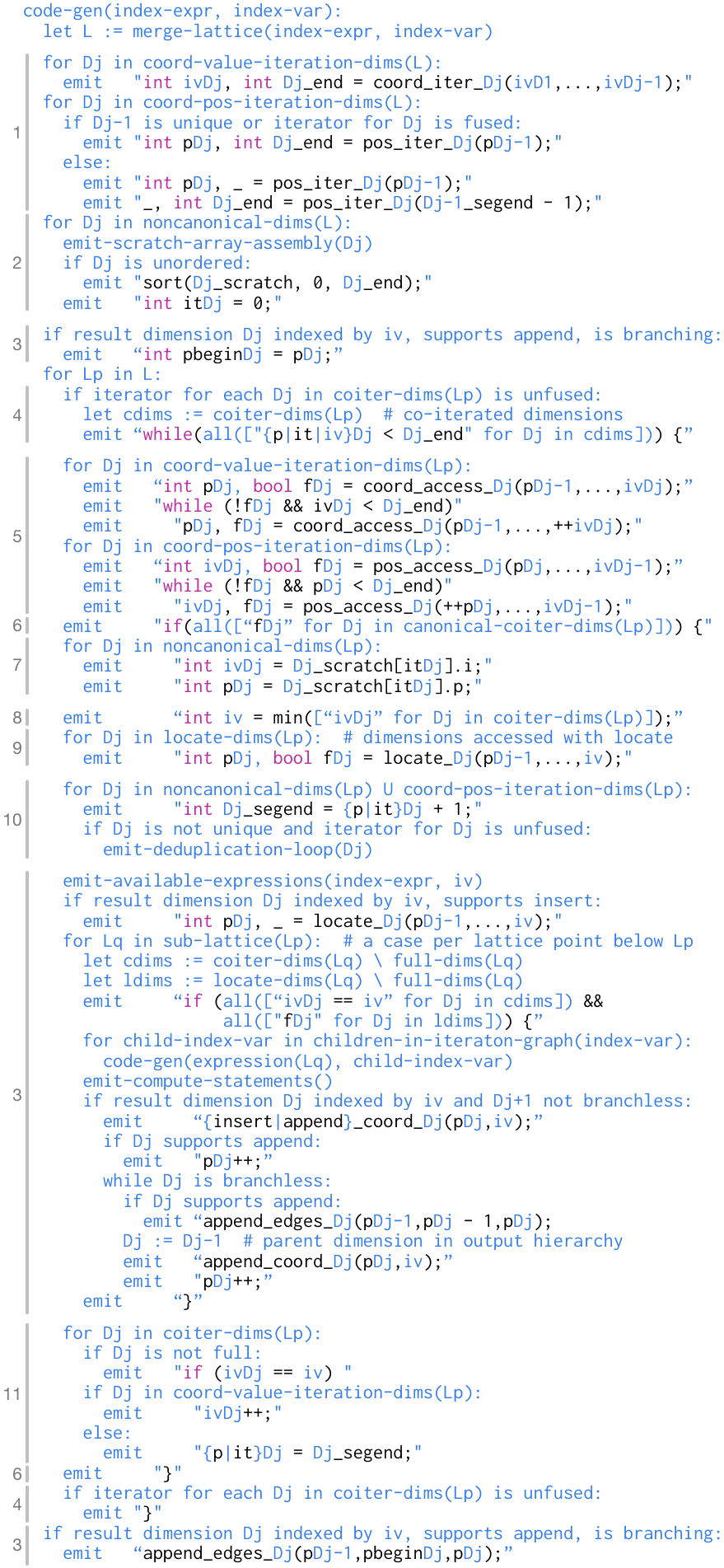}
    }
  \end{minipage}
  \hfill
  \begin{minipage}{0.34\linewidth}
    \centering
    \hspace*{-2.5mm}
    \subcaptionbox{Generated code, with level function calls inlined, for adding a CSR matrix to a COO matrix with no empty row, stored to a dense output matrix.\label{fig:codegen-example}}{
      \includegraphics{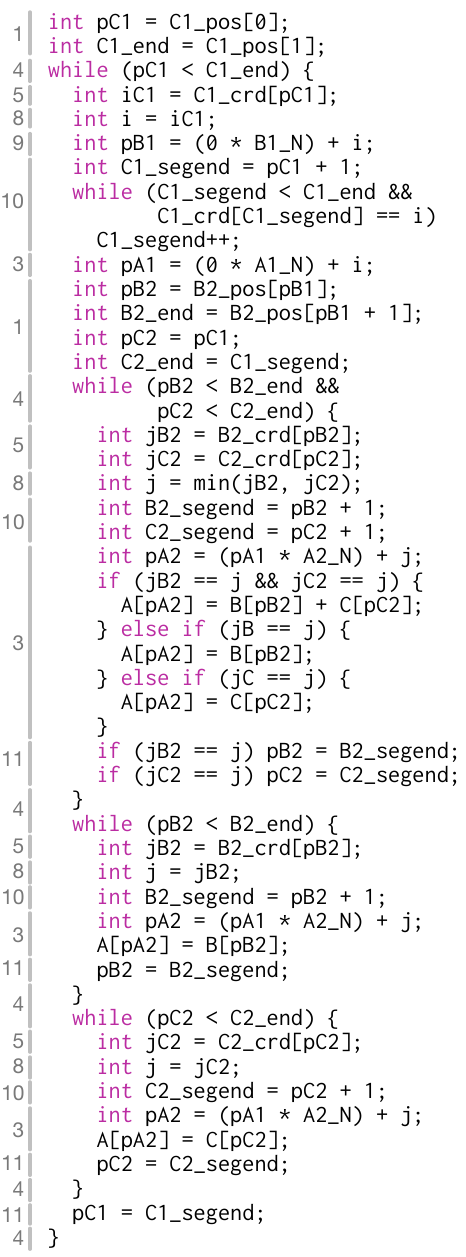}
    }
    \caption{
      Algorithm for generating code that computes tensor algebra expressions on
      operands represented as coordinate hierarchies, and an example of code it
      generates.  The sets \code{coord-value-iteration-dims} and
      \code{coord-pos-iteration-dims} exclude dimensions in
      \code{noncanonical-} \code{dims}.  Here, canonical dimensions refer to
      those that do not require a scratch array (as described
      in~\secref{iterator-conversion}) in order to be co-iterated.
    }
  \end{minipage}
\end{figure}

\figref{codegen-algorithm} shows our code generation algorithm, which
incorporates all of the concepts we presented in the previous subsections.
Each part of the algorithm is labeled from 1 to 11; throughout the discussion
of the algorithm in the rest of this section, we identify relevant parts using
these labels.  The algorithm emits code that iterates over the proper
intersections and unions of the input tensors by calling relevant access
capability level functions.  At points in the joint iteration space, the
emitted code computes result values and assembles the output tensor by calling
relevant assembly capability level functions.  The emitted code is then
specialized to compute with specific tensor formats by mechanically inlining
all level function calls.  This approach bounds the complexity of the code
generation mechanism, since it only needs to account for a finite and fixed set
of level capabilities and properties.  The result is an algorithm that supports
many disparate tensor formats and that does not need modification to add
support for more level types and tensor formats.  \figref{codegen-example}
shows an example of code that our algorithm generates, with level function
calls inlined.

Our algorithm takes as input a tensor algebra expression and recursively
calls itself on index variables in the expression, in the order given by the
corresponding iteration graph.  At each recursion level, it generates code for
one index variable \code{index-var} in the input expression.  The algorithm
begins by emitting code that initializes iterators over input coordinate
hierarchy levels, which entails calling their appropriate coordinate value or
position iteration level functions (1).  It also emits code to perform any
necessary iterator conversion described in~\secref{iterator-conversion} (1, 2).

The algorithm additionally constructs a merge lattice at every recursion level
for the corresponding index variable in the input tensor algebra expression.
This is done by applying the merge lattice construction algorithm proposed
by~\citet[Section 5.1]{kjolstad2017} and simplifying the resulting lattice with
the first optimization described in~\secref{merge-lattice-optimizations}.  For
every point $L_p$ in the simplified merge lattice, the algorithm then emits a
loop to merge the coordinate hierarchy levels that represent input tensor
dimensions that need to be merged by $L_p$ (4).  The subset of merged levels
that must be co-iterated by each loop (i.e., \code{coiter-dims($L_p$)}) is
determined by applying the recursive algorithm described
in~\secref{merging-hierarchy-levels} (with the sub-expression to be computed by
$L_p$ as input) and the second optimization described
in~\secref{merge-lattice-optimizations}.  Within each loop, the generated code
dereferences (potentially converted) iterators over the levels that must be
co-iterated (5, 7, 10), making sure to not inadvertently dereference any
iterator that has exceeded its ending bound (6).  The next coordinate to be
visited in the joint iteration space, $iv$, is then computed (8) and used to
index into the levels that can instead be accessed with the locate capability
(9).  At the end of each loop iteration, the generated code advances every
iterator that referenced the merged coordinate $iv$ (11), so that subsequent
iterations of the loop will not visit the same coordinate again.

Within each loop, the generated code must also actually compute the value of
the result tensor at each coordinate as well as assemble the output indices
(3).  The latter entails emitting code that calls the appropriate assembly
capability level functions to store result nonzeros in the output data
structures.  The algorithm emits specialized compute and assembly code for each
merge lattice point that is dominated by $L_p$, which handles the case where
the corresponding subset of inputs contain nonzeros at the same coordinate.

\paragraph*{Fusing Iterators}

By default, at every recursion level, the algorithm emits loops that iterate
over a single coordinate hierarchy level of each input tensor.  However, an
optimization that improves performance when computing with formats like COO
entails emitting code that simultaneously iterates over multiple coordinate
hierarchy levels of one tensor.  The algorithm implements this optimization by
fusing iterators over branchless levels with iterators over their preceding
levels.  This is legal as long as the fused iterators do not need to
participate in co-iteration (i.e., if the other levels to be merged can be
accessed with locate).  The algorithm then avoids emitting loops for levels
accessed by fused iterators (4), which eliminates unnecessary branching
overhead.  For some computations, however, this optimization transforms the
emitted kernel from a gather code that enumerates each result nonzero once to a
scatter code that accumulates into the output.  In such cases, the algorithm
ensures the output can also be accessed with locate.
\figref{elwise-multiplication-coo-dense} gives an example of code this
optimization generates, which iterates over two tensor dimensions with a single
loop.

\section{Evaluation}
\label{sec:evaluation}
To evaluate our contributions, we compare code that our technique generates to
five state-of-the-art sparse linear and tensor algebra libraries.  We find that
the sparse tensor algebra code our technique emits for many disparate formats
have performance competitive with hand-implemented kernels, which shows we can
get both performance and generality.  We further find that our technique's
ability to support disparate tensor formats can be crucial for performance in
practice.

\subsection{Experimental Setup}
\label{sec:experimental-setup}

We implemented our technique as an extension to the open-source
\taco{}\footnote{\protect\url{https://github.com/tensor-compiler/taco}} tensor
algebra compiler~\cite{kjolstad2017}.  To evaluate it, we used \taco{} with our
extension to generate kernels that compute various sparse linear and tensor
algebra operations with real-world application, including SpMV and MTTKRP.  We
compared the generated kernels against five other existing sparse libraries:
Intel MKL~\cite{mkl}, SciPy~\cite{scipy}, MTL4~\cite{mtl4}, the MATLAB Tensor
Toolbox~\cite{bader2007}, and TensorFlow~\cite{tensorflow}.  Intel MKL is a C
and Fortran math processing library that is heavily optimized for Intel
processors.  SciPy is a popular scientific computing library for Python.  MTL4
is a C++ library that specializes linear algebra operations for fast execution
using template metaprogramming.  The Tensor Toolbox is a MATLAB library that
implements many kernels and factorization algorithms for any-order dense and
sparse tensors.  TensorFlow is a machine learning library that supports some
basic sparse tensor operations.  We did not directly compare code generated by
our technique and the approach of~\citet{kjolstad2017}, since the two
techniques emit identical code for formats they both support.

All experiments were run on a two-socket, 12-core/24-thread 2.4 GHz Intel Xeon
E5-2695 v2 machine with 30 MB of L3 cache per socket and 128 GB of main memory,
using GCC 5.4.0 and MATLAB 2016b.  We ran each experiment between 10 (for
longer-running benchmarks) to 100 times (for shorter-running benchmarks), with
the cache cleared of input data before each run, and report average execution
times.  All results are for single-threaded execution.

We ran our experiments with real-world and synthetic tensors of varying sizes
and structures as input, inspired by similar collections of test matrices and
tensors from related works~\cite{bell2008, smith2015tensor}.
\tabref{input-summary} describes these tensors in more detail.  The real-world
tensors come from applications in many disparate domains and were obtained from
the Suite\-Sparse Matrix Collection~\cite{SparseSuite} and the FROSTT Tensor
Collection~\cite{frosttdataset}.  We stored tensor coordinates as integers and
component values as double-precision floats, except for the Tensor Toolbox's
TTM and INNERPROD kernels.  Those two kernels do not support integer
coordinates, so we evaluated them with double-precision floating-point
coordinates. 

\begin{table}
  \caption{
    Summary of matrices and tensors used in experiments.
  }
  \centering
  {\small
  \begin{tabularx}{\columnwidth}{lXlrrr}
    \toprule
    \multicolumn{1}{c}{Tensor} & \multicolumn{1}{c}{Domain} & \multicolumn{1}{c}{Dimensions} & \multicolumn{1}{c}{Nonzeros} & \multicolumn{1}{c}{Density} & \multicolumn{1}{c}{Diagonals}\\
    \midrule
    pdb1HYS & Protein data base & 36K $\times$ 36K & $\num[group-separator={,}]{4344765}$ & $\num{3e-3}$ & $\num[group-separator={,}]{25577}$ \\
    jnlbrng1 & Optimization & 40K $\times$ 40K & $\num[group-separator={,}]{199200}$ & $\num{1e-4}$ & $\num{5}$ \\
    obstclae & Optimization & 40K $\times$ 40K & $\num[group-separator={,}]{197608}$ & $\num{1e-4}$ & $\num{5}$ \\
    chem & Chemical master equation & 40K $\times$ 40K & $\num[group-separator={,}]{201201}$ & $\num{1e-4}$ & $\num{5}$ \\
    rma10 & 3D CFD & 46K $\times$ 46K & $\num[group-separator={,}]{2329092}$ & $\num{1e-3}$ & $\num[group-separator={,}]{17367}$ \\
    dixmaanl & Optimization & 60K $\times$ 60K & $\num[group-separator={,}]{299998}$ & $\num{8e-5}$ & $\num{7}$ \\
    cant & FEM/Cantilever & 62K $\times$ 62K & $\num[group-separator={,}]{4007383}$ & $\num{1e-3}$ & $\num{99}$ \\
    consph & FEM/Spheres & 83K $\times$ 83K & $\num[group-separator={,}]{6010480}$ & $\num{9e-4}$ & $\num[group-separator={,}]{13497}$ \\
    denormal & Counter-example problem & 89K $\times$ 89K & $\num[group-separator={,}]{1156224}$ & $\num{1e-4}$ & $\num{13}$ \\
    Baumann & Chemical master equation & 112K $\times$ 112K & $\num[group-separator={,}]{748331}$ & $\num{6e-5}$ & $\num{7}$ \\
    cop20k\_A & FEM/Accelerator & 121K $\times$ 121K & $\num[group-separator={,}]{2624331}$ & $\num{2e-4}$ & $\num[group-separator={,}]{221205}$ \\
    shipsec1 & FEM & 141K $\times$ 141K & $\num[group-separator={,}]{3568176}$ & $\num{2e-4}$ & $\num[group-separator={,}]{10475}$ \\
    scircuit & Circuit & 171K $\times$ 171K & $\num[group-separator={,}]{958936}$ & $\num{3e-5}$ & $\num[group-separator={,}]{159419}$ \\
    mac\_econ & Economics & 207K $\times$ 207K & $\num[group-separator={,}]{1273389}$ & $\num{9e-5}$ & $\num{511}$ \\
    pwtk & Wind tunnel & 218K $\times$ 218K & $\num[group-separator={,}]{11524432}$ & $\num{2e-4}$ & $\num[group-separator={,}]{19929}$ \\
    Lin & Structural problem & 256K $\times$ 256K & $\num[group-separator={,}]{1766400}$ & $\num{3e-5}$ & $\num{7}$ \\
    synth1 & Synthetic matrix & 500K $\times$ 500K & $\num[group-separator={,}]{1999996}$ & $\num{8e-6}$ & $\num{4}$ \\
    synth2 & Synthetic matrix & 1M $\times$ 1M & $\num[group-separator={,}]{1999999}$ & $\num{2e-6}$ & $\num{2}$ \\
    ecology1 & Animal movement & 1M $\times$ 1M & $\num[group-separator={,}]{4996000}$ & $\num{5e-6}$ & $\num{5}$ \\
    webbase & Web connectivity & 1M $\times$ 1M & $\num[group-separator={,}]{3105536}$ & $\num{3e-6}$ & $\num[group-separator={,}]{564259}$ \\
    atmosmodd & Atmospheric model & 1.3M $\times$ 1.3M & $\num[group-separator={,}]{8814880}$ & $\num{5e-6}$ & $\num{7}$ \\
    \midrule
    Facebook & Social media & 1.6K $\times$ 64K $\times$ 64K & $\num[group-separator={,}]{737934}$ & $\num{1e-7}$ & \\
    NELL-2 & Machine learning & 12K $\times$ 9.2K $\times$ 29K & $\num[group-separator={,}]{76879419}$ & $\num{2e-5}$ & \\
    NELL-1 & Machine learning & 2.9M $\times$ 2.1M $\times$ 25M & $\num[group-separator={,}]{143599552}$ & $\num{9e-13}$ & \\
    \bottomrule
  \end{tabularx}
  }
  \label{tab:input-summary}
\end{table}

\subsection{Sparse Matrix Computations}
\label{sec:sparse-matrix-computations}

Our technique generates efficient tensor algebra kernels that are specialized
to the layouts and attributes (e.g., sortedness) of the tensor operands.  In
this section, we compare the performance of \taco{}-generated kernels that
compute operations on COO, CSR, DIA, and ELL matrices with equivalent
implementations in MKL, SciPy, MTL4, and TensorFlow.

\begin{figure}
    \centering
    \hspace*{-3mm}
    \includegraphics[width=1.03\linewidth]{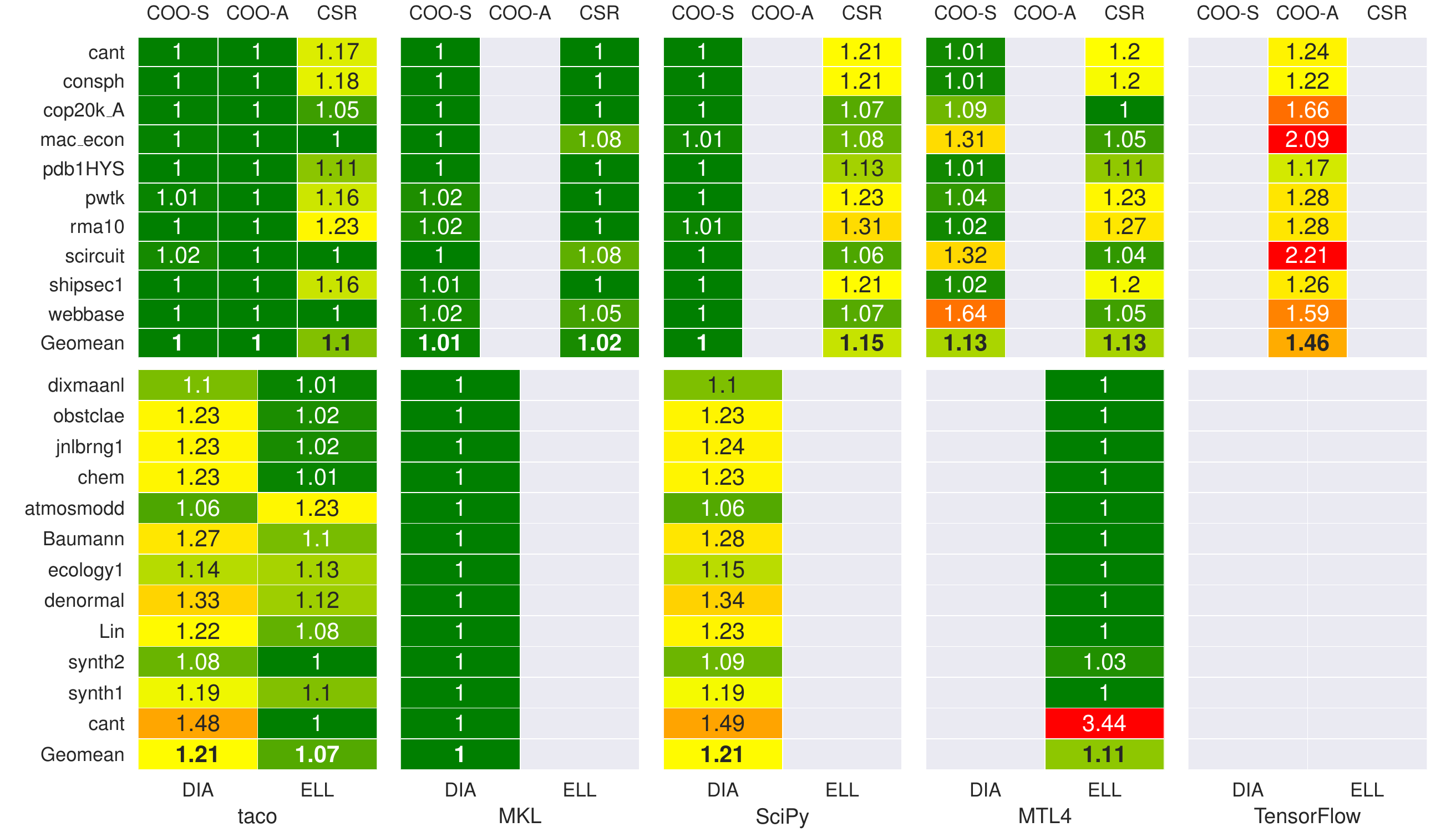}
    \caption{
      Normalized execution time of SpMV ($y = Ax$) with matrix $A$ stored in
      various formats, using \taco{} (with our extension) and other existing
      libraries.  Results are normalized to the fastest library for each matrix
      and format, and the geometric means of the results are shown in bold.
      Unlabeled cells in gray indicate a library does not support that format;
      \taco{} is the only library that supports SpMV for all the formats we
      evaluate.  COO-S and COO-A denote the struct-of-arrays and
      array-of-structs variants of COO respectively.
    }
    \label{fig:spmv-results}
\end{figure}

\figref{spmv-results} shows results for sparse matrix-vector multiplication
(SpMV), an important operation in many iterative methods for solving
large-scale linear systems from scientific and engineering
applications~\cite{bell2008}.  Our technique is the only one that supports all
the formats we survey; MTL4 does not support DIA, neither MKL nor SciPy
supports ELL, and TensorFlow only supports COO.  MKL and SciPy also only
support the struct of arrays (SoA) variant of the COO format, which is the
variant shown in~\figref{matrix-example-coo}.  TensorFlow, on the other hand,
only supports array of structs (AoS) COO, which uses a single \code{crd} array
to store coordinates for each individual tensor component contiguously in
memory.  By contrast, our technique supports both variants; supporting AoS COO
only requires defining variants of the compressed and singleton level formats
with slightly modified definitions of \code{pos_access} but the same abstract
interface.

\begin{figure}
\vspace{2.5mm}
\begin{minipage}[t]{.49\linewidth}
  \centering
  \includegraphics[width=\linewidth]{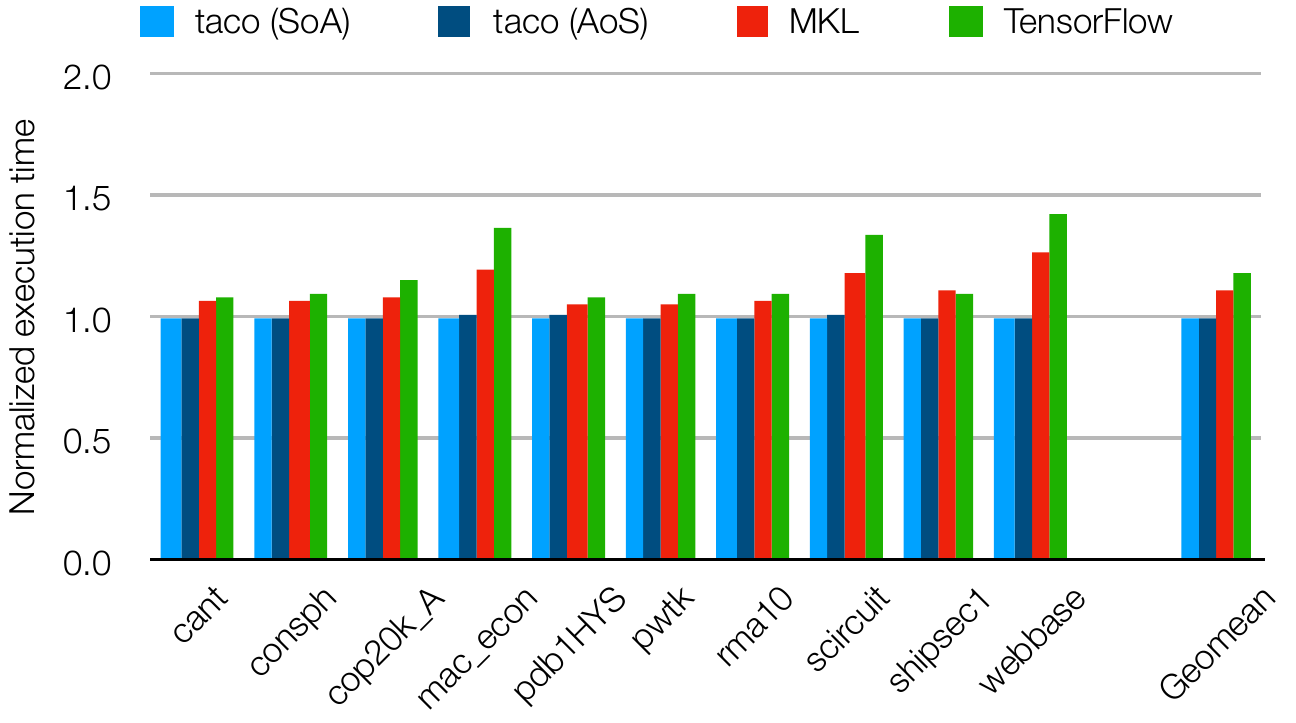}
  \captionof{figure}{
    Normalized execution time of COO SpDM ($A = BC$, where $B$ is in COO and
    $A$ and $C$ are dense matrices) with \taco{} and other libraries that
    support the operation, relative to \taco{} for each matrix.
  }
  \label{fig:coo-spdm}
\end{minipage}
\hfill
\begin{minipage}[t]{.49\linewidth}
  \centering
  \includegraphics[width=\linewidth]{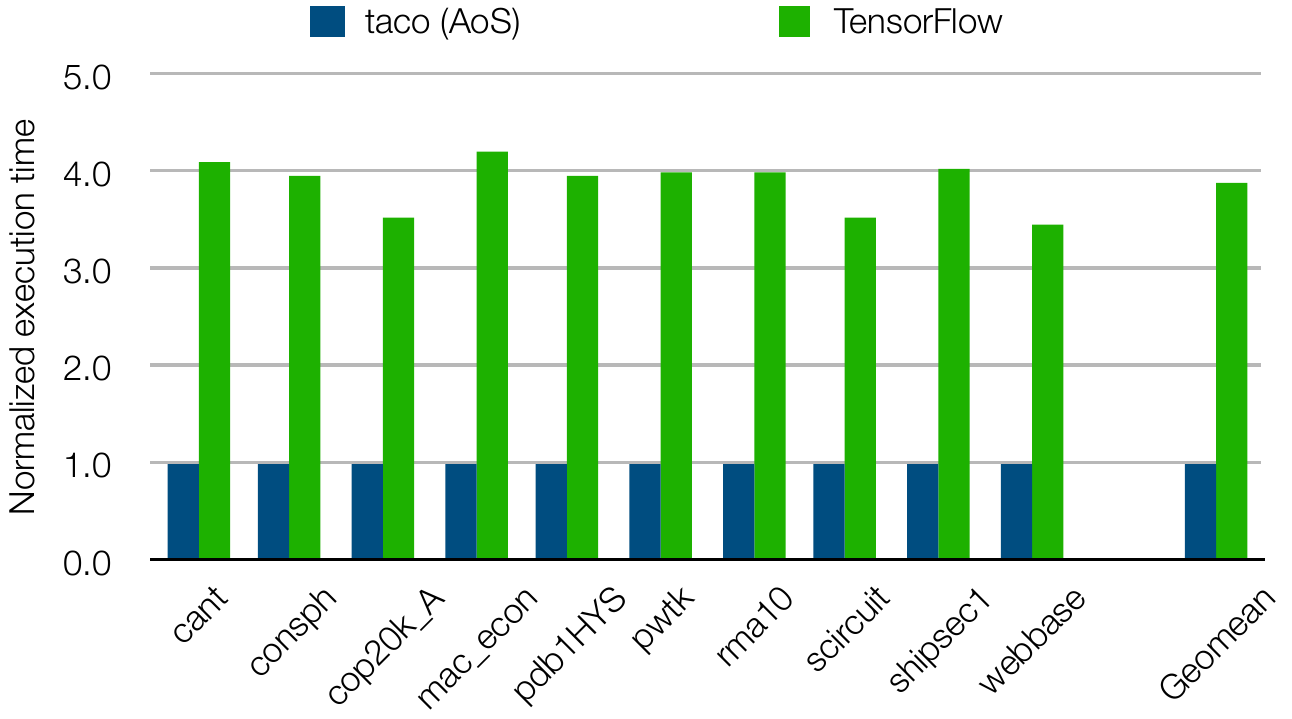}
  \captionof{figure}{
    Normalized execution time of COO matrix addition ($A = B + C$, where all
    matrices are stored in the AoS COO format) with \taco{} and TensorFlow,
    relative to \taco{} for each matrix.
  }
  \label{fig:coo-plus}
\end{minipage}
\end{figure}

\begin{wrapfigure}{R}{.49\linewidth}
  \vspace{-1.5mm}
  \centering
  \includegraphics[width=\linewidth]{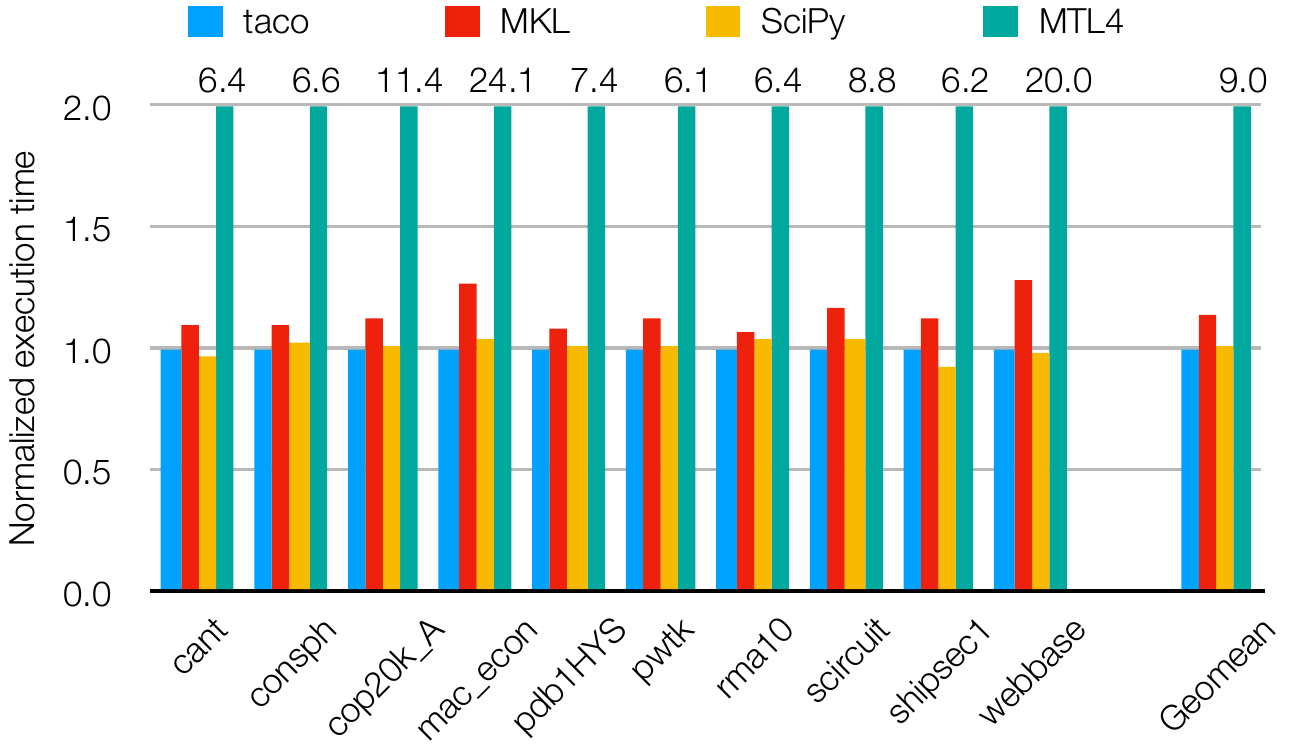}
  \caption{
    Normalized execution time of CSR matrix addition, relative to \taco{} for
    each matrix.
  }
  \label{fig:csr-plus}
\end{wrapfigure}

\figsref{coo-spdm}{coo-plus} show results for sparse matrix-dense matrix
multiplication (SpDM), another important operation in many data analytics and
machine learning applications~\cite{koanantakool2016}, and matrix addition with
sparse COO matrices.  \figref{csr-plus} shows results for CSR matrix addition.
Our technique is again the only one that supports all operations.  SciPy does
not support COO SpDM, while MTL4 only supports SpDM with sorted COO matrices.
Furthermore, only TensorFlow and \taco{} support computing COO matrix addition
with a COO output, and TensorFlow does not support CSR matrix addition.  These
omissions highlight the advantage of a compiler approach such as ours that does
not require every operation to be manually implemented.

Overall, the results show that our technique generates code that has
performance competitive with existing libraries.  For SpDM and sparse matrix
addition, \taco{}-emitted code consistently has performance equal to or better
than other libraries.  For SpMV, \taco{}-emitted code outperforms TensorFlow,
perform similar to SciPy and MTL4, and is competitive with MKL on the whole.
Even for DIA, \taco{} is only about 21\% slower than MKL on average.  These
results are explained below.

\subsubsection{COO Kernels}

The code that our technique generates to compute COO SpMV implements the same
algorithm as SciPy and MKL, and they therefore have the same performance.  MTL4
also implements this algorithm, but stores the result of the computation in a
temporary that is subsequently copied to the output.  This incurs additional
cache misses for matrices with larger dimensions, such as webbase.  TensorFlow,
on the other hand, does not implement a COO SpMV kernel and the operation must
be cast as an SpDM with a single-column matrix.  It therefore incurs overhead
because every input vector access requires a loop over its trivial column
dimension.

The COO matrix addition code that our technique emits is specialized to the
order of the tensor operands.  By contrast, TensorFlow has loops that iterate
over the coordinates of each component.  These loops let TensorFlow support
tensors of any order but again introduces unnecessary overhead.  TensorFlow's
sparse addition kernel is also hard-coded to compute with 64-bit coordinates,
whereas \taco{} can emit code with narrower-width coordinates, which reduces
memory traffic.

\subsubsection{CSR Kernels}

Code that our technique generates for CSR SpMV iterates over the rows of the
input matrix and, for each row, computes its dot product with the input vector.
SciPy and MTL4 implement the same algorithm and thus have similar performance
as \taco{}, while MKL vectorizes the dot product.  This, however, requires
vectorizing the input vector gathers, which most SIMD architectures cannot
handle very efficiently.  Thus, while this optimization is beneficial with many
of our test matrices (e.g., rma10), it is not always so (e.g., mac\_econ).

The generated code for CSR matrix addition also uses the same algorithm as
SciPy and MKL and thus has similar performance.  The emitted code exploits the
sortedness of the input matrices to enumerate result nonzeros in order.  This
lets it cheaply assemble the output \code{crd} array with appends.  By
contrast, MTL4 assigns one operand to a sparse temporary and then increments it
by the other operand.  This latter step can require significant data shuffling
to keep coordinates stored in order within the sparse temporary.  Finally,
converting the temporary back to CSR incurs yet more overhead, leading to
MTL4's poor performance.

\subsubsection{DIA and ELL SpMV}

The code that our technique generates for DIA SpMV iterates over
each matrix diagonal and, for each diagonal, accumulates the component-wise
product of the diagonal and the input vector into the result.  SciPy implements
the same algorithm that \taco{} emits and has the same performance.  MKL, by
contrast, tiles the computation to maximize cache utilization and thus
outperform both \taco{} and SciPy, particularly for large matrices with many
diagonals.  Future work includes generalizing our technique to support
iteration space tiling.

Similarly, our technique generates code for ELL SpMV that iterates over matrix
components in the order they are laid out in memory.  MTL4, on the other hand,
iterates over components row by row to maximize the cache hits for vector
accesses.  This lets MTL4 marginally outperform \taco{} for matrices with few
nonzeros per row.  When the number of nonzeros per row is large, however, this
approach reduces the cache hit rate for the matrix accesses, since components
adjacent in memory are not accessed consecutively.  The cost can be
significant, as the results for the cant matrix show.

\subsection{Sparse Tensor Computations}
\label{sec:tensor-computations}

We also compare the performance of the following higher-order tensor kernels
generated with our technique to hand-implemented kernels in the Tensor Toolbox
(TTB) and TensorFlow (TF):

\smallskip
\hspace*{-2.5mm}
\begin{minipage}[t]{.26\linewidth}
\begin{description}[style=multiline,leftmargin=0.95cm,labelindent=0cm]
    \item[TTV] $A_{ij} = \sum_{k} B_{ijk}c_{k}$
    \item[TTM] $A_{ijk} = \sum_{l} B_{ijl}C_{kl}$ 
\end{description}
\end{minipage}%
\hspace*{0.9mm}
\begin{minipage}[t]{.35\linewidth}
\begin{description}[style=multiline,leftmargin=1.65cm,labelindent=0cm]
    \item[PLUS] $A_{ijk} = B_{ijk} + C_{ijk}$ 
    \item[MTTKRP] $A_{ij} = \sum_{k,l} B_{ikl}C_{kj}D_{lj}$ 
\end{description}
\end{minipage}%
\hspace*{0.9mm}
\begin{minipage}[t]{.38\linewidth}
\begin{description}[style=multiline,leftmargin=2.2cm,labelindent=0cm]
	  \item[INNERPROD] $\alpha = \sum_{i,j,k} B_{ijk}C_{ijk}$  
\end{description}
\end{minipage}
\smallskip

\noindent
where 3rd-order tensors and the outputs of TTV, TTM, and PLUS are stored in the
COO format with ordered coordinates, while all other operands are dense.  All
these operations have real-world applications.  The TTM and MTTKRP operations,
for example, are building blocks of widely used algorithms for computing Tucker
and CP decompositions~\cite{fcoo, smith2015splatt}.  These same operations were
also evaluated in~\citet[Section 8.4]{kjolstad2017}, though that work measured
the performance of \taco{}-generated code that compute with the more efficient 
CSF format.

\begin{table*}
\setlength\tabcolsep{4.5pt}
\caption{
  Execution times of sparse COO tensor algebra kernels in milliseconds.
  Figures in parentheses are slowdowns relative to \taco{}.  A missing entry
  means a library does not support an operation, while OOM means the kernel
  runs out of memory.  NELL-2 and NELL-1 are too large for TensorFlow's
  protocol buffers.
}
{\small
\begin{tabularx}{\columnwidth}{Xlllllll}
\toprule
& \multicolumn{3}{c}{Facebook} &
\multicolumn{2}{c}{NELL-2} &
\multicolumn{2}{c}{NELL-1} \\
\cmidrule(r){2-4} \cmidrule{5-6} \cmidrule(l){7-8}
& \multicolumn{1}{c}{\taco} & \multicolumn{1}{c}{TTB} & \multicolumn{1}{c}{TF}
& \multicolumn{1}{c}{\taco} & \multicolumn{1}{c}{TTB} &
\multicolumn{1}{c}{\taco} & \multicolumn{1}{c}{TTB} \\
\midrule
TTV & 13 & 55 (4.1$\times$) & & 337 & 4797 (14.3$\times$) & 2253 & 11239 (5.0$\times$) \\
TTM & 444 & 18063 (40.7$\times$) & & 5350 & 48806 (9.1$\times$) & 56478 & OOM \\
PLUS & 37 & 539 (14.6$\times$) & 60 (1.6$\times$) & 3085 & 73380 (23.8$\times$) & 6289 & 123387 (19.6$\times$) \\
MTTKRP & 44 & 364 (8.4$\times$) & & 3819 & 43102 (11.3$\times$) & 21042 & 110502 (5.3$\times$) \\
INNERPROD & 12 & 670 (57.1$\times$) & & 416 & 82748 (199.0$\times$) & 985 & 148592 (150.9$\times$) \\
\bottomrule
\end{tabularx}
}
\label{tab:tensor-results}
\end{table*}

\tabref{tensor-results} shows the results of this experiment, with Intel MKL,
SciPy, and MTL4 omitted as they do not support sparse higher-order tensor
algebra. The Tensor Toolbox and TensorFlow are on opposite sides of the
trade-off space for hand-written sparse libraries.  The Tensor Toolbox supports
all the operations in our benchmark but has poor performance, while TensorFlow
supports only one operation but is more efficient than the Tensor Toolbox.  Our
technique, by contrast, emits efficient code for all five operations, showing
generality and performance are not mutually exclusive.

As with sparse matrix addition, the code that \taco{} with our extension
generates for adding 3rd-order COO tensors has better performance than
TensorFlow's generic sparse tensor addition kernel.  Furthermore, \taco{}
generates code that significantly outperforms the Tensor Toolbox kernels, often
by more than an order of magnitude.  This is because the Tensor Toolbox relies
on MATLAB functionalities that cannot directly operate on tensor indices or
exploit tensor properties to optimize the computation.  To add two sparse
tensors, for instance, the Tensor Toolbox computes the set of output nonzero 
coordinates by calling a MATLAB built-in function that computes the union of
the sets of input nonzero coordinates.  MATLAB's implementation of set union,
however, cannot exploit the fact that the inputs are already individually
sorted and must sort the concatenation of the two input indices.  By contrast,
\taco{} emits code that directly iterates over and merges the two input indices
without first re-sorting them, reducing the asymptotic complexity of the
computation.  Additionally, \taco{} emits code that directly assembles sparse
output indices, whereas for computations such as TTM the Tensor Toolbox stores
the results in intermediate dense structures.

\subsection{Benefits of Supporting Disparate Formats}
\label{sec:format-comparison}

\begin{table}
  \caption {
    Support for various sparse tensor formats by \taco{} with our extension
    (this work) and without (\cite{kjolstad2017}) as well as by other existing
    sparse linear and tensor algebra libraries.  \choice{} identifies tensor
    formats that our technique can support by defining additional level
    formats.
  }
  {\small
  \begin{tabularx}{\linewidth}{lXccccccc}
    \toprule
    \multicolumn{1}{c}{\multirow{2}{*}{\shortstack{Tensor Type}}} & \multicolumn{1}{c}{\multirow{2}{*}{Format}} & \multicolumn{2}{c}{\taco{}} & \multicolumn{1}{c}{\multirow{2}{*}{MKL}} & \multicolumn{1}{c}{\multirow{2}{*}{SciPy}} & \multicolumn{1}{c}{\multirow{2}{*}{MTL4}} & \multicolumn{1}{c}{\multirow{2}{*}{TTB}} & \multicolumn{1}{c}{\multirow{2}{*}{TF}} \\
    \cmidrule(lr){3-4}
    & & \multicolumn{1}{c}{This Work} & \multicolumn{1}{c}{\cite{kjolstad2017}} & & & & & \\
    \midrule
    \multirow{2}{*}{Vector} & Sparse vector & \yes & \yes & \yes & \yes & \yes & \yes & \yes \\
    & Hash map & \yes & & & \yes & & & \\
    \midrule
    \multirow{12}{*}{Matrix} & COO & \yes & & \yes & \yes & \yes & \yes & \yes \\
    & CSR & \yes & \yes & \yes & \yes & \yes & \yes & \\
    & DCSR & \yes & \yes & & & & & \\
    & ELL & \yes & & & & \yes & & \\
    & DIA & \yes & & \yes & \yes & & & \\
    & BCSR & \yes & \yes & \yes & \yes & \yes & & \\
    & CSB & \yes & & & & & & \\
    & DOK & & & & \yes & & & \\
    & LIL & & & & \yes & & & \\
    & Skyline & \choice & & \yes & & & & \\
    & Banded & \choice & & & & \yes & & \\
    \midrule
    \multirow{3}{*}{\shortstack[l]{3rd-Order \\ Tensor}} & COO & \yes & & & & & \yes & \yes \\
    & CSF & \yes & \yes & & & & & \\
    & Mode-generic & \yes & & & & & & \\
    \bottomrule
  \end{tabularx}
  }
  \label{tab:supported-formats}
\end{table}

\tabref{supported-formats} more comprehensively shows, for a wide variety of
sparse tensor formats, which formats are supported by our technique, the
approach of~\citet{kjolstad2017}, and the other existing sparse linear and
tensor algebra libraries we evaluate.  In addition to the formats described
in~\secref{tensor-formats}, we also consider all the other formats for storing
unfactorized, non-symmetric sparse tensors that are supported by at least one
existing library.  This includes DOK~\cite{dok}, which uses a hash map indexed
by both row and column coordinates to encode both dimensions of a matrix in a
shared data structure, and LIL~\cite{lil}, which uses linked lists to store the
nonzeros of each matrix row.  Additionally, the skyline
format~\cite{remington1996} is a format designed for storing variably-banded
triangular matrices, while the (sparse) banded matrix format is similar to DIA
but instead stores components in each row contiguously.  Our technique, on the
whole, supports a much larger and more diverse set of sparse tensor
formats than other existing libraries and the approach of~\citet{kjolstad2017}.
Furthermore, our technique can be extended to support the skyline and banded
matrix formats by defining additional level formats that use their data
structures.  Fully supporting DOK and LIL, however, requires extensions to the
coordinate hierarchy abstraction, which we believe is interesting future work.
To be able to emit code that randomly access DOK matrices, the locate
capability would need to effectively permit traversing multiple coordinate
hierarchy levels with a single call to \code{locate}.  Additionally, to support
LIL would require our abstraction to support storing component values
non-contiguously in memory.

\begin{figure}
\begin{minipage}[t]{.49\linewidth}
  \centering
  \includegraphics[width=\linewidth]{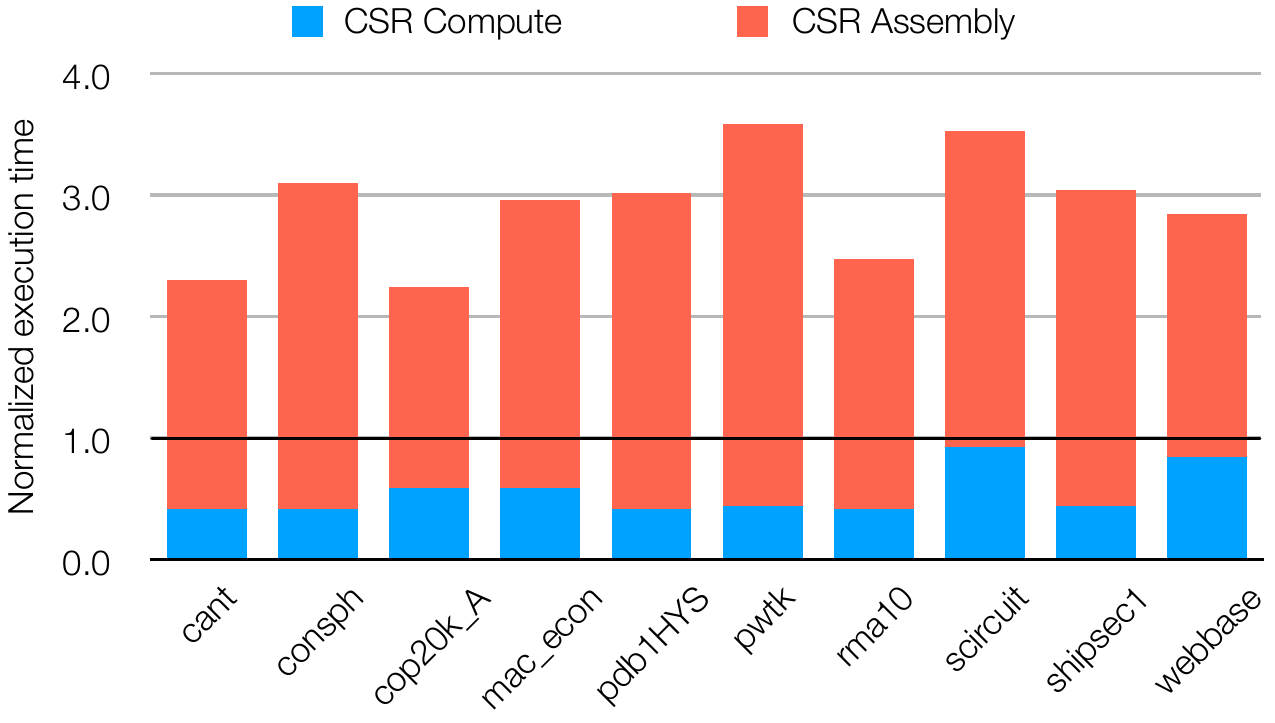}
  \captionof{figure}{
    Normalized execution time of CSR SpMV relative to COO SpMV, taking into
    account the cost of assembling CSR indices for the input matrices.  These
    results show that computing with CSR is faster than with COO (black line)
    only if the cost of assembling CSR indices can be amortized over multiple
    iterations.
  }
  \label{fig:csr-overhead}
\end{minipage}
\hfill
\begin{minipage}[t]{.49\linewidth}
  \centering
  \includegraphics[width=\linewidth]{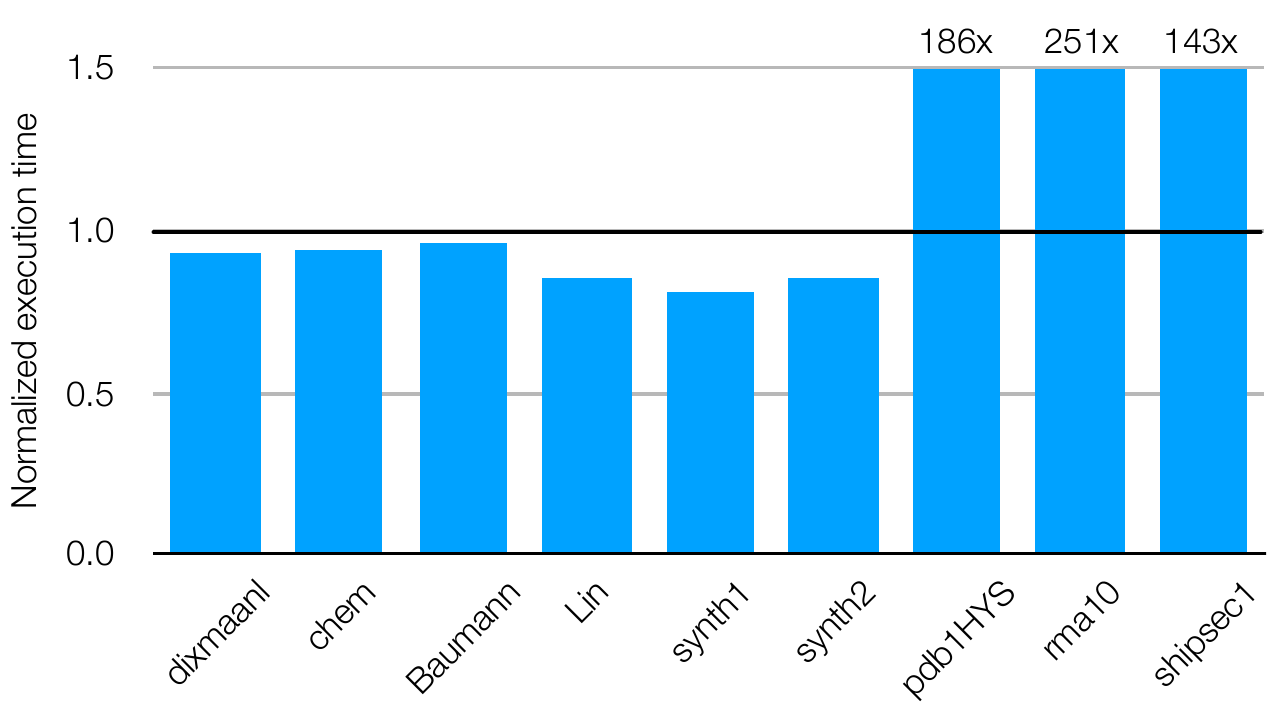}
  \captionof{figure}{
    Normalized execution time of \taco{}'s DIA SpMV kernel relative to
    \taco{}'s CSR SpMV kernel.  Storing the input matrix in the DIA format can
    accelerate SpMV if all the nonzeros in the matrix are confined to a few
    densely-filled diagonals, but can drastically degrade performance if that
    is not the case.
  }
  \label{fig:dia-gain}
\end{minipage}
\end{figure}

Our technique's support for more disparate formats can enable it to outperform
the approach of~\citet{kjolstad2017} in practical use, depending on
characteristics of the computation and data.  The COO format, for instance, is
the intuitive way to represent sparse tensors and is used by many file formats
to encode sparse tensors.  Thus, it is a natural format for importing and
exporting sparse tensors into and out of an application.  As the blue bars
in~\figref{csr-overhead} show, computing matrix-vector products directly on COO
matrices can take up to twice as much time as with CSR matrices due to higher
memory traffic.  If a matrix is imported into the application in the COO format
though, then it must be converted to a CSR matrix before the more efficient CSR
SpMV kernel can be used.  This preprocessing step incurs significant overhead
that, as the red bars in~\figref{csr-overhead} show, exceeds the cost of
computing on the original COO matrix.  For non-iterative applications that
cannot amortize this conversion overhead, our technique offers better
end-to-end performance by enabling SpMV to be computed directly on the COO
input matrix, thereby eliminating the overhead.

Which format is most performant also depends on the tensor's sparsity
structure.  To show this, we compare the performance of SpMV computed on CSR
and DIA matrices using \taco{}-generated kernels.  For matrices like Lin and
synth1 whose nonzeros are all in a few densely-filled diagonals, storing
them in DIA exposes opportunities for vectorization.  As \figref{dia-gain}
shows, our technique can exploit them to improve SpMV
performance by up to 22\% relative to the approach of~\citet{kjolstad2017},
which only supports CSR SpMV.  However, DIA SpMV performs significantly worse
for matrices like rma10 whose nonzeros are spread amongst many sparsely-filled
diagonals, since it has to unnecessarily compute with all the zeros in the
nonempty diagonals.

\begin{figure}
  \vspace{1mm}
  \centering
  \includegraphics[width=0.855\linewidth]{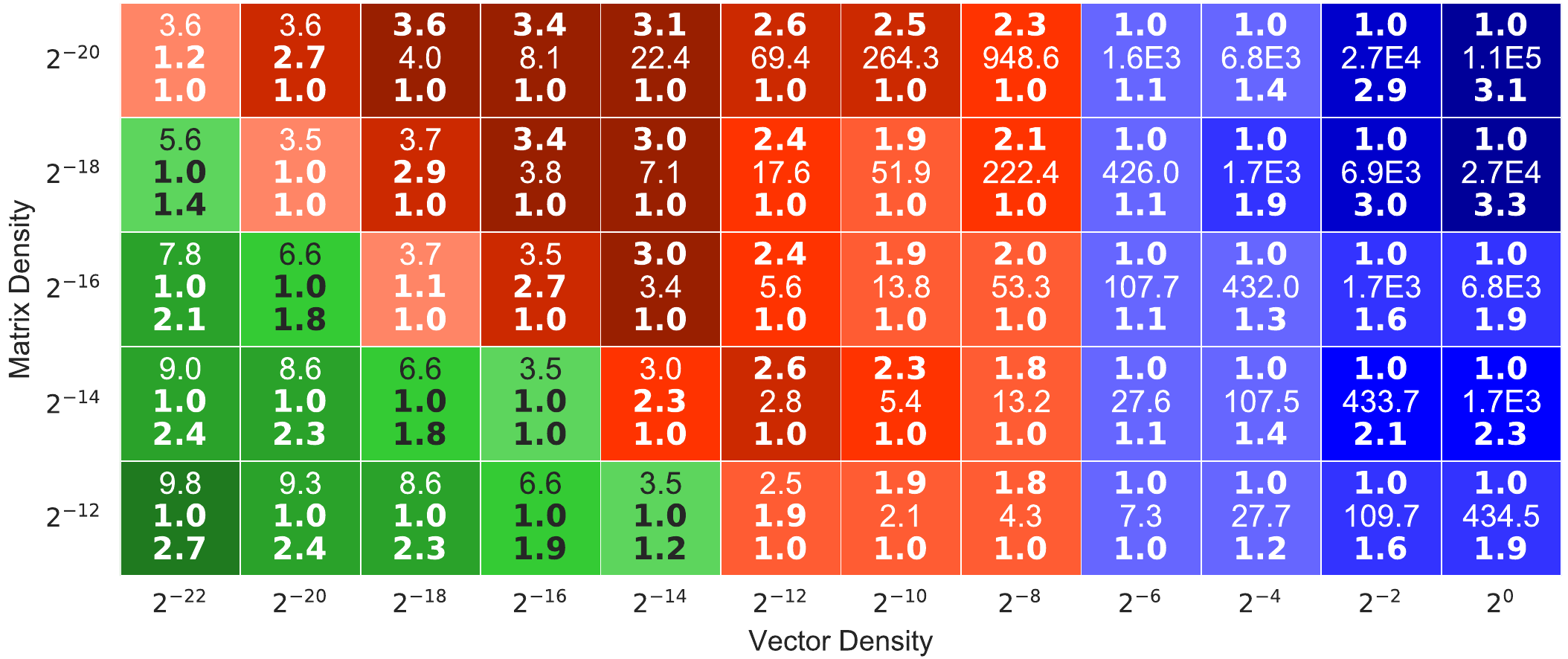}
  \caption{
    Normalized execution time of \taco{}'s CSR SpMV with inputs of varying 
    density and input vectors stored in different formats, relative to the 
    most performant format for each configuration.  Each cell shows results 
    for dense arrays (top), sparse vectors (middle), and hash maps (bottom).  
    Cells are highlighted based on which vector format is most performant 
    (blue for dense arrays, green for sparse vectors, red for hash maps).
  }
  \label{fig:vector-format-tradeoffs}
\end{figure}

We further compare the performance of CSR SpMV with input vectors stored as
dense arrays, sparse vectors, and hash maps, for operands of varying density.
\figref{vector-format-tradeoffs} shows the results.  When the input vector
contains mostly nonzeros, dense arrays are suitable for SpMV as they provide
efficient random access without needing to explicitly store coordinates of
nonzeros.  Conversely, when the input vector contains mostly zeros and the
matrix is much denser, sparse vectors offer better performance for SpMV as it
can be computed without accessing all matrix nonzeros.  However, when the input
vector is large and sparse but still denser than the matrix, computing SpMV
with hash map vectors reduces the number of accesses that go out of cache.  At
the same time, the hash map format's random access capability makes it possible
to compute SpMV without accessing the full input vector.  This enables the our
technique, with its support for hash maps, to outperform the approach
of~\citet{kjolstad2017}, which only supports the dense array and sparse vector
formats.

\section{Related Works}
\label{sec:related-works}
Related work in this area can be categorized into explorations of sparse tensor
formats and prior work on abstractions for sparse tensor storage and code
generation for sparse tensor computations.

\subsection{Sparse Tensor Formats}
\label{sec:sparse-tensor-formats}

There is a large body of work on sparse matrix and higher-order tensor formats.
Sparse matrix data structures were first introduced by
\citeauthor{csr}~\shortcite{csr}, who appear to have implemented the CSR data
structure.  \citeauthor{mcnamee1971}~\shortcite{mcnamee1971} described an early
library that supports computations on sparse matrices stored in a compressed
variant of CSR.  The analogous CSC format is also commonly used partly because
it is convenient for direct solves~\cite{davis2006}.  Many other formats for
storing sparse matrices and higher-order tensors have since been proposed;
\secref{tensor-formats} describes them in detail.  This work shows that all
these tensor formats can be represented within the same framework with just six
composable level formats that share a common interface. 

Other proposed sparse tensor formats include BICRS~\cite{bicrs}, which extends
the CSR format to efficiently store nonzeros of a sparse matrix in Hilbert
order.  For efficiently computing SpMV on GPUs,
\citeauthor{sell}~\shortcite{sell} proposed sliced ELLPACK, which generalizes
ELL by partitioning the matrix into strips of a fixed number of adjacent rows,
with each strip possibly storing a different number of nonzeros per row so as
to minimize the number of stored zeros.  \citeauthor{fcoo}~\shortcite{fcoo}
also proposed F-COO, which extends COO for enabling efficient computation of
sparse higher-order tensor kernels on GPUs.
\citeauthor{bell2008}~\shortcite{bell2008} described the HYB format, which
stores most components of a matrix in an ELL submatrix and the remaining
components in another COO submatrix.  The HYB format is useful for computing
SpMV on vector architectures with matrices that contain similar numbers of
nonzeros in most rows but have a few rows that contain many more nonzeros.
The Cocktail Format in clSpMV~\cite{clspmv}, generalizes HYB to support any
number of submatrices stored in one of nine fixed sparse matrix formats.

\subsection{Tensor Storage Abstractions and Code Generation}
\label{sec:storage-abstractions-and-code-generation}

Researchers have also explored approaches to describing sparse vector and
matrix storage for sparse linear algebra.
\citeauthor{thibault1994}~\shortcite{thibault1994} proposed a technique that
describes regular geometric partitions in arrays and automatically generates
corresponding indexing functions.  The technique compresses matrices with
regular structure, but does not generalize to unstructured matrices.

In the context of compilers for sparse linear and tensor algebra,
\citeauthor{kjolstad2017}~\shortcite{kjolstad2017} proposed a formulation for
tensor formats that designates each dimension as either dense or sparse, which
are stored using the same data structures as dense and compressed level types
in our abstraction.  However, their formulation can only describe formats that
are composed strictly of those two specific types of data structures, which
precludes their technique from generating tensor algebra kernels that compute
on many other common formats like COO and DIA.  The Bernoulli
project~\cite{bernoulli, stodghill-thesis,kotlyar-thesis}, which adopted a
relational database approach to sparse linear algebra compilation, proposed a
black-box protocol with access paths that describe how matrices map to physical
storage.  The black-box protocol is similar to our level format interface, but
they only address linear algebra and only computations involving
multiplications and not additions.  The black-box protocol also does not
support on-the-fly assembly of sparse indices, which is often essential for
applications with sparse high-order outputs.  SIPR~\cite{sipr}, a framework
that transforms dense linear algebra code to sparse code, represents sparse
vectors and matrices with hard-coded element stores that provide enumerators
and accessors that are analogous to level capabilities.  The framework provides
just two types of element store and cannot be readily extended to support new
types of element store for representing other formats.
\citeauthor{Arnold:2010:SVS:1863543.1863581}~\shortcite{Arnold:2010:SVS:1863543.1863581,ll}
proposed LL, a verifiable functional language for sparse matrix programs in
which a sparse matrix format is defined as some nesting of lists and pairs that
encode components of a dense matrix.  How an LL format should be interpreted is
described as part of the computation in LL, so the same computation with
different matrix formats can require completely different definitions.

\citeauthor{dutch1}~\shortcite{dutch1, dutch2} developed an early compiler that
transforms dense linear algebra code to equivalent sparse code by moving
nonzero guards into sparse data structures.  More recently,
\citeauthor{chill}~\shortcite{chill} proposed a technique for generating
inspector/executor code that may, at runtime, transform input matrices from one
format to another.  Both techniques support a fixed set of standard sparse
matrix formats and only generate code that work with matrices stored in those
formats.  Finally, \citeauthor{sparso}~\shortcite{sparso} proposed a technique
that discovers and exploits invariant properties of matrices in a sparse linear
algebra program to optimize the program as a whole.

Much work has also been done on compilers~\cite{lgen,btoblas} and loop
transformation
techniques~\cite{wolfe1982optimizing,wolf1991data,mckinley1996improving} for
dense linear algebra.  An early effort for dense higher-order tensor algebra
was the Tensor Contraction Engine~\cite{tensor-contraction-engine}.
libtensor~\cite{epifanovsky2013}, CTF~\cite{solomonik2014}, and
GETT~\cite{springer2016} are examples of systems and techniques that transform
tensor contractions into dense matrix multiplications by transposing tensor
operands.  TBLIS~\cite{matthews2017} and InTensLi~\cite{li2015} avoid explicit
transpositions by computing tensor contractions in-place.  All of these systems
and techniques deal exclusively with tensors stored as dense arrays.

\section{Conclusion and Future Work}
\label{sec:conclusion}
We have described and implemented a new technique for generating tensor algebra
kernels that efficiently compute on tensors stored in disparate formats.  Our
technique's modularity enables it to support many formats and to be
extended to new formats without having to modify the code generator.  This
makes our technique practical for disparate domains that need to efficiently
perform different types of computations with dissimilar data.  Large-scale
applications that need performance may also have to use multiple formats that
are each optimized for a different subcomputation.  A technique like ours that
shapes computation to data makes it possible to work with different formats
without incurring excessive data translation costs, thus optimizing
whole-program performance.

Future work includes extending our technique to support even more disparate
tensor formats, including DOK and LIL as well as custom formats designed to
take advantage of power-law structures in social graphs for graph analytics or
specialized hardware accelerator capabilities for deep learning
architectures~\cite{eyeriss}.  Additional examples include formats that exploit
structural and value symmetries, which are common in many scientific and
engineering domains, to reduce memory footprint.  Another direction of future
work is to extend the code generation algorithm to emit shared-memory parallel
code and perform further optimizations like iteration space tiling, as well as
specifically target accelerators (e.g., GPUs) and distributed memory systems.
Our modular approach enables these lines of research to be pursued
independently.

\begin{acks}
  We thank Shoaib Kamil, Vladimir Kiriansky, David Lugato, Charith Mendis,
  Yunming Zhang, and the anonymous reviewers for helpful reviews and
  suggestions.  This work was supported by the Application Driving
  Architectures (ADA) Research Center, a JUMP Center co-sponsored by SRC and
  DARPA; the Toyota Research Institute; the U.S. Department of Energy, Office
  of Science, Office of Advanced Scientific Computing Research under Award
  Numbers DE-SC0008923 and DE-SC0018121; the National Science Foundation under
  Grant No. CCF-1533753; and DARPA under Award Number HR0011-18-3-0007.  Any
  opinions, findings, and conclusions or recommendations expressed in this
  material are those of the authors and do not necessarily reflect the views of
  the funding agencies.
\end{acks}

\bibliography{paper} 

\end{document}